\documentclass{article}%
\usepackage{amsfonts}
\usepackage{amsmath}%
\usepackage{amssymb}%
\usepackage{graphicx}
\usepackage{hyperref}%
\newtheorem{theorem}{Theorem}

\newtheorem{condition}[theorem]{Condition}

\newtheorem{definition}[theorem]{Definition}

\newtheorem{proposition}[theorem]{Proposition}

\newenvironment{proof}[1][Proof]{\noindent\textbf{#1.} }{\ \rule{0.5em}{0.5em}}
\begin{document}

\title{Quantum Mechanics Revisited}
\author{Jean Claude Dutailly\\Paris (France)}
\maketitle

\begin{abstract}
The paper proposes a new approach in the foundations of Quantum Mechanics. It
does not make any assumption about the physical world, but looks at the
consequences of the formalism used in models. Whenever a system is represented
by variables which meet precise, but common, mathematical properties, one can
prove theorems which are very close to the axioms of Quantum Mechanics
(Hilbert spaces, observables, eigen values,...). It is then possible to
explore the conditions of the validity of these axioms and to give a firm
ground to the usual computations. Moreover this approach sheds a new ligth on
the issues of determism, and interacting systems.

In the third edition of this paper developments have been added about the
statistical procedures used to detect anomalies in Physical Laws.

\end{abstract}

\section{Introduction}

Quantum Physics encompasses several theories, with three distinct areas:

i) Quantum Mechanics (QM) proper, which can be expressed as a collection of
axioms, such as summarized by Weinberg :

- Physical states of a system are represented by vectors $\psi$ in a Hilbert
space $H$, defined up to a complex number (a ray in a projective Hilbert space)

- Observables are represented by Hermitian operators

- The only values that can be observed for an operator are one of its eigen
values $\lambda_{k}$ corresponding to the eigen vector $\psi_{k}$

- The probability to observe $\lambda_{k}$ if the system is in the state
$\psi$ is proportional to $\left\vert \left\langle \psi,\psi_{k}\right\rangle
\right\vert ^{2}$

- If two systems with Hilbert space $H_{1},H_{2}$ interact, the states of the
total system are represented in $H_{1}\otimes H_{2}$

and, depending on the authors, the Schr\"{o}dinger's equation.

ii) Wave Mechanics, which states that particles can behave like fields which
propagate, and conversely force fields can behave like pointwise particles.
Moreover particles are endowed with a spin. In itself it constitutes a new
theory, with the introduction of new concepts, for which QM is the natural
formalism. Actually this is essentially a theory of electromagnetism, and is
formalized in Quantum Electrodynamics (QED).

iii) The Quantum Theory of Fields (QTF) is a theory which encompasses
theoretically all the phenomena at the atomic or subatomic scale, but has been
set up mainly to deal with the other forces (weak and strong interactions) and
the organization of elementary particles. It uses additional concepts (such as
gauge fields) and formalism and computation rules (Feynman diagrams, path integrals).

I will address in this paper \textit{QM only}.

\bigskip

The status of the axioms listed above is special. They are not Physical laws,
they do not define any physical object, or physical property (if we except the
Schr\"{o}dinger's equation which is or not part of the corpus). they are
deemed valid for any system at a scale which is not even mentioned but they
are not falsifiable (how could we check that an observable is a Hermitian
operator ?). This strange status, quite unique in Science, is at the origin of
the search for interpretations, and for the same reason, makes so difficult
any sensible discussion on the topic. Actually these axioms have emerged
slowly from the practices of great physicists, kept without any change in the
last decenniums, and endorsed by the majority, mostly because it is part of
their environment.

Most of these interpretations (there are hundreds of them) look for what one
can tell \textquotedblleft physical interpretations\textquotedblright\ : the
axioms reflect the physical world, so we must explain their statements through
the properties of natural objects or phenomena, how bizarre they could seem. I
will not enter into this debate : my proposals do not assume anything about
the physical properties of the real world.

Others have proposed a different direction : these axioms come from
foundamental limitations in our capability to know the real world. The main
endeavour has been, since seminal books and articles of von Neumann and
Birkhoff, to set up a formal system in which the assertions done in Physics
can be expressed and used in the predictions of experiments, and so granting
to Physics a status which would be less speculative and more respecting of the
facts as they can actually be established. This is actually similar to what is
done in Mathematics for Arithmetic or Sets Theory. The work has been pursued,
notably by Jauch, Haag, Varadarajan and Francis in the recent years.\ An
extension which accounts for Relativity has been proposed by Wightman and has
been developed as an Axiomatic Quantum Field Theory (Haag, Araki, Halvorson,
Borchers, Doplicher, Roberts, Schroer, Fredenhagen, Buchholz, Summers,
Longo,...).\ It assumes the existence of the formalism of Hilbert space
itself, so the validity of most of the axioms, and emphasizes the role to the
algebra of operators. Since all the information which can be extracted from a
system goes through operators, it can be conceived to define the system itself
as the set of these operators. This is a more comfortable venue, as it is
essentially mathematical, which has been studied by several authors (Bratelli
and others). Recently this approach has been completed by attempts to link QM
with Information Theory, either in the framework of Quantum Computing, or
through the use of the Categories Theory.

These works share some philosophical convictions, supported with a strength
depending on the authors, but which are nonetheless present :

i) A deep mistrust with regard to realism, the idea that there is a real
world, which can be understood and described through physical concepts such as
particles, location,...At best they are useless, at worst they are misleading.

ii) A great faith in the mathematical formalism, which should ultimately
replace the concepts.

iii) The preeminence of experimentation over theories : experimental facts are
seen as the unique source of innovation, physical laws are essentially the
repeated occurrences of events whose correlation must be studied by
statistical methods, the imperative necessity to consider the conditions in
which the experiments can or cannot be made.

As any formal system, the axiomatic QM defines its own objects, which are
basically the assertions that a physicist can make from the results of
experiments (\textquotedblleft the yes-no experiments\textquotedblright\ of
Jauch), and set up a system of rules of inference according to which other
assertions can be made, with a special attention given to the possibility to
make simultaneous measures, and the fact that any measure is the product of a
statistical estimation. With the addition of some axioms, which obviously
cannot reflect any experimental work (it is necessary to introduce infinity),
the formal system is then identified, by a kind of structural isomorphism,
with the usual Hilbert space and its operators of Mathematics. And from there
the axioms of QM are deemed to be safely grounded.

One can be satisfied or not by this approach. But some remarks can be done.

In many ways this attempt is similar to the one by which mathematicians tried
to give an ultimate, consistent and logical basis to Mathematics. Their
attempt has not failed, but have shown the limits of what can be achieved :
the necessity to detach the objects of the formal system from any idealization
of physical objects, the non unicity of the axioms, and the fact that they are
justified by experience and efficiency and not by a logical necessity. The
same limits are obvious in axiomatic QM.\ If to acknowledge the role of
experience and efficiency in the foundations of the system should not be
disturbing, the pretense to enshrine them in axioms, not refutable and not
subject to verification, places a great risk to the possibility of any
evolution. And indeed the axioms have not changed for more than 50 years,
without stopping the controversies about their meaning. The unavoidable
replacement of physical concepts, identification of physical objects and their
properties, by formal and abstract objects, which is consistent with the
philosophical premises, is specially damaging in Physics.\ Because there is
always a doubt about the meaning of the objects (for instance it is quite
impossible to find the definition of a \textquotedblleft
state\textquotedblright) the implementation of the system sums up practically
to a set of \textquotedblleft generally accepted
computations\textquotedblright, it makes its learning and teaching perilous
(the Feynmann's affirmation that it cannot be understood), and eventually to
the recurring apparitions of \textquotedblleft unidentified physical
objects\textquotedblright\ whose existence is supposed to fill the gap. In
many ways the formal system has replaced the Physical Theories, that is a set
of objects, properties and behaviors, which can be intuitively identified and
understood.\ The Newton's laws of motion are successful, not only because they
can be checked, but also because it is easy to understand them.\ This is not
the case for the decoherence of the wave function...

Nevertheless, this attempt is right in looking for the origin of these axioms
in the critique (in the Kantian meaning) of the method specific to Physics.
But it is aimed at the wrong target : the concepts are not the source of the
problems, they are and will stay necessary because they make the link between
formalism and real world, and are the field in which new ideas can
germinate.\ And the solution is not in a sanctification of the experiments,
which are too diverse to be submitted to any analytical method. Actually these
attempts have missed a step, which always exists between the concepts and the
collection of data : the mathematical formalization itself, in models. Models,
because they use a precise formalism, can be easily analyzed and it is
possible to show that, indeed, they have specific properties of their own,
which do not come from the reality they represent, but from their mathematical
properties and the way they are used. The objects of an axiomatic QM, if one
wishes to establish one, are then clearly identified, without disturbing the
elaboration or the implementation of theories. The axioms can then be proven,
they can also be safely used, as we will show in this paper.

QM is about the representation of physical phenomena, and not a representation
of these phenomena (as can be Wave Mechanics, QED or QTF). It expresses
properties of the data which can be extracted from measures of physical
phenomena but not properties of physical objects. To sum up : QM is not about
how the physical world works, it is about how it looks.

\bigskip

The results presented here are theorems : they are proven, as consequence of
some basic assumptions about the mathematical properties of the models used
commonly in Physics.\ They state precise conditions, and use common
Mathematics Theorems, which can be be found in my book "Mathematics for
Theoretical Physics") which is freely available.\ They will be referred to as (Maths.XXX).

In the first section we will introduce the Hilbert space and its mandatory
tool, the Hilbertian basis.

In the second section we will introduce Observables, and their main properties.

In the third section we will see how we can understand the introduction of
probability in QM.

In the fourth section we will prove a theorem, similar to the Wigner's
theorem, about the change of variables and the use of QM with the
representation of groups.

In the fifth section we will introduce two theorems, similar to the
Schr\"{o}dinger's equation, for the evolution of systems.

In the sixth section we will consider interacting systems.

The comparison between the results and the usual axioms is done in the last section.

\newpage

\section{HILBERT\ SPACE}

\subsection{Representation of a system}

To implement a scientific law, either to check it or to use it for practical
purpose (to predict an outcome), scientists and engineers use models. A
\textbf{model} can be seen as the general representation of the law.\ It
comprises :

- a system : the area in which the system is located and the time frame during
which it is observed, the list of the objects and of their properties which
are considered

- the circumstances if they are specific (temperature, interference with the
exterior of the system,...)

- the variables representing the properties, associated each to a mathematical
object with more specific mathematical properties if necessary (a scalar can
be positive, a function can be continuous,...)

- the procedures used to collect and analyze the data, notably if statistical
methods are used.

Building and using models are a crucial part of the scientific work. Any
engineer or theoretical physicist use them, either to compute solutions of a
problem from well established laws, or to explore the consequences of more
general hypotheses. A model is a representation, usually simplified, of part
of the reality, built from concepts, assumptions and accepted laws. The
simplification helps to focus on the purpose, trading accuracy for
efficiency.\ Models provide both a framework in which to make the
computations, using some formalism in an ideal representation, and a practical
procedure to organize the collection and analysis of the data. They are the
embodiment of scientific laws, implemented in more specific circumstances, but
still with a large degree of generality which enables to transpose the results
from one realization to another. Actually most, if not all, scientific laws
can be expressed in the framework of a model.

Models use a formalism, that is a way to represent the properties in terms of
variables, which can take different values according to the specific
realizations of the model, and which are used to make computations to predict
a result. The main purpose of the formalism is efficiency, because it enables
to use rules and theorems well established in a more specific field.\ In
Physics the formalism is mathematical, but other formalisms exist (for
instance the atomic representation used in Chemistry).

The most elaborate models are in Analytic Mechanics and, indeed, they stand at
the heart of QM. A system, meaning a delimited area of space comprising
material bodies, is represented by scalar generalized coordinates $q=\left(
q_{1},...,q_{N}\right)  $ its evolution by the derivatives $q^{\prime}=\left(
q_{1}^{\prime},...,q_{N}^{\prime}\right)  $ . By extension q can be the
coordinates of a point $Q$ of some manifold M to account for additional
constraints, and then the state of the system at a given time is fully
represented by a point of the vector bundle $TM$ : $W=\left(  Q,V_{Q}\right)
$ . By mathematical transformations the derivatives $q^{\prime}$ can be
exchanged with conjugate momenta, and the state of the system is then
represented in the phase space, with a symplectic structure. But we will not
use this addition and stay at the very first step, that is the representation
of the system by $\left(  q,q^{\prime}\right)  $ .

Trouble arises when one considers the other fundamental objects of Physics :
force fields. By definition their value is defined all over the space x time.
So in the previous representation one should account, at a given time, for the
value of the fields at each point, and introduce unaccountably infinitely many
coordinates. This issue has been at the core of many attempts to improve
Analytic Mechanics.

But let us consider two facts :

- Analytic Mechanics, as it is usually used, is aimed at representing the
evolution of the system over a whole period of time $[0,T]$, as it is clear in
the Lagrangian formalism : the variable are accounted, together, for the
duration of the experiment;

- the state of the system is represented by a map $W:\left[  0,T\right]
\rightarrow\left(  Q,V_{Q}\right)  $ : the knowledge of this map sums up all
that can be said on the system, the map itself represents the state of the system.

Almost all the problems in Physics involve a model which comprises the
following :

i) a set of physical objects (material bodies or particles, force fields) in a
delimited area $\Omega$ of space x time (it can be in the classical or the
relativist framework) called the system;

ii) the state of the system is represented by a fixed finite number N of
variables $X=\left(  X_{k}\right)  _{k=1}^{N}$ which can be maps defined on
$\Omega$ , with their derivatives;

so that the state of the system is defined by a finite number of maps, which
usually belong themselves to infinite dimensional vector spaces.

And it is legitimate to substitute the maps to the coordinates in $\Omega$. We
still have infinite dimensional vector spaces, but by proceeding first to an
aggregation by maps, the vector space is more manageable, and we have some
mathematical tools to deal with it. But we need to remind the definition of a
manifold (more in Maths.15.1.1).

\subsection{Manifold}

Let M be a set, E a topological vector space, an atlas, denoted $A=\left(
O_{i},\varphi_{i},E\right)  _{i\in I}$ is a collection of :

\qquad subsets $\left(  O_{i}\right)  _{i\in I}$ of M such that $\cup_{i\in
I}O_{i}=M$ (this is a cover of M)

\qquad maps $\left(  \varphi_{i}\right)  _{i\in I}$ called \textbf{charts},
such that :

i) $\varphi_{i}:O_{i}\rightarrow U_{i}::\xi=\varphi_{i}\left(  m\right)  $ is
bijective and $\xi$ are the coordinates of M in the chart

ii) $U_{i}$ is an open subset of E

iii) $\forall i,j\in I:O_{i}\cap O_{j}\neq\varnothing:$

$\varphi_{i}\left(  O_{i}\cap O_{j}\right)  ,\varphi_{j}\left(  O_{i}\cap
O_{j}\right)  $ are open subsets of E, and there is a bijective, continuous
map, called a transition map :

$\varphi_{ij}:\varphi_{i}\left(  O_{i}\cap O_{j}\right)  \rightarrow
\varphi_{j}\left(  O_{i}\cap O_{j}\right)  $

Notice that no mathematical structure of any kind is required on M. A
topological structure can be imported on M, by telling that all the charts are
continuous, and conversely if there is a topological structure on M the charts
must be compatible with it.\ But the set M has no algebraic structure : a
combination such as $am+bm^{\prime}$ has no meaning.

Two atlas $A=\left(  O_{i},\varphi_{i},E\right)  _{i\in I},A^{\prime}=\left(
O_{j}^{\prime},\varphi_{j}^{\prime},E\right)  _{j\in J}$ of M are said to be
compatible if their union is still an atlas. Which implies that :

$\forall i\in I,j\in J:O_{i}\cap O_{j}^{\prime}\neq\varnothing:\exists
\varphi_{ij}:\varphi_{i}\left(  O_{i}\cap O_{j}^{\prime}\right)
\rightarrow\varphi_{j}^{\prime}\left(  O_{i}\cap O_{j}^{\prime}\right)  $
which is a homeomorphism

The relation $A,A^{\prime}$are compatible atlas of M, is a relation of
equivalence. A class of equivalence is a \textbf{structure of manifold} on the
set M.

The key points are :

- there can be different structures of manifold on the same set.\ On $%
\mathbb{R}
^{4}$ there are unaccountably many non equivalent structures of smooth
manifolds (this is special to $%
\mathbb{R}
^{4}$: on $%
\mathbb{R}
^{n},n\neq4$ all the smooth structures are equivalent !).

- all the interesting properties on M come from E : the dimension of M is the
dimension of E (possibly infinite); if E is a Fr\'{e}chet space we have a
Fr\'{e}chet manifold, if E is a Banach space we have a Banach manifold and
then we can have differentials, if E is a Hilbert space we have a Hilbert
manifold, but these additional properties require that the transition maps
$\varphi_{ij}$ meet additional properties.

- for many sets several charts are required (a sphere requires at least two
charts) but an atlas can have only one chart, then the manifold structure is
understood as the same point M will be defined by a set of compatible charts.

The usual, euclidean, 3 dimensional space of Physics is an affine space.\ It
has a structure of manifold, which can use an atlas with orthonormal frames,
or with curved coordinates (spherical or cylindrical). Passing from one system
of coordinates to another is a change of charts, and represented by transition
maps $\varphi_{ij}.$

\subsection{Fundamental theorem}

We will consider models which meet the following conditions:

\begin{condition}
\textit{i)} \textit{The system is represented by a fixed finite number N of
variables} $\left(  X_{k}\right)  _{k=1}^{N}$

\textit{ii) Each variable belongs to an open subset }$O_{k}$\textit{\ of a
separable Fr\'{e}chet real vector space }$V_{k}$

\textit{iii) At least one of the vector spaces }$\left(  V_{k}\right)
_{k=1}^{N}$\textit{\ is infinite dimensional}

iv) \textit{For any other model of the system using N variables} $\left(
X_{k}^{\prime}\right)  _{k=1}^{N}$ \textit{belonging to open subset }%
$O_{k}^{\prime}$\textit{\ of }$V_{k},$\textit{\ and for} $X_{k},X_{k}^{\prime
}\in O_{k}\cap O_{k}^{\prime}$ \textit{there is a continuous map :}
$X_{k}^{\prime}=\digamma_{k}\left(  X_{k}\right)  $
\end{condition}

\textbf{Remarks :}

i) The \textit{variables must be vectorial}.\ This condition is similar to the
superposition principle which is assumed in QM. This is one of the most
important condition. By this we mean that the associated physical phenomena
can be represented as vectors (or tensors, or scalars). The criterion, to
check if this is the case, is : if the physical phenomenon can be represented
by $X$ and $X^{\prime}$, does the phenomenon corresponding to any linear
combination $\alpha X+\beta X^{\prime}$ has a physical meaning ?

Are usually vectorial variables : the speed of a material point, the electric
or magnetic field, a force, a moment,...and the derivatives, which are, by
definition, vectors.

Are not usually vectorial variables : qualitative variables (which take
discrete values), a point in the euclidean space or on a circle, or any
surface. The point can be represented by coordinates, but these coordinates
are not the physical object, which is the material point. For instance in
Analytic Mechanics the coordinates $q=\left(  q_{1},...,q_{N}\right)  $ are
not a geometric quantity : usually a linear combination $\alpha q+\beta
q^{\prime}$ has no physical meaning. The issue arises because physicists are
used to think in terms of coordinates (in euclidean or relativist Lorentz
frame) which leads to forget that the coordinates are just a representation of
an object which, even in its mathematical form (a point in an affine space) is
not vectorial.

So this condition, which has a simple mathematical expression, has a deep
physical meaning : it requires to understand clearly why the properties of the
physical phenomena can be represented by a vectorial variable, and reaches the
most basic assumptions of the theory. The status, vectorial or not, of a
quantity is not something which can be decided at will by the Physicist : it
is part of the Theory which he uses to build his model.

However we will see that the addition of a variable which is not a vector can
be very useful (Theorem \ref{QMMAnifold}).

ii) The variables are assumed to be independent, in the meaning that there is
no given relation such that $\sum_{k}X_{k}=1.$ Of course usually the model is
used with the purpose to compute or check relations between the variables, but
these relations do not matter here.\ Actually to check the validity of a model
one considers all the variables, those which are given and those which can be
computed, they are all subject to measures and this is the comparison, after
the experiment, between computed values and measured values which provides the
validation.\ So in this initial stage of specification of the model there is
no distinction between the variables, which are on the same footing.

Similarly there is no distinction between variables internal and external to
the system : if the evolution of a variable is determined by the observer or
by phenomena out of the system (it is external) its value must be measured to
be accounted for in the model, so it is on the same footing as any other
variable. And it is assumed that the value of all variables can be measured
(we will come back on this point in the next section).

The derivative $\frac{dX_{k}}{dt}$ (or partial derivative at any order) of a
variable $X_{k}$ is considered as an independent variable, as it is usually
done in Analytic Mechanics and in the mathematical formalism of r-jets.

iii) The variables can be restricted to take only some range (for instance it
must be positive). The vector spaces are infinite dimensional whenever the
variables are functions. The usual case is when they represent the evolution
of the system with the time t : then $X_{k}$ is the function itself : $X_{k}:%
\mathbb{R}
\rightarrow O_{k}::X_{k}\left(  t\right)  .$ What we consider here are
variables which cover the whole evolution of the system over the time, and not
only just a snapshot $X_{k}\left(  t\right)  $ at a given time. But the
condition encompasses other cases, notably fields $F$ which are defined over a
domain $\Omega$.\ \textit{The variables are the maps} $F_{k}:\Omega\rightarrow
O_{k}$ \textit{and not their values} $F_{k}\left(  \xi\right)  $ \textit{at a
given point} $\xi\in\Omega.$

iv) A Fr\'{e}chet space is a Hausdorff, complete, topological space endowed
with a countable family of semi-norms (Maths.971). It is locally convex and
metric. This, quite complicated, mathematical definition is required because
we will prove a theorem, and as usual in Mathematics we need to be precise in
stating the conditions of its validity.

Are Fr\'{e}chet spaces :

- any Banach vector space : the spaces of bounded functions, the spaces
$L^{p}\left(  E,\mu,%
\mathbb{C}
\right)  $ of integrable functions on a measured space $\left(  E,\mu\right)
$ (Maths.2270), the spaces $L^{p}\left(  M,\mu,E\right)  $ of integrable
sections of a vector bundle (valued in a Banach E) (Maths.2276)

- the spaces of continuously differentiable sections on a vector bundle
(Maths.2310), the spaces of differentiable functions on a manifold (Maths.2314).

A topological vector space is separable if it has a dense countable subset
(Maths.590) which, for a Fr\'{e}chet space, is equivalent to be second
countable (Maths.698). A totally bounded ($\forall r>0$ there is a finite
number of balls which cover V), or a connected locally compact Fr\'{e}chet
space, is separable (Maths.702, 703). The spaces $L^{p}\left(
\mathbb{R}
^{n},dx,%
\mathbb{C}
\right)  $ of integrable functions for $1$ $\leq p<\infty$, the spaces of
continuous functions on a compact domain, are separable (Lieb).

Thus this somewhat complicated specification encompasses most of the usual cases.

A case which appears quite often in Physics is the following : maps :
$X:\Omega\rightarrow E$ from a relatively compact subset $\Omega$ of a
manifold M to a finite dimensional vector space, endowed with a norm (it is
important to notice that a definite positive scalar product is not required, a
norm suffices).\ Then the space of maps such that $\int_{\Omega}\left\Vert
X\left(  m\right)  \right\Vert \varpi\left(  m\right)  <\infty$ where $\varpi$
is a measure on M (a volume measure) is an infinite dimensional, separable,
Fr\'{e}chet space.

v) The condition iv addresses the case when the variables are defined over
connected domains.\ But it implicitly tells that any other set of variables
which represent the same phenomena are deemed compatible with the model. This
point is addressed more precisely in another section, with the change of variables.

\bigskip

The set of all potential states of the system is then given by the set
$S=\left\{  \left(  X_{k}\right)  _{k=1}^{N},X_{k}\in O_{k}\right\}  .$ If
there is some relation between the variables, stated by a physical law or
theory, its consequence is to restrict the domain in which the state of the
system will be found, but as said before we stay at the step before any
experiment, so $O_{k}$ represents the set of all possible values of $X_{k}.$

\begin{theorem}
\label{Th1Hilbert}For any system represented by a model meeting the conditions
1, there is a separable, infinite dimensional, Hilbert space H, defined up to
isomorphism, such that $\mathcal{S}$ can be embedded as an open subset
$\Omega\subset H$ which contains 0 and a convex subset.
\end{theorem}

\begin{proof}
i) Each value of the set $\mathcal{S}$ of variables defines a state of the
system, denoted $X$, belonging to the product $O=%
{\displaystyle\prod\limits_{k=1}^{N}}
O_{k}\subset V=%
{\displaystyle\prod\limits_{k=1}^{N}}
V_{k}.$ The couple $(O,X)$, together with the property iv) defines the
structure of a Fr\'{e}chet manifold M on the set $\mathcal{S}$, modelled on
the Fr\'{e}chet space $V=%
{\displaystyle\prod\limits_{k1}^{N}}
V_{k}$.\ The coordinates are the values $\left(  x_{k}\right)  _{k=1}^{N}$ of
the functions $X_{k}.$\ This manifold is infinite dimensional. Any Fr\'{e}chet
space is metric, so V is a metric space, and M is metrizable.

ii) As M is a metrizable manifold, modelled on an infinite dimensional
separable Fr\'{e}chet space, the Henderson's theorem (Henderson - corollary 5,
Maths.1386) states that it can be embedded as a open subset $\Omega$ of an
infinite dimensional separable Hilbert space $H$, defined up to isomorphism.
Moreover this structure is smooth, the set $H-\Omega$ is homeomorphic to $H$,
the border $\partial\Omega$ is homeomorphic to $\Omega$ and its closure
$\overline{\Omega}$.

iii) Translations by a field vector are isometries.\ Let us denote
$\left\langle {}\right\rangle _{H}$ the scalar product on $H$ (this is a
bilinear symmetric positive definite form). The map : $\Omega\rightarrow%
\mathbb{R}
::\left\langle \psi,\psi\right\rangle _{H}$ is bounded from below and
continuous, so it has a minimum (possibly not unique) $\psi_{0}$ in $\Omega.$
By translation of $H$ with $\psi_{0}$ we can define an isomorphic structure,
and then assume that 0 belongs to $\Omega$. There is a largest convex subset
of $H$ contained in $\Omega$, defined as the intersection of all the convex
subset contained in $\Omega$. Its interior is an open convex subset C. It is
not empty : because $0$ belongs to $\Omega$ which is open in $H$, there is an
open ball $B_{0}=(0,r)$ contained in $\Omega$.
\end{proof}

\bigskip

So the state of the system can be represented by a single vector $\psi$ in a
Hilbert space.

From a practical point of view, often V itself can be taken as the product of
Hilbert spaces, notably of square summable functions such as $L^{2}\left(
\mathbb{R}
,dt\right)  $ which are separable Hilbert spaces and then the proposition is obvious.

If the variables belong to an open $O^{\prime}$ such that $O\subset O^{\prime
}$ we would have the same Hilbert space, and an open $\Omega^{\prime}$ such
that $\Omega\subset\Omega^{\prime}.$ V is open so we have a largest open
$\Omega_{V}\subset H$ which contains all the $\Omega.$

Notice that this is a real vector space.

The interest of Hilbert spaces lies with Hilbertian basis, and we now see how
to relate such basis of H with a basis of the vector space V. It will enable
us to show a linear chart of the manifold M.

\subsection{Basis}

\begin{theorem}
\label{Th2Basis}For any basis $\left(  e_{i}\right)  _{i\in I}$ of V contained
in $O$, there are unique families $\left(  \varepsilon_{i}\right)  _{i\in
I},\left(  \phi_{i}\right)  _{i\in I}$ of independent vectors of H, a linear
isometry $\Upsilon:V\rightarrow H$ such that :

$\forall X\in O:\Upsilon\left(  X\right)  =\sum_{i\in I}\left\langle \phi
_{i},\Upsilon\left(  X\right)  \right\rangle _{H}\varepsilon_{i}\in\Omega$

$\forall i\in I:\varepsilon_{i}=\Upsilon\left(  e_{i}\right)  $

$\forall i,j\in I:\left\langle \phi_{i},\varepsilon_{j}\right\rangle
_{H}=\delta_{ij}$

and $\Upsilon$ is a compatible chart of M$.$
\end{theorem}

\begin{proof}
i) Let $\left(  e_{i}\right)  _{i\in I}$ be a basis of V such that $e_{i}\in
O$ and $V_{0}=Span\left(  e_{i}\right)  _{i\in I}$. Thus $O\subset V_{0}.$

Any vector of $V_{0}$ reads : $X=\sum_{i\in I}x_{i}e_{i}$ where only a finite
number of $x_{i}$ are non null. Or equivalently the following map is bijective :

$\pi_{V}:V_{0}\rightarrow%
\mathbb{R}
_{0}^{I}::\pi_{V}\left(  \sum_{i\in I}x_{i}e_{i}\right)  =x=\left(
x_{i}\right)  _{i\in I}$

where the set $%
\mathbb{R}
_{0}^{I}\subset%
\mathbb{R}
^{I}$ is the subset of maps $I\rightarrow%
\mathbb{R}
$ such that only a finite number of components $x_{i}$ are non null.

$(O,X)$ is an atlas of the manifold M and M is embedded in $H$, let us denote
$\Xi:O\rightarrow\Omega$ a homeomorphism accounting for this embedding.

The inner product on $H$ defines a positive kernel :

$K:H\times H\rightarrow%
\mathbb{R}
::K\left(  \psi_{1},\psi_{2}\right)  =\left\langle \psi_{1},\psi
_{2}\right\rangle _{H}$

Then $K_{V}:O\times O\rightarrow%
\mathbb{R}
::K_{V}\left(  X,Y\right)  =K\left(  \Xi\left(  X\right)  ,\Xi\left(
Y\right)  \right)  $ defines a positive kernel on $O$ (Math.1196).

$K_{V}$ defines a definite positive symmetric bilinear form on $V_{0}$,
denoted $\left\langle {}\right\rangle _{V},$ by :

$\left\langle \sum_{i\in I}x_{i}e_{i},\sum_{i\in I}y_{i}e_{i}\right\rangle
_{V}=\sum_{i,j\in I}x_{i}y_{j}K_{ij}$ with $K_{ij}=K_{V}\left(  e_{i}%
,e_{j}\right)  $

which is well defined because only a finite number of monomials $x_{i}y_{j}$
are non null. It defines a norm on $V_{0}.$

ii) Let : $\varepsilon_{i}=\Xi\left(  e_{i}\right)  \in\Omega$ and
$H_{0}=Span\left(  \varepsilon_{i}\right)  _{i\in I}$ the set of finite linear
combinations of vectors $\left(  \varepsilon_{i}\right)  _{i\in I}.$ It is a
vector subspace (Math.901) of H. The family $\left(  \varepsilon_{i}\right)
_{i\in I}$ is linearly independent, because, for any finite subset J of I, the determinant

$\det\left[  \left\langle \varepsilon_{i},\varepsilon_{j}\right\rangle
_{H}\right]  _{i,j\in J}=\det\left[  K_{V}\left(  e_{i},e_{j}\right)  \right]
_{i,j\in J}\neq0.$

Thus $\left(  \varepsilon_{i}\right)  _{i\in I}$ is a non Hilbertian basis of
$H_{0}.$

$H_{0}$ can be defined similarly by the bijective map :

$\pi_{H}:H_{0}\rightarrow%
\mathbb{R}
_{0}^{I}::\pi_{H}\left(  \sum_{i\in I}y_{i}\varepsilon_{i}\right)  =y=\left(
y_{i}\right)  _{i\in I}$

iii) By the Gram-Schmidt procedure (which works for infinite sets of vectors)
it is always possible to built an orthonormal basis $\left(  \widetilde
{\varepsilon}_{i}\right)  _{i\in I}$ of $H_{0}$ starting with the vectors
$\left(  \varepsilon_{i}\right)  _{i\in I}$ indexed on the same set I (as H is
separable I can be assimilated to $%
\mathbb{N}
).$

$\ell^{2}\left(  I\right)  \subset%
\mathbb{R}
^{I}$ is the set of families $y=\left(  y_{i}\right)  _{i\in I}\subset%
\mathbb{R}
^{I}$ such that :

$\sup\left(  \sum_{i\in J}\left(  y_{i}\right)  ^{2}\right)  <\infty$ for any
countable subset J of I.

$%
\mathbb{R}
_{0}^{I}\subset\ell^{2}\left(  I\right)  $

The map : $\chi:\ell^{2}\left(  I\right)  \rightarrow H_{1}::\chi\left(
y\right)  =\sum_{i\in I}y_{i}\widetilde{\varepsilon}_{i}$ is an isomorphism to
the closure $H_{1}=\overline{Span\left(  \widetilde{\varepsilon}_{i}\right)
_{i\in I}}=\overline{H_{0}}$ of $H_{0}$ in $H$ (Math.1121). $H_{1}$ is a
closed vector subspace of $H$, so it is a Hilbert space. The linear span of
$\left(  \widetilde{\varepsilon}_{i}\right)  _{i\in I}$ is dense in $H_{1}$,
so it is a Hilbertian basis of $H_{1}$ (Math.1122).

Let $\pi:H\rightarrow H_{1}$ be the orthogonal projection on $H_{1}:\left\Vert
\psi-\pi\left(  \psi\right)  \right\Vert _{H}=\min_{u\in H_{1}}\left\Vert
\psi-u\right\Vert _{H}$ then :

$\psi=\pi\left(  \psi\right)  +o\left(  \psi\right)  $ with $o\left(
\psi\right)  \in H_{1}^{\perp}$ which implies : $\left\Vert \psi\right\Vert
^{2}=\left\Vert \pi\left(  \psi\right)  \right\Vert ^{2}+\left\Vert o\left(
\psi\right)  \right\Vert ^{2}$

There is a open convex subset, containing $0$, which is contained in $\Omega$
so there is $r>0$ such that :

$\left\Vert \psi\right\Vert <r\Rightarrow\psi\in\Omega$ and as $\left\Vert
\psi\right\Vert ^{2}=\left\Vert \pi\left(  \psi\right)  \right\Vert
^{2}+\left\Vert o\left(  \psi\right)  \right\Vert ^{2}<r^{2}$

then $\left\Vert \psi\right\Vert <r\Rightarrow\pi\left(  \psi\right)
,o\left(  \psi\right)  \in\Omega$

$o\left(  \psi\right)  \in H_{1}^{\perp},H_{0}\subset H_{1}\Rightarrow
o\left(  \psi\right)  \in H_{0}^{\perp}$

$\Rightarrow\forall i\in I:\left\langle \varepsilon_{i},o\left(  \psi\right)
\right\rangle _{H}=0=K_{V}\left(  \Xi^{-1}\left(  \varepsilon_{i}\right)
,\Xi^{-1}\left(  o\left(  \psi\right)  \right)  \right)  =K_{V}\left(
e_{i},\Xi^{-1}\left(  o\left(  \psi\right)  \right)  \right)  $

$\Rightarrow\Xi^{-1}\left(  o\left(  \psi\right)  \right)  =0\Rightarrow
o\left(  \psi\right)  =0$

$H_{1}^{\perp}=0$ thus $H_{1}$ is dense in $H$ (Math.1115), and as it is
closed : $H_{1}=H$

$\left(  \widetilde{\varepsilon}_{i}\right)  _{i\in I}$ is a Hilbertian basis
of $H$ and

$\forall\psi\in H:\psi=\sum_{i\in I}\left\langle \widetilde{\varepsilon}%
_{i},\psi\right\rangle _{H}\widetilde{\varepsilon}_{i}$ with $\sum_{i\in
I}\left\vert \left\langle \widetilde{\varepsilon}_{i},\psi\right\rangle
_{H}\right\vert ^{2}<\infty$

$\Leftrightarrow\left(  \left\langle \widetilde{\varepsilon}_{i}%
,\psi\right\rangle _{H}\right)  _{i\in I}\in\ell^{2}\left(  I\right)  $

$H_{0}$ is the interior of H, it is the union of all open subsets contained in
H, so $\Omega\subset H_{0}$

$H_{0}=Span\left(  \left(  \widetilde{\varepsilon}_{i}\right)  _{i\in
I}\right)  $ thus the map :

$\widetilde{\pi}_{H}:H_{0}\rightarrow%
\mathbb{R}
_{0}^{I}::\widetilde{\pi}_{H}\left(  \sum_{i\in I}\widetilde{y}_{i}%
\widetilde{\varepsilon}_{i}\right)  =\widetilde{y}=\left(  \widetilde{y}%
_{i}\right)  _{i\in I}$

is bijective and : $\widetilde{\pi}_{H}\left(  H_{0}\right)  =\widetilde
{R}_{0}\subset%
\mathbb{R}
_{0}^{I}\subset\ell^{2}\left(  I\right)  $

Moreover : $\forall\psi\in H_{0}:\widetilde{\pi}_{H}\left(  \psi\right)
=\left(  \left\langle \widetilde{\varepsilon}_{i},\psi\right\rangle
_{H}\right)  _{i\in I}\in%
\mathbb{R}
_{0}^{I}$

Thus :

$\forall X\in O:\Xi\left(  X\right)  =\sum_{i\in I}\left\langle \widetilde
{\varepsilon}_{i},\Xi\left(  X\right)  \right\rangle _{H}\widetilde
{\varepsilon}_{i}\in\Omega$

and $\widetilde{\pi}_{H}\left(  \Xi\left(  X\right)  \right)  =\left(
\left\langle \widetilde{\varepsilon}_{i},\Xi\left(  X\right)  \right\rangle
_{H}\right)  _{i\in I}\in\widetilde{R}_{0}$

$\forall i\in I,e_{i}\in O\Rightarrow\Xi\left(  e_{i}\right)  =\varepsilon
_{i}=\sum_{j\in I}\left\langle \widetilde{\varepsilon}_{j},\varepsilon
_{i}\right\rangle _{H}\widetilde{\varepsilon}_{j}$

and $\widetilde{\pi}_{H}\left(  \varepsilon_{i}\right)  =\left(  \left\langle
\widetilde{\varepsilon}_{j},\varepsilon_{i}\right\rangle _{H}\right)  _{j\in
I}\in\widetilde{R}_{0}$

iv) Let be : $\widetilde{e}_{i}=\Xi^{-1}\left(  \widetilde{\varepsilon}%
_{i}\right)  \in V_{0}$ and $%
\mathcal{L}%
_{V}\in GL\left(  V_{0};V_{0}\right)  ::%
\mathcal{L}%
_{V}\left(  e_{i}\right)  =\widetilde{e}_{i}$

We have the following diagram :

\bigskip%

\begin{tabular}
[c]{cccccc}
& $\Xi$ &  & $%
\mathcal{L}%
_{H}^{-1}$ &  & \\
$e_{i}$ & $\rightarrow$ & $\varepsilon_{i}$ & $\rightarrow$ & $\widetilde
{\varepsilon}_{i}$ & \\
& $\searrow$ &  &  & $\downarrow$ & \\
& $%
\mathcal{L}%
_{V}$ & $\searrow$ &  & $\downarrow$ & $\Xi^{-1}$\\
&  &  & $\searrow$ & $\downarrow$ & \\
&  &  &  & $\widetilde{e}_{i}$ &
\end{tabular}

\bigskip

$\left\langle \widetilde{e}_{i},\widetilde{e}_{j}\right\rangle _{V}%
=\left\langle \Xi\left(  \widetilde{e}_{i}\right)  ,\Xi\left(  \widetilde
{e}_{j}\right)  \right\rangle _{H}=\left\langle \widetilde{\varepsilon}%
_{i},\widetilde{\varepsilon}_{j}\right\rangle _{H}=\delta_{ij}$

So $\left(  \widetilde{e}_{i}\right)  _{i\in I}$is an orthonormal basis of
$V_{0}$ for the scalar product $K_{V}$

$\forall X\in V_{0}:X=\sum_{i\in I}\widetilde{x}_{i}\widetilde{e}_{i}%
=\sum_{i\in I}\left\langle \widetilde{e}_{i},X\right\rangle _{V}\widetilde
{e}_{i}$ and $\left(  \left\langle \widetilde{e}_{i},X\right\rangle
_{V}\right)  _{i\in I}\in%
\mathbb{R}
_{0}^{I}$

The coordinates of $X\in O$ in the basis $\left(  \widetilde{e}_{i}\right)
_{i\in I}$ are $\left(  \left\langle \widetilde{e}_{i},X\right\rangle
_{V}\right)  _{i\in I}\in%
\mathbb{R}
_{0}^{I}$

The coordinates of $\Xi\left(  X\right)  \in H_{0}$ in the basis $\left(
\widetilde{\varepsilon}_{i}\right)  _{i\in I}$ are $\left(  \left\langle
\widetilde{\varepsilon}_{i},\Xi\left(  X\right)  \right\rangle _{H}\right)
_{i\in I}\in%
\mathbb{R}
_{0}^{I}$

$\left\langle \widetilde{\varepsilon}_{i},\Xi\left(  X\right)  \right\rangle
_{H}=\left\langle \Xi\left(  \widetilde{e}_{i}\right)  ,\Xi\left(  X\right)
\right\rangle _{H}=\left\langle \widetilde{e}_{i},X\right\rangle _{V}$

Define the maps :

$\widetilde{\pi}_{V}:V_{0}\rightarrow%
\mathbb{R}
_{0}^{I}::\widetilde{\pi}_{V}\left(  \sum_{i\in I}\widetilde{x}_{i}%
\widetilde{e}_{i}\right)  =\widetilde{x}=\left(  \widetilde{x}_{i}\right)
_{i\in I}$

$\Upsilon:V_{0}\rightarrow H_{0}::\Upsilon=\widetilde{\pi}_{H}^{-1}%
\circ\widetilde{\pi}_{V}^{-1}$

which associates to each vector of V the vector of $H$ with the same
components in the orthonormal bases, then :

$\forall X\in O:\Upsilon\left(  X\right)  =\Xi\left(  X\right)  $

and $\Upsilon$ is a bijective, linear map, which preserves the scalar product,
so it is continuous and is an isometry.

v) There is a bijective linear map : $%
\mathcal{L}%
_{H}\in GL\left(  H_{0};H_{0}\right)  $ such that : $\forall i\in
I:\varepsilon_{i}=%
\mathcal{L}%
_{H}\left(  \widetilde{\varepsilon}_{i}\right)  $ .\ 

$\left(  \widetilde{\varepsilon}_{i}\right)  _{i\in I}$ is a basis of $H_{0}$
thus $\varepsilon_{i}=\sum_{j\in I}\left[
\mathcal{L}%
_{H}\right]  _{i}^{j}\widetilde{\varepsilon}_{j}$ where only a finite number
of coefficients $\left[
\mathcal{L}%
_{H}\right]  _{i}^{j}$ is non null.

Let us define : $\varpi_{i}:H_{0}\rightarrow%
\mathbb{R}
::\varpi_{i}\left(  \sum_{j\in I}\psi_{j}\varepsilon_{j}\right)  =\psi_{i}$

This map is continuous at $\psi=0$ on $H_{0}:$

take $\psi\in H_{0},\left\Vert \psi\right\Vert \rightarrow0$

then $\psi=\sum_{i\in I}\left\langle \widetilde{\varepsilon}_{i}%
,\psi\right\rangle _{H}\widetilde{\varepsilon}_{i}$ and $\widetilde{\psi}%
_{j}=\left\langle \widetilde{\varepsilon}_{i},\psi\right\rangle _{H}%
\rightarrow0$

so if $\left\Vert \psi\right\Vert <r$ then $\left\Vert \psi\right\Vert
^{2}=\sum_{j\in I}\left\vert \widetilde{\psi}_{j}\right\vert ^{2}<r^{2}$ and
$\forall j\in I:$ $\left\vert \widetilde{\psi}_{j}\right\vert <r$

$\psi_{i}=\sum_{j\in J}\left[
\mathcal{L}%
_{H}\right]  _{i}^{j}\widetilde{\psi}_{j}\Rightarrow\left\vert \psi
_{i}\right\vert <\varepsilon\sum_{j\in I}\max\left\vert \left[
\mathcal{L}%
_{H}\right]  _{i}^{j}\right\vert $ and $\left(  \left\vert \left[
\mathcal{L}%
_{H}\right]  _{i}^{j}\right\vert \right)  _{j\in I}$ is bounded $\Rightarrow
\left\vert \psi_{i}\right\vert \rightarrow0$

Thus $\varpi_{i}$ is continuous and belongs to the topological dual
$H_{0}^{\prime}$ of $H_{0}.$ It can be extended as a continuous map
$\overline{\varpi}_{i}\in H^{\prime}$ according to the Hahn-Banach theorem
(Maths.958). Because $H$ is a Hilbert space, there is a vector $\phi_{i}\in H
$ such that : $\forall\psi\in H:\overline{\varpi}_{i}\left(  \psi\right)
=\left\langle \phi_{i},\psi\right\rangle _{H}$ so that :

$\forall X\in O:\Upsilon\left(  X\right)  =\Xi\left(  X\right)  =\sum_{i\in
I}\psi_{i}\varepsilon_{i}$

$=\sum_{i\in I}\left\langle \phi_{i},\psi\right\rangle _{H}\varepsilon
_{i}=\sum_{i\in I}\left\langle \phi_{i},\Xi\left(  X\right)  \right\rangle
_{H}\varepsilon_{i}$

$\forall i\in I:$

$\Xi\left(  e_{i}\right)  =\varepsilon_{i}=\Upsilon\left(  e_{i}\right)
=\sum_{j\in I}\left\langle \phi_{j},\varepsilon_{i}\right\rangle
_{H}\varepsilon_{j}\Rightarrow\left\langle \phi_{j},\varepsilon_{i}%
\right\rangle _{H}=\delta_{ij}$

$\Xi\left(  \widetilde{e}_{i}\right)  =\sum_{j\in I}\left\langle \phi_{j}%
,\Xi\left(  \widetilde{e}_{i}\right)  \right\rangle _{H}\varepsilon
_{j}=\widetilde{\varepsilon}_{i}=\sum_{j\in I}\left\langle \phi_{j}%
,\widetilde{\varepsilon}_{i}\right\rangle _{H}\varepsilon_{j}$

vi) The map $\Upsilon:O\rightarrow\Omega$ is a linear chart of M, using two
orthonormal bases : it is continuous, bijective so it is an homeomorphism, and
is obviously compatible with the chart $\Xi.$
\end{proof}

\bigskip

\subsubsection{Remarks}

i) Because $\left(  \widetilde{\varepsilon}_{i}\right)  _{i\in I}$ is a
Hilbertian basis of the separable infinite dimensional Hilbert space H, I is a
countable set which can be identified to $%
\mathbb{N}
.$ The assumption about $\left(  e_{i}\right)  _{i\in I}$ is that it is a
Hamel basis, which is the most general because any vector space has one. From
the proposition above we see that this basis must be of cardinality
$\aleph_{0}$ .\ Hamel bases of infinite dimensional normed vector spaces must
be uncountable, however our assumption about V is that it is a Fr\'{e}chet
space, which is a metrizable but not a normed space. If V is a Banach vector
space then, according to the Mazur theorem, it implies that there it has an
infinite dimensional vector subspace W which has a Shauder basis $:\forall
X\in W:X=\sum_{i\in I}x_{i}e_{i}$ where the sum is understood in the
topological limit. Then the same reasoning as above shows that the closure of
W is itself a Hilbert space. Moreover it has been proven that any separable
Banach space is homeomorphic to a Hilbert space, and most of the applications
will concern spaces of integrable functions (or sections of vector bundle
endowed with a norm) which are separable Fr\'{e}chet spaces.

One interesting fact is that we assume that the variables belong to an open
subset $O$ of V. The main concern is to allow for variables which can take
values only in some bounded domain. But this assumption addresses also the
case of a Banach vector space which is \textquotedblleft hollowed
out\textquotedblright\ : $O$ can be itself a vector subspace (in an infinite
dimensional vector space a vector subspace can be open), for instance
generated by a countable subbasis of a Hamel basis, and we assume explicitly
that the basis $\left(  e_{i}\right)  _{i\in I}$ belongs to $O$.

ii) For $O=V$ we have a largest open $\Omega_{V}$ and a linear map
$\Upsilon:V\rightarrow\Omega_{V}$ with domain V.

iii) To each (Hamel) basis on V is associated a linear chart $\Upsilon$ of the
manifold, such that a point of M has the same coordinates both in V and H. So
$\Upsilon$ depends on the choice of the basis, and similarly the positive
kernel $K_{V}$ depends on the basis.

iv) In the proof we have introduced a map : $K_{V}:O\times O\rightarrow%
\mathbb{R}
::K_{V}\left(  X,Y\right)  $ which is not bilinear, but is definite positive
in a precise way.\ It plays an important role in several following
demonstrations.\ From a physical point of view it can be seen as related to
the probability of transition between two states $X,Y$ often used in QM.

\subsection{Complex structure}

The variables X and vector space V are real and $H$ is a real Hilbert space.
The condition that the vector space V is real is required only in Theoorem 2
to prove the existence of a Hilbert space, because the Henderson's theorem
holds only for real structures.\ However, as it is easily checked, if $H$
exists, all the following theorems hold even if $H$ is a complex Hilbert
space. This is specially useful when the space V over which the maps $X$ are
defined is itself a complex Hilbert space, as this is often the case.

Moreover it can be useful to endow $H$ with the structure of a complex Hilbert
space : the set does not change but one distinguishes real and imaginary
components, and the scalar product is given by a Hermitian form. Notice that
this is a convenience, not a necessity.

\bigskip

\begin{theorem}
\label{Th3Complex}Any real separable infinite dimensional Hilbert space can be
endowed with the structure of a complex separable Hilbert space
\end{theorem}

\begin{proof}
$H$ has a infinite countable Hilbertian basis $\left(  \varepsilon_{\alpha
}\right)  _{\alpha\in%
\mathbb{N}
}$ because it is separable.

A complex structure is defined by a linear map : $J\in%
\mathcal{L}%
\left(  H;H\right)  $ such that $J^{2}=-Id.$ Then the operation : $i\times
\psi$ is defined by : $i\psi=J\left(  \psi\right)  .$

Define :

$J\left(  \varepsilon_{2\alpha}\right)  =\varepsilon_{2\alpha+1};J\left(
\varepsilon_{2\alpha+1}\right)  =-\varepsilon_{2\alpha}$

$\forall\psi\in H:i\psi=J\left(  \psi\right)  $

So : $i\left(  \varepsilon_{2\alpha}\right)  =\varepsilon_{2\alpha+1};i\left(
\varepsilon_{2\alpha+1}\right)  =-\varepsilon_{2\alpha}$

The bases $\varepsilon_{2\alpha}$ or $\varepsilon_{2\alpha+1}$\ are complex
bases of $H$ :

$\psi=\sum_{\alpha}\psi^{2\alpha}\varepsilon_{2\alpha}+\psi^{2\alpha
+1}\varepsilon_{2\alpha+1}=\sum_{\alpha}\left(  \psi^{2\alpha}-i\psi
^{2\alpha+1}\right)  \varepsilon_{2\alpha}$

$=\sum_{\alpha}\left(  -i\psi^{2\alpha}+\psi^{2\alpha+1}\right)
\varepsilon_{2\alpha+1}$

$\left\Vert \psi\right\Vert ^{2}=\sum_{\alpha}\left\vert \psi^{2\alpha}%
-i\psi^{2\alpha+1}\right\vert ^{2}$

$=\sum_{\alpha}\left\vert \psi^{2\alpha}\right\vert ^{2}+\left\vert
\psi^{2\alpha+1}\right\vert ^{2}+i\left(  -\overline{\psi}^{2\alpha}%
\psi^{2\alpha+1}+\psi^{2\alpha}\overline{\psi}^{2\alpha+1}\right)  $

$=\sum_{\alpha}\left\vert \psi^{2\alpha}\right\vert ^{2}+\left\vert
\psi^{2\alpha+1}\right\vert ^{2}+i\left(  -\psi^{2\alpha}\psi^{2\alpha+1}%
+\psi^{2\alpha}\psi^{2\alpha+1}\right)  $

Thus $\varepsilon_{2\alpha}$ is a Hilbertian complex basis

$H$ has a structure of complex vector space that we denote $H_{%
\mathbb{C}
}$

The map : $T:H\rightarrow H_{%
\mathbb{C}
}:T\left(  \psi\right)  =\sum_{\alpha}\left(  \psi^{2\alpha}-i\psi^{2\alpha
+1}\right)  \varepsilon_{2\alpha}$ is linear and continuous

The map : $\overline{T}:H\rightarrow H_{%
\mathbb{C}
}:\overline{T}\left(  \psi\right)  =\sum_{\alpha}\left(  \psi^{2\alpha}%
+i\psi^{2\alpha+1}\right)  \varepsilon_{2\alpha}$ is antilinear and continuous

Define : $\gamma\left(  \psi,\psi^{\prime}\right)  =\left\langle \overline
{T}\left(  \psi\right)  ,T\left(  \psi^{\prime}\right)  \right\rangle _{H}$

$\gamma$ is sesquilinear

$\gamma\left(  \psi,\psi^{\prime}\right)  =\left\langle \sum_{\alpha}\left(
\psi^{2\alpha}+i\psi^{2\alpha+1}\right)  \varepsilon_{2\alpha},\sum_{\alpha
}\left(  \psi^{\prime2\alpha}-i\psi^{\prime2\alpha+1}\right)  \varepsilon
_{2\alpha}\right\rangle _{H}$

$=\sum_{\alpha}\left(  \psi^{2\alpha}+i\psi^{2\alpha+1}\right)  \left(
\psi^{\prime2\alpha}-i\psi^{\prime2\alpha+1}\right)  $

$=\sum_{\alpha}\psi^{2\alpha}\psi^{\prime2\alpha}+\psi^{2\alpha+1}\psi
^{\prime2\alpha+1}+i\left(  \psi^{2\alpha+1}\psi^{\prime2\alpha}-\psi
^{2\alpha}\psi^{\prime2\alpha+1}\right)  $

$\gamma\left(  \psi,\psi\right)  =0\Rightarrow\left\langle \psi,\psi
\right\rangle _{H}=0\Rightarrow\psi=0$

Thus $\gamma$ is definite positive
\end{proof}

\subsection{Decomposition of the Hilbert space}

V is the product $V=V_{1}\times V_{2}...\times V_{N}$ of vector spaces, thus
the proposition implies that the Hilbert space $H$ is also the direct product
of Hilbert spaces $H_{1}\times H_{2}...\times H_{N}$ or equivalently
$H=\oplus_{k=1}^{N}H_{k}$ where $H_{k}$ are Hilbert vector subspaces of H.
More precisely :

\bigskip

\begin{theorem}
\label{Th4SeveralHilbert}If the model is comprised of N continuous variables
$\left(  X_{k}\right)  _{k=1}^{N},$ each belonging to a separable Fr\'{e}chet
vector space $V_{k},$ then the real Hilbert space $H$ of states of the system
is the Hilbert sum of N Hilbert space $H=\oplus_{k=1}^{N}H_{k}$ and any vector
$\psi$ representing a state of the system is uniquely the sum of N vectors
$\psi_{k},$ each image of the value of one variable $X_{k}$ in the state
$\psi$
\end{theorem}

\begin{proof}
By definition $V=%
{\displaystyle\prod\limits_{k=1}^{N}}
V_{k}$ .The set $V_{k}^{0}=\left\{  0,..,V_{k},...0\right\}  \subset V$\ is a
vector subspace of V.\ A\ basis of $V_{k}^{0}$ is a subfamily $\left(
e_{i}\right)  _{i\in I_{k}}$\ of a basis $\left(  e_{i}\right)  _{i\in I}$ of
V.\ $V_{k}^{0}$\ has for image by the continuous linear map $\Upsilon$\ a
closed vector subspace $H_{k}$\ of $H$. \ Any vector $X$ of V reads : $X\in%
{\displaystyle\prod\limits_{k=1}^{N}}
V_{k}:X=\sum_{k=1}^{N}\sum_{i\in I_{k}}x^{i}e_{i}$ and it has for image by
$\Upsilon:\psi=\Upsilon\left(  X\right)  =\sum_{k=1}^{N}\sum_{i\in I_{k}}%
x^{i}\varepsilon_{i}=\sum_{k=1}^{N}\psi_{k}$ with $\psi_{k}\in H_{k}$ .This
decomposition of $\Upsilon\left(  X\right)  $ is unique.

Conversely, the family $\left(  e_{i}\right)  _{i\in I_{k}}$ has for image by
$\Upsilon$ the set $\left(  \varepsilon_{i}\right)  _{i\in I_{k}}$ which are
linearly independent vectors of $H_{k}.$It is always possible to build an
orthonormal basis $\left(  \widetilde{\varepsilon}_{i}\right)  _{i\in I_{k}}$
from these vectors as done previously. $H_{k}$ is a closed subspace of $H$, so
it is a Hilbert space. The map : $\widehat{\pi}_{k}:\ell^{2}\left(
I_{k}\right)  \rightarrow H_{k}::\widehat{\pi}_{k}\left(  x\right)
=\sum_{i\in I_{k}}x^{i}\widetilde{\varepsilon}_{i}$ is an isomorphism of
Hilbert spaces and :$\forall\psi\in H_{k}:\psi=\sum_{i\in I_{k}}\left\langle
\widetilde{\varepsilon}_{i},\psi\right\rangle _{H}\widetilde{\varepsilon}_{i}$.

$\forall\psi_{k}\in H_{k},\psi_{l}\in H_{l},k\neq l:\left\langle \psi_{k}%
,\psi_{l}\right\rangle _{H}=\left\langle \Upsilon^{-1}\left(  \psi_{k}\right)
,\Upsilon^{-1}\left(  \psi_{l}\right)  \right\rangle _{E}=0$

Any vector $\psi\in H$ reads : $\psi=\sum_{k=1}^{N}\pi_{k}\left(  \psi\right)
$ with the orthogonal projection $\pi_{k}:H\rightarrow H_{k}::\pi_{k}\left(
\psi\right)  =\sum_{i\in I_{k}}\left\langle \widetilde{\varepsilon}_{i}%
,\psi\right\rangle _{H}\widetilde{\varepsilon}_{i}$ so $H$ is the Hilbert sum
of the $H_{k}$
\end{proof}

\bigskip

As a consequence the definite positive kernel of $(V,\Upsilon)$ decomposes as :

$K\left(  \left(  X_{1},...X_{N}\right)  ,\left(  X_{1}^{\prime}%
,...X_{N}^{\prime}\right)  \right)  $

$=\sum_{k=1}^{N}K_{k}\left(  X_{k},X_{k}^{\prime}\right)  $

$=\sum_{k=1}^{N}\left\langle \Upsilon\left(  X_{k}\right)  ,\Upsilon\left(
X_{k}^{\prime}\right)  \right\rangle _{H_{k}}$

This decomposition comes handy when we have to translate relations between
variables into relations between vector states, notably it they are linear.
But it requires that we keep the real Hilbert space structure.

\subsection{Discrete variables}

It is common in a model to have discrete variables $\left(  D_{k}\right)
_{k=1}^{K},$ taking values in a finite discrete set. They correspond to
different cases:

i) the discrete variables identify different elementary systems (such as
different populations of particles) which coexist simultaneously in the same
global system, follow different rules of behavior, but interact together. We
will see later how to deal with these cases (tensorial product).

ii) the discrete variables identify different populations, whose interactions
are not relevant. Actually one could consider as many different systems but,
by putting them together, one increases the size of the samples of data and
improve the statistical estimations. They are not of great interest here, in a
study of formal models.

iii) the discrete variables represent different kinds of behaviors, which
cannot be strictly identified with specific populations. Usually a discrete
variable is then used as a proxy for a quantitative parameter which tells how
close the system is from a specific situation.

We will focus on this third case. The system is represented as before by
quantitative variables $X$, whose possible values belong to some set M, which
has the structure of an infinite dimensional manifold. The general idea in the
third case is that the possible states of the system can be regrouped in two
distinct subsets. That we formalize in the following assumption : the set $O$
of possible states of the system has two connected components $O_{1},O_{2}$

\bigskip

\begin{theorem}
\label{QMDiscrete}If the condition of the theorem 2 are met, and the set $O$
of possible states of the system has two connected components $O_{1},O_{2}$
then there is a continuous function $f:H\rightarrow\left[  0,1\right]  $ such
that $f\left(  \Upsilon\left(  X\right)  \right)  =1$ in $O_{1}$ and $f\left(
\Upsilon\left(  X\right)  \right)  =0$ in $O_{2}$
\end{theorem}

\begin{proof}
The connected components $O_{1},O_{2}$ of a topological space are closed, so
$O_{1},O_{2}$ are disjoint and both open and closed in V (Maths.624). Using a
linear continuous map $\Upsilon$ then $\Omega$ has itself two connected
components, $\Omega_{1}=\Upsilon^{-1}\left(  O_{1}\right)  ,\Omega
_{2}=\Upsilon^{-1}\left(  O_{2}\right)  $ both open and closed, and disjoint.
$H$ is metric, so it is normal (Maths.705). $\Omega_{1},\Omega_{2}$ are
disjoint and closed in $H$. Then, by the Urysohn's Theorem (Maths.596) there
is a continuous function $f$ on H valued in [0,1] such that $f\left(
\psi\right)  =1$ in $H_{1}$ and $f\left(  \psi\right)  =0$ in $H_{2}$.
\end{proof}

\bigskip

The set of continuous, bounded functions is a Banach vector space, so it is
always possible, in these conditions, to replace a discrete variable by a
quantitative variable with the same features.

\newpage

\section{OBSERVABLES}

\bigskip

The key point in the conditions 1 above is that the variables are maps, which
take an infinite number of values (usually non countable). So the variables
would require the same number of data to be totally known, which is
impossible. The physicist estimates the variable by statistical methods. But
any practical method involves a first step : the scope of all maps is reduced
from V to a smaller subset W, so that any map of W can be characterized by a
finite number of parameters. The procedure sums up to replace $X$ by another
variable $\Phi\left(  X\right)  $ that we will call an observable, which is
then estimated from a finite batch of data. The mechanism of estimating the
variables $X\subset V$ is then the following :

- the observer collects data, as a set $Y=\left\{  x_{p}\right\}  _{p=1}^{N}$
of values assumed to be taken by the variable $X$, in the mathematical format
fitted to $X$ (scalars, vectors,..for different values of the arguments)

- he proceeds to the estimation $\widehat{X}$ of the map $\Phi\left(
X\right)  $ by statistical adjustment to the data $\left\{  x_{p}\right\}
_{p=1}^{N}$ . Because there are a finite number of parameters (the coordinates
of $\Phi\left(  X\right)  $ in W) this is possible

- the estimation is : $\widehat{X}=\varphi\left(  Y\right)  \in W:$ this is a
map which is a simplified version of X.

The procedure of the replacement of $X$ by $\Phi\left(  X\right)  $, called
the choice of a \textbf{specification}, is done by the physicist, and an
observable is not unique. However we make three general assumptions about
$\Phi:$

\bigskip

\begin{definition}
\label{DefObserv}

\textit{i) an \textbf{observable} is a linear map : }$\Phi\in L\left(
V;V\right)  $

\textit{ii) the range of an observable is a finite dimensional vector W
subspace of V :} $W\subset V,\dim\Phi\left(  W\right)  <\infty$

\textit{iii)} $\forall X\in O,\Phi\left(  X\right)  $ \textit{is an admissible
value, that is} $\Phi\left(  O\right)  \subset O.$
\end{definition}

\bigskip

Using the linear chart $\Upsilon$ given by any basis$,$ to $\Phi$ one can
associate a map :%

\begin{equation}
\widehat{\Phi}:H\rightarrow H::\widehat{\Phi}=\Upsilon\circ\Phi\circ
\Upsilon^{-1}%
\end{equation}

and $\widehat{\Phi}$ is an operator on $H$. And conversely.

The image of $W$ by $\Upsilon$ is a finite dimensional vector subspace
$H_{\Phi}=\Upsilon\left(  W\right)  $ of $H$, so it is closed and a Hilbert
space : $\widehat{\Phi}\in%
\mathcal{L}%
\left(  H;H_{\Phi}\right)  $

\bigskip

$%
\begin{array}
[c]{ccccccc}
&  &  & \Phi &  &  & \\
& V & \rightarrow & \rightarrow & \rightarrow & W & \\
& \downarrow &  &  &  & \downarrow & \\
\Upsilon & \downarrow &  &  &  & \downarrow & \Upsilon\\
& \downarrow &  & \widehat{\Phi} &  & \downharpoonleft & \\
& H & \rightarrow & \rightarrow & \rightarrow & H_{\Phi} &
\end{array}
$

\bigskip

\subsection{Primary observables}

The simplest specification for an observable is, given a basis $\left(
e_{i}\right)  _{i\in I}$ , to define $\Phi$ as the projection on the subspace
spanned by a finite number of vectors of the basis. For instance if $X$ is a
function $X(t)$ belonging to some space such as : $X\left(  t\right)
=\sum_{n\in%
\mathbb{N}
}a_{n}e_{n}\left(  t\right)  $ where $e_{n}\left(  t\right)  $ are fixed
functions, then a primary observable would be $Y_{J}\left(  X\left(  t\right)
\right)  =\sum_{n=0}^{N}a_{n}e_{n}\left(  t\right)  $ meaning that the
components $\left(  a_{n}\right)  _{n>N}$ are discarded and the data are used
to compute $\left(  a_{n}\right)  _{n=0}^{N}.$ To stay at the most general
level, we define :

\bigskip

\begin{definition}
\label{DefPrimObs}\textit{A \textbf{primary observable} }$\Phi=Y_{J}%
$\textit{\ is the projection of }$X=\left\{  X_{k},k=1...N\right\}
$\textit{\ on the vector subspace }$V_{J}$\textit{\ spanned by the vectors
}$\left(  e_{i}\right)  _{i\in J}\equiv\left(  e_{i}^{k}\right)  _{i\in J_{k}%
}$\textit{\ where }$J=%
{\displaystyle\prod\limits_{k=1}^{N}}
J_{k}\subset I=%
{\displaystyle\prod\limits_{k=1}^{N}}
I_{k}$\textit{\ is a finite subset of I and }$\left(  \varepsilon_{i}\right)
_{i\in I}=%
{\displaystyle\prod\limits_{k=1}^{N}}
\left(  e_{i}^{k}\right)  _{i\in I_{k}}$\textit{is a basis of V. }
\end{definition}

\bigskip

So the procedure can involve simultaneously several variables. It
requires\textit{\ the choice of a finite set of independent vectors of V.}

\bigskip

\begin{theorem}
To any primary observable $Y_{J}$ is associated uniquely a self-adjoint,
compact, trace-class operator $\widehat{Y}_{J}$ on H : $Y_{J}=\Upsilon
^{-1}\circ\widehat{Y}_{J}\circ\Upsilon$ such that the measure $Y_{J}\left(
X\right)  $ of the primary observable $Y_{J}$, if the system is in the state
$X\in O,$ is

$Y_{J}\left(  X\right)  =\sum_{i\in I}\left\langle \phi_{i},\widehat{Y}%
_{J}\left(  \Upsilon\left(  X\right)  \right)  \right\rangle _{H}e_{i}$
\end{theorem}

\begin{proof}
i) We use the notations and definitions of the previous section. The family of
variables $X=\left(  X_{k}\right)  _{k=1}^{N}$ define the charts :
$\Xi:O\rightarrow\Omega$ and the basis $\left(  e_{i}\right)  _{i\in I}$
defines the bijection $\Upsilon:V\rightarrow H$

$\forall X=\sum_{i\in I}x_{i}e_{i}\in O:$

$\Upsilon\left(  X\right)  =\sum_{i\in I}x_{i}\Upsilon\left(  e_{i}\right)
=\sum_{i\in I}x_{i}\varepsilon_{i}=\sum_{i\in I}\left\langle \phi_{i}%
,\Upsilon\left(  X\right)  \right\rangle _{H}\varepsilon_{i}$

$\Leftrightarrow x_{i}=\left\langle \phi_{i},\Upsilon\left(  X\right)
\right\rangle _{H}$

$\forall i,j\in I:\left\langle \phi_{i},\varepsilon_{j}\right\rangle
_{H}=\delta_{ij}$

ii) The primary observable $Y_{J}$ is the map :

$Y_{J}:V\rightarrow V_{J}::Y_{J}\left(  X\right)  =\sum_{j\in J}x_{j}e_{j}$

This is a projection : $Y_{J}^{2}=Y_{J}$

$Y_{J}\left(  X\right)  \in O$ so it is associated to a vector of $H$ :

$\Upsilon\left(  Y_{J}\left(  X\right)  \right)  =\Upsilon\left(  \sum_{j\in
J}x_{j}e_{j}\right)  =\sum_{j\in J}\left\langle \phi_{j},\Upsilon\left(
Y_{J}\left(  X\right)  \right)  \right\rangle _{H}\varepsilon_{j}$

$=\sum_{j\in J}\left\langle \phi_{j},\Upsilon\left(  X\right)  \right\rangle
_{H}\varepsilon_{j}$

iii) $\forall X\in O:\Upsilon\left(  Y_{J}\left(  X\right)  \right)  \in
H_{J}$ where $H_{J}$ is the vector subspace of $H$ spanned by $\left(
\varepsilon_{j}\right)  _{j\in J}.$ It is finite dimensional, thus it is
closed in $H$. There is a unique map (Math.1111) :

$\widehat{Y}_{J}\in%
\mathcal{L}%
\left(  H;H\right)  ::\widehat{Y}_{J}^{2}=\widehat{Y}_{J},\widehat{Y}%
_{J}=\widehat{Y}_{J}^{\ast}$

$\widehat{Y}_{J}$ is the orthogonal projection from $H$ onto $H_{J}.$ It is
linear, self-adjoint, and compact because its range is a finite dimensional
vector subspace. As a projection : $\left\Vert \widehat{Y}_{J}\right\Vert =1.
$

$\widehat{Y}_{J}$ is a Hilbert-Schmidt operator (Maths.1143) : take the
Hilbertian basis $\widetilde{\varepsilon}_{i}$ in $H$:

$\sum_{i\in I}\left\Vert \widehat{Y}_{J}\left(  \widetilde{\varepsilon}%
_{i}\right)  \right\Vert ^{2}=\sum_{ij\in J}\left\vert \left\langle \phi
_{j},\widetilde{\varepsilon}_{i}\right\rangle \right\vert ^{2}\left\Vert
\varepsilon_{j}\right\Vert ^{2}=\sum_{j\in J}\left\Vert \phi_{j}\right\Vert
^{2}\left\Vert \varepsilon_{j}\right\Vert ^{2}<\infty$

$\widehat{Y}_{J}$ is a trace class operator (Maths.1147) with trace $\dim
H_{J}$

$\sum_{i\in I}\left\langle \widehat{Y}_{J}\left(  \widetilde{\varepsilon}%
_{i}\right)  ,\widetilde{\varepsilon}_{i}\right\rangle =\sum_{ij\in
J}\left\langle \phi_{j},\widetilde{\varepsilon}_{i}\right\rangle \left\langle
\varepsilon_{j},\widetilde{\varepsilon}_{i}\right\rangle =\sum_{j\in
J}\left\langle \phi_{j},\varepsilon_{j}\right\rangle =\sum_{j\in J}\delta
_{jj}=\dim H_{J}$

iv) $\forall\psi\in H_{J}:\widehat{Y}_{J}\left(  \psi\right)  =\psi$

$\forall X\in O:\Upsilon\left(  Y_{J}\left(  X\right)  \right)  \in H_{J}$

$\forall X\in O:\Upsilon\left(  Y_{J}\left(  X\right)  \right)  =\widehat
{Y}_{J}\left(  \Upsilon\left(  X\right)  \right)  \Leftrightarrow Y_{J}\left(
X\right)  =\Upsilon^{-1}\circ\widehat{Y}_{J}\left(  \Upsilon\left(  X\right)
\right)  \Leftrightarrow Y_{J}=\Upsilon^{-1}\circ\widehat{Y}_{J}\circ\Upsilon$

v) The value of the observable reads : $Y_{J}\left(  X\right)  =\sum_{i\in
I}\left\langle \phi_{i},\widehat{Y}_{J}\left(  \Upsilon\left(  X\right)
\right)  \right\rangle _{H}e_{i}$
\end{proof}

\subsection{von Neumann algebras}

There is a bijective correspondence between the projections, meaning the maps
$P\in%
\mathcal{L}%
\left(  H;H\right)  :P^{2}=P,P=P^{\ast}$ and the closed vector subspaces of
$H$ (Maths.1111). Then $P$ is the orthogonal projection on the vector
subspace. So the operators $\widehat{Y}_{J}$ for any finite subset $J$ of $I$
are the orthogonal projections on the finite dimensional, and thus closed,
vector subspace $H_{J}$ spanned by $\left(  \varepsilon_{j}\right)  _{j\in
J}.$

We will enlarge the family of primary observables in several steps, in keeping
the same basis $\left(  e_{i}\right)  _{i\in I}$ of V.

\bigskip

1. For any given basis $\left(  e_{i}\right)  _{i\in I}$ of V, we extend the
definition of these operators $\widehat{Y}_{J}$ to any finite or infinite,
subset of $I$ by taking $\widehat{Y}_{J}$ as the orthogonal projection on the
closure $\overline{H_{J}}$ in $H$ of the vector subspace $H_{J}$ spanned by
$\left(  \varepsilon_{j}\right)  _{j\in J}$: $\overline{H_{J}}=\overline
{Span\left(  \varepsilon_{j}\right)  _{j\in J}}$.

\bigskip

\begin{theorem}
The operators $\left\{  \widehat{Y}_{J}\right\}  _{J\subset I}$ are
self-adjoint and commute
\end{theorem}

\begin{proof}
Because they are projections the operators $\widehat{Y}_{J}$ are such that :
$\widehat{Y}_{J}^{2}=\widehat{Y}_{J},\widehat{Y}_{J}^{\ast}=\widehat{Y}_{J}$

$\widehat{Y}_{J}$ has for eigen values :

1 for $\psi\in\overline{H_{J}}$

0 for $\psi\in\left(  \overline{H_{J}}\right)  ^{\perp}$

For any subset J of I, by the Gram-Schmidt procedure one can built an
orthonormal basis $\left(  \widetilde{\varepsilon}_{i}\right)  _{i\in J}$ of
$H_{J}$ starting with the vectors $\left(  \varepsilon_{i}\right)  _{i\in J}$
and an orthonormal basis $\left(  \widetilde{\varepsilon}_{i}\right)  _{i\in
J^{c}}$ of $H_{J^{c}}$ starting with the vectors $\left(  \varepsilon
_{i}\right)  _{i\in J^{c}}$

Any vector $\psi\in H$ can be written :

$\psi=\sum_{j\in I}x_{j}\widetilde{\varepsilon}_{j}=\sum_{j\in J}%
x_{j}\widetilde{\varepsilon}_{j}+\sum_{j\in J^{c}}x_{j}\widetilde{\varepsilon
}_{j}$ with $\left(  x_{j}\right)  _{j\in I}\in\ell^{2}\left(  I\right)  $

$\overline{H_{J}}$ is defined as $\sum_{j\in J}x_{j}\widetilde{\varepsilon
}_{j}$ with $\left(  x_{j}\right)  _{j\in J}\in\ell^{2}\left(  J\right)  $ and
similarly $\overline{H_{J^{c}}}$ is defined as $\sum_{j\in J^{c}}%
x_{j}\widetilde{\varepsilon}_{j}$ with $\left(  x_{j}\right)  _{j\in J^{c}}%
\in\ell^{2}\left(  J^{c}\right)  $

So $\widehat{Y}_{J}$ can be defined as : $\widehat{Y}_{J}\left(  \sum_{j\in
I}x_{j}\widetilde{\varepsilon}_{j}\right)  =\sum_{j\in J}x_{j}\widetilde
{\varepsilon}_{j}$

For any subsets $J_{1},J_{2}\subset I:$

$\widehat{Y}_{J_{1}}\circ\widehat{Y}_{J_{2}}=\widehat{Y}_{J_{1}\cap J_{2}%
}=\widehat{Y}_{J_{2}}\circ\widehat{Y}_{J_{1}}$

$\widehat{Y}_{J_{1}\cup J_{2}}=\widehat{Y}_{J_{1}}+\widehat{Y}_{J_{2}%
}-\widehat{Y}_{J_{1}\cap J_{2}}=\widehat{Y}_{J_{1}}+\widehat{Y}_{J_{2}%
}-\widehat{Y}_{J_{1}}\circ\widehat{Y}_{J_{2}}$

So the operators commute.
\end{proof}

\bigskip

2. Let us define $W=Span\left\{  \widehat{Y}_{i}\right\}  _{i\in I}$ the
vector subspace of $%
\mathcal{L}%
\left(  H;H\right)  $ comprised of finite linear combinations of $\widehat
{Y}_{i}$ (as defined in 1 above). The elements $\left\{  \widehat{Y}%
_{i}\right\}  _{i\in I}$ are linearly independent and constitute a basis of W.

The operators $\widehat{Y}_{j},\widehat{Y}_{k}$ are mutually orthogonal for
$j\neq k:$

$\widehat{Y}_{j}\circ\widehat{Y}_{k}\left(  \psi\right)  =\left\langle
\phi_{k},\psi\right\rangle \left\langle \phi_{j},\varepsilon_{k}\right\rangle
\varepsilon_{j}=\left\langle \phi_{k},\psi\right\rangle \delta_{jk}%
=\delta_{jk}\widehat{Y}_{j}\left(  \psi\right)  $

Let us define the scalar product on W :

$\left\langle \sum_{i\in I}a_{i}\widehat{Y}_{i},\sum_{i\in I}b_{i}\widehat
{Y}_{i}\right\rangle _{W}=\sum_{i\in I}a_{i}b_{i}$

$\left\Vert \sum_{i\in I}a_{i}\widehat{Y}_{i}\right\Vert _{W}^{2}=\sum_{i\in
I}a_{i}^{2}\left\Vert \widehat{Y}_{i}\right\Vert _{W}^{2}=\sum_{i\in I}%
a_{i}^{2}$

$W$ is isomorphic to $%
\mathbb{R}
_{0}^{I}$ and its closure in $%
\mathcal{L}%
\left(  H;H\right)  $ : $\overline{W}=\overline{Span\left\{  \widehat{Y}%
_{i}\right\}  _{i\in I}}$ is isomorphic to $\ell^{2}\left(  I\right)  ,$ and
has the structure of a Hilbert space with :

$\overline{W}=\left\{  \sum_{i\in I}a_{i}\widehat{Y}_{i},\left(  a_{i}\right)
_{i\in I}\in\ell^{2}\left(  I\right)  \right\}  $

\bigskip

3. Let us define $A$ as the algebra generated by any finite linear combination
or products of elements $\widehat{Y}_{J},J$ finite or infinite, and
$\overline{A}$ as the closure of $A$ in $%
\mathcal{L}%
\left(  H;H\right)  $ : $\overline{A}=\overline{Span\left\{  \widehat{Y}%
_{J}\right\}  _{J\subset I}}$ with respect to the strong topology, that is in norm.

\bigskip

\begin{theorem}
$\overline{A}$ is a commutative von Neumann algebra of $%
\mathcal{L}%
\left(  H,H\right)  $
\end{theorem}

\begin{proof}
It is obvious that A is a *subalgebra of $%
\mathcal{L}%
\left(  H,H\right)  $ with unit element $Id=\widehat{Y}_{I}.$

Because its generators are projections, $\overline{A}$ is a von Neumann
algebra (Maths.1190).

The elements of $A=Span\left\{  \widehat{Y}_{J}\right\}  _{J\subset I}$ that
is of finite linear combination of $\widehat{Y}_{J}$ commute

$Y,Z\in\overline{A}\Rightarrow\exists\left(  Y_{n}\right)  _{n\in%
\mathbb{N}
},\left(  Z_{n}\right)  _{n\in%
\mathbb{N}
}\in A^{%
\mathbb{N}
}:Y_{n}\rightarrow_{n\rightarrow\infty}Y,Z_{n}\rightarrow_{n\rightarrow\infty
}Z$

The composition is a continuous operation.

$Y_{n}\circ Z_{n}=Z_{n}\circ Y_{n}\Rightarrow\lim\left(  Y_{n}\circ
Z_{n}\right)  =\lim\left(  Z_{n}\circ Y_{n}\right)  =\lim Y_{n}\circ\lim
Z_{n}=\lim Z_{n}\circ\lim Y_{n}=Z\circ Y=Y\circ Z$

So $\overline{A}$ is commutative.

$\overline{A}$ is identical to the bicommutant of its projections, that is to
$\overline{A}"$ (Maths.1189)
\end{proof}

\bigskip

This result is of interest because commutative von Neumann algebras are
classified : they are isomorphic to the space of functions $f\in L^{\infty
}\left(  E,\mu\right)  $ acting by pointwise multiplication $\varphi
\rightarrow f\varphi$ on functions $\varphi\in L^{2}\left(  E,\mu\right)  $
for some set E and measure $\mu$ (not necessarily absolutely continuous). They
are the topic of many studies, notably in ergodic theory. The algebra
$\overline{A}$ depends on the choice of a basis $\left(  e_{i}\right)  _{i\in
I} $ and, as can be seen in the formulation through $\left(  \widetilde
{\varepsilon}_{i}\right)  _{i\in I},$ is defined up to a unitary transformation.

\bigskip

In the axiomatisation of QM, it is usual to define a system by a von Neumann
algebra of operators on a Hilbert space.\ We see here how such algebras appear
naturally. However the algebra $\overline{A}$ is commutative, and this
property is the consequence of the choice of a unique basis $\left(
e_{i}\right)  _{i\in I}.$ It would not hold for primary observables defined
through different bases : they do not even constitute an algebra. Any von
Neumann algebra is the closure of the linear span of its projections
(Maths.1190), and any projection can be defined through a basis, thus one can
say that the \textquotedblleft observables\textquotedblright\ (with their
usual definition) of a system are the collection of all primary observables
(as defined here) for all bases of $V$. This is a crucial issue in the
axiomatic interpretation of QM, but the concept of observables introduced here
allows us to deal with this issue and understand how probabilities enter the
picture. But, before that, we need to see what can be said about more general
observables, which are not just primary.

\subsection{Secondary observables}

Beyond primary observables, general observables $\Phi$ can be studied using
spectral theory.

\bigskip

1. A spectral measure defined on a measurable space $E$ with $\sigma-$algebra
$\sigma_{E}$ and acting on the Hilbert space $H$ is a map : $P:\sigma
_{E}\rightarrow%
\mathcal{L}%
\left(  H;H\right)  $ such that (Maths.1240) :

i) $P\left(  \varpi\right)  $ is a projection

ii) $P(E)=Id$

iii) $\forall\psi\in H$ the map: $\varpi\rightarrow\left\langle P\left(
\varpi\right)  \psi,\psi\right\rangle _{H}=\left\Vert P\left(  \varpi\right)
\psi\right\Vert ^{2}$ is a finite positive measure on $(E,\sigma_{E})$.

One can show (Maths.1242) that there is a bijective correspondence between the
spectral measures on $H$ and the maps : $\chi:\sigma_{E}\rightarrow H$ such
that :

i) $\chi\left(  \varpi\right)  $ is a closed vector subspace of $H$

ii) $\chi(E)=H$

iii) $\forall\varpi,\varpi^{\prime}\in\sigma_{E},\varpi\cap\varpi^{\prime
}=\varnothing:\chi\left(  \varpi\right)  \cap\chi\left(  \varpi^{\prime
}\right)  =\left\{  0\right\}  $

then $P(\varpi)$ is the orthogonal projection on $\chi(\varpi),$ denoted
$:\widehat{\pi}_{\chi(\varpi)}$

Thus, for any fixed $\psi\neq0\in H$ the function $\widehat{\chi}_{\psi
}:\sigma_{E}\rightarrow%
\mathbb{R}
::\widehat{\chi}_{\psi}\left(  \varpi\right)  =\frac{\left\langle \widehat
{\pi}_{\chi\left(  \varpi\right)  }\psi,\psi\right\rangle }{\left\Vert
\psi\right\Vert ^{2}}=\frac{\left\Vert \widehat{\pi}_{\chi\left(
\varpi\right)  }\psi\right\Vert ^{2}}{\left\Vert \psi\right\Vert ^{2}}$ is a
probability law on $(E,\sigma_{E})$.

\bigskip

2. An application of standard theorems on spectral measures (Maths.1243, 1245)
tells that, for any bounded measurable function $f:E\rightarrow%
\mathbb{R}
$ , the spectral integral : $\int_{E}f\left(  \xi\right)  \widehat{\pi}%
_{\chi(\xi)}$\ defines a continuous operator $\widehat{\Phi}_{f}$ on $H$.
$\widehat{\Phi}_{f}$ is such that :

$\forall\psi,\psi^{\prime}\in H:\left\langle \widehat{\Phi}_{f}\left(
\psi\right)  ,\psi^{\prime}\right\rangle =\int_{E}f\left(  \xi\right)
\left\langle \widehat{\pi}_{\chi\left(  \xi\right)  }\left(  \psi\right)
,\psi^{\prime}\right\rangle $

And conversely (Math.1252), for any continuous normal operator $\widehat{\Phi
}$ on $H$, that is such that :

$\widehat{\Phi}\in%
\mathcal{L}%
\left(  H;H\right)  :\widehat{\Phi}\circ\widehat{\Phi}^{\ast}=\widehat{\Phi
}^{\ast}\circ\widehat{\Phi}$ with the adjoint $\widehat{\Phi}^{\ast}$

there is a unique spectral measure $P$ on $(%
\mathbb{R}
,\sigma_{%
\mathbb{R}
})$ such that $\widehat{\Phi}=\int_{Sp\left(  \widehat{\Phi}\right)
}sP\left(  s\right)  $ where $Sp(\widehat{\Phi})\subset%
\mathbb{R}
$ is the spectrum of $\widehat{\Phi}.$

So there is a map $\chi:\sigma_{%
\mathbb{R}
}\rightarrow H$ where $\sigma_{%
\mathbb{R}
}$ is the Borel algebra of $%
\mathbb{R}
$ such that :

$\chi\left(  \varpi\right)  $ is a closed vector subspace of $H$

$\chi\left(
\mathbb{R}
\right)  =Id$

$\forall\varpi,\varpi^{\prime}\in\sigma_{%
\mathbb{R}
},\varpi\cap\varpi^{\prime}=\varnothing\Rightarrow\chi\left(  \varpi\right)
\cap\chi\left(  \varpi^{\prime}\right)  =\left\{  0\right\}  $

and $\widehat{\Phi}=\int_{Sp(\widehat{\Phi})}s\widehat{\pi}_{\chi\left(
s\right)  }$

The spectrum $Sp(\widehat{\Phi})$ is a non empty compact subset of $%
\mathbb{R}
.$ If $\widehat{\Phi}$ is normal then $\lambda\in Sp(\widehat{\Phi
})\Leftrightarrow\overline{\lambda}\in Sp(\widehat{\Phi}^{\ast}).$

For any fixed $\psi\neq0\in H$ the function $\widehat{\mu}_{\psi}:\sigma_{%
\mathbb{R}
}\rightarrow%
\mathbb{R}
::\widehat{\mu}_{\psi}\left(  \varpi\right)  =\frac{\left\langle \widehat{\pi
}_{\chi\left(  \varpi\right)  }\psi,\psi\right\rangle }{\left\Vert
\psi\right\Vert ^{2}}=\frac{\left\Vert \widehat{\pi}_{\chi\left(
\varpi\right)  }\psi\right\Vert ^{2}}{\left\Vert \psi\right\Vert ^{2}}$ is a
probability law on $(%
\mathbb{R}
,\sigma_{%
\mathbb{R}
})$.

\bigskip

3. We will define :

\bigskip

\begin{definition}
\textit{A \textbf{secondary observable} is a linear map }$\Phi\in L\left(
V;V\right)  $\textit{\ valued in a finite dimensional vector subspace of V,
such that }$\widehat{\Phi}=\Upsilon\circ\Phi\circ\Upsilon^{-1}$\textit{\ is a
normal operator : }$\widehat{\Phi}\circ\widehat{\Phi}^{\ast}=\widehat{\Phi
}^{\ast}\circ\widehat{\Phi}$ with the adjoint $\widehat{\Phi}^{\ast}$
\end{definition}

\bigskip

\begin{theorem}
Any secondary observable $\Phi$ is a compact, continuous map, its associated
map $\widehat{\Phi}=\Upsilon\circ\Phi\circ\Upsilon^{-1}$ is a compact,
self-adjoint, Hilbert-Schmidt and trace class operator.

$\Phi=\sum_{p=1}^{n}\lambda_{p}Y_{J_{p}}$ where $\left(  Y_{J_{p}}\right)
_{p=1}^{N}$ are primary observables associated to a basis $\left(
e_{i}\right)  _{i\in I}$ of V and $\left(  J_{p}\right)  _{p=1}^{n}$ are
disjoint finite subsets of I
\end{theorem}

\begin{proof}
i) $\widehat{\Phi}\left(  H\right)  $ is a finite dimensional vector subspace
of $H$. So :

$\widehat{\Phi}$ has $0$ for eigen value, with an infinite dimensional eigen
space $H_{c}.$

$\Phi,\widehat{\Phi}$ are compact and thus continuous (Maths.912).

ii) As $\widehat{\Phi}$ is continuous and normal, there is a unique spectral
measure P on $(%
\mathbb{R}
,\sigma_{%
\mathbb{R}
})$ such that $\widehat{\Phi}=\int_{Sp\left(  \widehat{\Phi}\right)
}sP\left(  s\right)  $ where $Sp(\widehat{\Phi})\subset%
\mathbb{R}
$ is the spectrum of $\widehat{\Phi}.$ As $\widehat{\Phi}$ is compact, by the
Riesz theorem (Maths.1142) its spectrum is either finite or is a countable
sequence converging to 0 (which may or not be an eigen value) and, except
possibly for $0$, is identical to the set $\left(  \lambda_{p}\right)  _{p\in%
\mathbb{N}
}$ of its eigen values (Maths.1020). For each distinct eigen value the eigen
spaces $H_{p}$ are orthogonal and $H$ is the direct sum $H=\oplus_{p\in%
\mathbb{N}
}H_{p}$. For each non null eigen value $\lambda_{p}$ the eigen space $H_{p}$
is finite dimensional.

Let $\lambda_{0}$ be the eigen value 0 of $\widehat{\Phi}.$ So :
$\widehat{\Phi}=\sum_{p\in%
\mathbb{N}
}\lambda_{p}\widehat{\pi}_{H_{p}}$and any vector of $H$ reads : $\psi
=\sum_{p\in%
\mathbb{N}
}\psi_{p}$ with $\psi_{p}=\widehat{\pi}_{H_{p}}\left(  \psi\right)  $

Because $\widehat{\Phi}\left(  H\right)  $ is finite dimensional, the spectrum
is finite and the non null eigen values are $\left(  \lambda_{p}\right)
_{p=1}^{n}$, the eigen space corresponding to 0 is $H_{c}=\left(  \oplus
_{p=1}^{n}H_{p}\right)  ^{\perp}$

$\forall\psi\in H:\psi=\psi_{c}+\sum_{p=1}^{n}\psi_{p}$ with $\psi
_{p}=\widehat{\pi}_{H_{p}}\left(  \psi\right)  ,\psi_{c}=\widehat{\pi}_{H_{c}%
}\left(  \psi\right)  $

$\widehat{\Phi}=\sum_{p=1}^{n}\lambda_{p}\widehat{\pi}_{H_{p}}$

Its adjoint reads : $\widehat{\Phi}^{\ast}=\sum_{p\in%
\mathbb{N}
}\overline{\lambda}_{p}\widehat{\pi}_{H_{p}}=\sum_{p\in%
\mathbb{N}
}\lambda_{p}\widehat{\pi}_{H_{p}}$ because $H$ is a real Hilbert space

$\widehat{\Phi}$ is then self-adjoint, Hilbert-Schmidt and trace class, as the
sum of the trace class operators $\widehat{\pi}_{H_{p}}.$

iii) The observable reads :

$\Phi=\sum_{p=1}^{n}\lambda_{p}\pi_{p}$ where $\pi_{p}=\Upsilon^{-1}%
\circ\widehat{\pi}_{H_{p}}\circ\Upsilon$ is the projection on a finite
dimensional vector subspace of V :

$\pi_{p}\circ\pi_{q}=\Upsilon^{-1}\circ\widehat{\pi}_{H_{p}}\circ\Upsilon
\circ\Upsilon^{-1}\circ\widehat{\pi}_{H_{q}}\circ\Upsilon=\Upsilon^{-1}%
\circ\widehat{\pi}_{H_{p}}\circ\widehat{\pi}_{H_{q}}\circ\Upsilon=\delta
_{pq}\Upsilon^{-1}\circ\widehat{\pi}_{H_{p}}\circ\Upsilon=\delta_{pq}\pi_{p}$

$\Phi\circ\pi_{p}=\lambda_{p}\pi_{p}$ so $\pi_{p}\left(  V\right)  =V_{p}$ is
the eigen space of $\Phi$ for the eigen value $\lambda_{p}$ and the subspaces
$\left(  V_{p}\right)  _{p=1}^{n}$ are linearly independent.

By choosing any basis $\left(  e_{i}\right)  _{i\in J_{p}}$ of $V_{p},$ and
$\left(  e_{i}\right)  _{i\in J^{c}}$ with $J^{c}=\complement_{I}\left(
\oplus_{p=1}^{n}J_{n}\right)  $ for the basis of $V_{c}=Span\left(  \left(
e_{i}\right)  _{i\in J^{c}}\right)  $

$X=Y_{J^{c}}\left(  X\right)  +\sum_{p=1}^{n}Y_{J_{p}}\left(  X\right)  $

the observable $\Phi$ reads : $\Phi=\sum_{p=1}^{n}\lambda_{p}Y_{J_{p}}$
\end{proof}

\bigskip

We have :

$Y_{J_{p}}\left(  X\right)  =\sum_{i\in J_{p}}\left\langle \phi_{i}%
,\widehat{Y}_{Jp}\left(  \Upsilon\left(  X\right)  \right)  \right\rangle
_{H}e_{i}$

$\Phi\left(  X\right)  =\sum_{p=1}^{n}\lambda_{p}\sum_{i\in J_{p}}\left\langle
\phi_{i},\widehat{Y}_{Jp}\left(  \Upsilon\left(  X\right)  \right)
\right\rangle _{H}e_{i}$

$=\sum_{i\in I}\left\langle \phi_{i},\sum_{p=1}^{n}\lambda_{p}\widehat{Y}%
_{Jp}\left(  \Upsilon\left(  X\right)  \right)  \right\rangle _{H}e_{i}$

$=\sum_{i\in I}\left\langle \phi_{i},\widehat{\Phi}\left(  \Upsilon\left(
X\right)  \right)  \right\rangle _{H}e_{i}$

$\Phi,\widehat{\Phi}$ have invariant vector spaces, which correspond to the
direct sum of the eigen spaces.

The probability law $\widehat{\mu}_{\psi}:\sigma_{%
\mathbb{R}
}\rightarrow%
\mathbb{R}
$ reads :

$\widehat{\mu}_{\psi}\left(  \varpi\right)  =\Pr\left(  \lambda_{p}\in
\varpi\right)  =\frac{\left\Vert \widehat{\pi}_{H_{p}}\left(  \psi\right)
\right\Vert ^{2}}{\left\Vert \psi\right\Vert ^{2}}$

To sum up :

\bigskip

\begin{theorem}
\label{QMEigenValue}For any primary or secondary observable $\Phi$, there is a
basis $\left(  e_{i}\right)  _{i\in I}$ of V, a compact, self-adjoint,
Hilbert-Schmidt and trace class operator $\widehat{\Phi}$ on the associated
Hilbert space H such that :

$\widehat{\Phi}=\Upsilon\circ\Phi\circ\Upsilon^{-1}$

if the system is in the state $X=\sum_{i\in I}\left\langle \phi_{i}%
,\Upsilon\left(  X\right)  \right\rangle _{H}e_{i}$ the value of the
observable is : $\Phi\left(  X\right)  =\sum_{i\in I}\left\langle \phi
_{i},\widehat{\Phi}\left(  \Upsilon\left(  X\right)  \right)  \right\rangle
_{H}e_{i}$
\end{theorem}

\bigskip

$\widehat{\Phi}$ has a finite set of eigen values, whose eigen spaces (except
possibly for 0) are finite dimensional and orthogonal. The vectors
corresponding to the eigen value 0 are never observed, so it is convenient to
represent the Hilbert space $H$ through a basis of eigen vectors, each of them
corresponding to a definite state, which usually can be identified. This is a
method commonly used in Quantum Mechanics, however the vector has also a
component in the eigen space corresponding to the null eigen value, which is
not observed but exists. Conversely any observable (on V) can be defined
through an operator on $H$ with the required properties (compact, normal, it
is then self-adjoint). We will come back on this point in the following, when
a group is involved.

\subsection{Efficiency of an observable}

A crucial factor for the quality and the cost of the estimation procedure is
the number of parameters to be estimated, which is closely related to the
dimension of the vector space $\Phi\left(  V\right)  ,$ which is finite. The
error made by the choice of $\Phi\left(  X\right)  $ when the system is in the
state $X$ is : $o_{\Phi}\left(  X\right)  =X-\Phi\left(  X\right)  .$ If two
observables $\Phi,\Phi^{\prime}$ are such that $\Phi\left(  V\right)
,\Phi^{\prime}\left(  V\right)  $ have the same dimension, one can say that
$\Phi$ is more efficient than $\Phi^{\prime}$ if : $\forall X:\left\Vert
o_{\Phi}\left(  X\right)  \right\Vert _{V}\leq\left\Vert o_{\Phi^{\prime}%
}\left(  X\right)  \right\Vert _{V}$

To assess the efficiency of a secondary observable $\Phi$ it is legitimate to
compare $\Phi$ to the primary observable $Y_{J}$ with a set $J$ which has the
same cardinality as the dimension of $\oplus_{p=1}^{n}H_{p}.$

The error with the choice of $\Phi$ is :

$o_{\Phi}\left(  X\right)  =X-\Phi\left(  X\right)  =Y_{c}\left(  \psi\right)
+\sum_{p=1}^{n}\left(  1-\lambda_{p}\right)  Y_{p}\left(  \psi\right)  $

$\left\Vert o_{\Phi}\left(  X\right)  \right\Vert _{V}^{2}=\left\Vert
Y_{c}\left(  \psi\right)  \right\Vert _{V}^{2}+\sum_{p=1}^{n}\left(
1-\lambda_{p}\right)  ^{2}\left\Vert Y_{p}\left(  \psi\right)  \right\Vert
^{2}$

$\widehat{o}_{\Phi}\left(  \Upsilon\left(  X\right)  \right)  =\Upsilon\left(
X\right)  -\widehat{\Phi}\left(  \Upsilon\left(  X\right)  \right)
=\widehat{\pi}_{H_{c}}\left(  \psi\right)  +\sum_{p=1}^{n}\left(
1-\lambda_{p}\right)  \widehat{\pi}_{H_{p}}\left(  \psi\right)  $

$\left\Vert \widehat{o}_{\Phi}\left(  \Upsilon\left(  X\right)  \right)
\right\Vert ^{2}=\left\Vert \widehat{\pi}_{H_{c}}\left(  \psi\right)
\right\Vert ^{2}+\sum_{p=1}^{n}\left(  1-\lambda_{p}\right)  ^{2}\left\Vert
\widehat{\pi}_{H_{p}}\left(  \psi\right)  \right\Vert ^{2}=\left\Vert o_{\Phi
}\left(  X\right)  \right\Vert _{V}^{2}$

And for $Y_{J}:\left\Vert \widehat{o}_{Y_{J}}\left(  \Upsilon\left(  X\right)
\right)  \right\Vert ^{2}=\left\Vert \widehat{\pi}_{H_{c}}\left(  \psi\right)
\right\Vert ^{2}$ because $\lambda_{p}=1$

So :

\bigskip

\begin{theorem}
\label{QMEfficiency}For any secondary observable there is always a primary
observable which is at least as efficient.
\end{theorem}

\bigskip

This result justifies the restriction, in the usual formalism, of observables
to operators belonging to a von Neumann algebra.

\subsection{Statistical estimation and primary observables}

At first the definition of a primary observable seems naive, and the previous
results will seem obvious to the specialists of Axiomatic QM.\ After all the
definition of a primary observable requires only the choice of a finite number
of orthonormal vectors of V. We have already seen that a primary observable is
always better than a, more sophisticated, secondary observable.\ But we have
also to compare a primary observable to what is practically done in an
experiment, where we have to estimate a map from a batch of data.

\bigskip

Consider a model with variables $X$, maps, belonging to a Hilbert space $H$
(to keep it simple), from a set M to a normed vector space $E$, endowed with a
scalar product $\left\langle {}\right\rangle _{E}$.\ The physicist has a batch
of data, that is a finite set $\left\{  x_{p}\in E,p=1...N\right\}  $ of N
measures of $X$ done at different points $\Omega=\left\{  m_{p}\in
M,p=1...N\right\}  $ : of M : $x_{p}=X\left(  m_{p}\right)  .$ The estimated
map $\widehat{X}$ should be a solution of the collection of equations :
$x_{p}=X\left(  m_{p}\right)  $ where $x_{p},m_{p}$ are known.

The \textbf{evaluation maps}, that we will encounter several times, is the
collection of maps $\mathcal{E}\left(  m\right)  $ on H :

$\mathcal{E}\left(  m\right)  :H\rightarrow E::\mathcal{E}\left(  m\right)
Y=Y\left(  m\right)  $

Because $H$ and $E$ are vector spaces $\mathcal{E}\left(  m\right)  $ is a
linear map : $\mathcal{E}\left(  m\right)  \in L\left(  H;E\right)  $,
depending on both $H$ and $E$.\ It can be continuous or not.

The set of solutions of the equations, that is of maps $Y$ of $H$ such that
$\forall m_{p}\in\Omega:Y\left(  m_{p}\right)  =x_{p}$ is :

$A=\cap_{m_{p}\in\Omega}\mathcal{E}\left(  m_{p}\right)  ^{-1}\left(
x_{p}\right)  $

$Y\in A\Leftrightarrow\forall m\in\Omega:Y\left(  m\right)  =X\left(
m\right)  $

It is not empty because it contains at least $X$. Its closed convex hull is
the set $B$ in $H$ (Maths.361) :

$\forall Z\in B:\exists\alpha\in\left[  0.1\right]  ,Y,Y^{\prime}\in
A:Z=\alpha Y+\left(  1-\alpha\right)  Y^{\prime}$

$\Rightarrow\forall m\in\Omega:Z\left(  m\right)  =x_{p}$

$B$ is the smallest closed set of $H$ such that all its elements $Z$ are
solutions of the equations : $\forall p=1..N:Z\left(  m_{p}\right)  =x_{p}.$

\bigskip

If we specify an observable, we restrict $X$ to a finite dimensional subspace
$H_{J}\subset H.$With the evaluation map $\mathcal{E}_{J}$ on $H_{J} $ we can
consider the same procedure, but then usually $A_{J}=\varnothing.$ The
simplification of the map to be estimated as for consequence that there is no
solution to the equations. So the physicist uses a statistical method, that is
a map which associates to each batch of data $X\left(  \Omega\right)  $ a map
$\varphi\left(  X\left(  \Omega\right)  \right)  =\widehat{X}\in H_{J}.$
Usually $\widehat{X}$ is such that it minimizes the sum of the distance
between points in $E$ : $\sum_{m\in\Omega}\left\Vert \widehat{X}\left(
m\right)  -x_{p}\right\Vert _{E} $ (other additional conditions can be required).

The primary observable $\Phi$ gives another solution : $\Phi\left(  X\right)
$ is the orthogonal projection of $X$ on the Hilbert space $H_{J},$ it is such
that it minimizes the distance between maps :

$\forall Z\in H_{J}:\left\Vert X-Z\right\Vert _{H}\geq\left\Vert X-\Phi\left(
X\right)  \right\Vert _{H}.$

$\Phi\left(  X\right)  $ always exist, and does not depend on the choice of an
estimation procedure $\varphi.$ $\Phi\left(  X\right)  $ minimizes the
distance between maps in H, meanwhile $\varphi\left(  X\left(  \Omega\right)
\right)  $ minimizes distance between points in E. Usually $\varphi\left(
X\left(  \Omega\right)  \right)  $ is different from $\Phi\left(  X\right)  $
and $\Phi\left(  X\right)  $ is a better estimate than $\widehat{X}$ :
\textit{a primary observable is actually the best statistical estimator} for a
given size of the sample. But it requires the explicit knowledge of the scalar
product and $H_{J}.$ This can be practically done in some significant cases
(see for an example J.C.Dutailly \textit{Estimation of the probability of
transitions between phases).}

\bigskip

Knowing the estimate $\widehat{X}$ provided by a statistical method $\varphi,
$ we can implement the previous procedure to the set $\widehat{X}\left(
\Omega\right)  $ and compute the set of solutions : $\widehat{A}=\cap
_{m_{p}\in\Omega}\mathcal{E}_{J}\left(  m_{p}\right)  ^{-1}\left(  \widehat
{X}\left(  m\right)  \right)  .$ It is not empty.\ Its closed convex hull
$\widehat{B}$ in $H_{J}$ can be considered as the domain of confidence of
$\widehat{X}:$ they are maps which take the same values as $\widehat{X}$ in
$\Omega$ and as a consequence give the same value to $\sum_{m\in\Omega
}\left\Vert \widehat{X}\left(  m\right)  -x_{p}\right\Vert _{E}.$

Because $\widehat{B}$ is closed and convex there is a unique orthogonal
projection $Y$ of $X$ on $\widehat{B}$ (Maths.1107) and :

$\forall Z\in\widehat{B}:\left\Vert X-Z\right\Vert _{H}\geq\left\Vert
X-Y\right\Vert _{H}\Rightarrow\left\Vert X-\widehat{X}\right\Vert _{H}%
\geq\left\Vert X-Y\right\Vert _{H}$

so $Y$ is a better estimate than $\varphi\left(  X\left(  \Omega\right)
\right)  ,$ and can be computed if we know the scalar product on H.

We see clearly the crucial role played by the choice of a specification. But
it leads to a more surprising result, of deep physical meaning.

\subsubsection{Quantization of singularities}

A classic problem in Physics is to prove the existence of a singular
phenomenon, appearing only for some values of the parameters $m$. To study
this problem we use a model similar to the previous one, with the same
notations. But here the variable $X$ is comprised of two maps, $X_{1},X_{2}$
with unknown, disconnected, domains $M_{1},M_{2}:M=M_{1}+M_{2}.$ The first
problem is to estimate $X_{1},X_{2}.$

\bigskip

With a statistical process $\varphi\left(  X\left(  \Omega\right)  \right)  $
it is always possible to find estimations $\widehat{X}_{1},\widehat{X}_{2}$ of
$X_{1},X_{2}.$ The key point is to distinguish in the set $\Omega$ the points
which belong to $M_{1}$ and $M_{2}.$ There are $\frac{1}{2}\left(
2^{N}-2\right)  =2^{N-1}-1$ distinct partitions of $\Omega$ in two subsets
$\Omega_{1}+\Omega_{2},$ on each subset the statistical method $\varphi$ gives
the estimates :

$\widehat{Y}_{1}=\varphi\left(  X\left(  \Omega_{1}\right)  \right)
,\widehat{Y}_{2}=\varphi\left(  X\left(  \Omega_{2}\right)  \right)  $

Denote : $\rho\left(  \Omega_{1},\Omega_{2}\right)  $

$=\sum_{m_{p}\in\Omega_{1}}\left\Vert X\left(  m_{p}\right)  -\varphi\left(
X\left(  \Omega_{1}\right)  \right)  \left(  m_{p}\right)  \right\Vert
+\sum_{m_{p}\in\Omega_{2}}\left\Vert X\left(  m_{p}\right)  -\varphi\left(
X\left(  \Omega_{2}\right)  \right)  \left(  m_{p}\right)  \right\Vert $

A partition $\left(  \Omega_{1},\Omega_{2}\right)  $ is said to be a better
fit than $\left(  \Omega_{1}^{\prime},\Omega_{2}^{\prime}\right)  $ if :

$\rho\left(  \Omega_{1},\Omega_{2}\right)  \leq\rho\left(  \Omega_{1}^{\prime
},\Omega_{2}^{\prime}\right)  $

Then $\widehat{X}_{1}=\varphi\left(  X\left(  \Omega_{1}\right)  \right)
,\widehat{X}_{2}=\varphi\left(  X\left(  \Omega_{2}\right)  \right)  $ is the
solution for the best partition.

So there is a procedure, which provides always the best solution given the
data and $\varphi,$ but it does not give $M_{1},M_{2}$ precisely, their
estimation depends on the structure of M.

However it is a bit frustrating, if we want to test a law, because the
procedure provides always a solution, even if actually there is no such
partition of $X$. And this can happen. If we define the sets as above with the
evaluation map : $\mathcal{E}_{J}\left(  m\right)  :H_{J}\rightarrow
E::\mathcal{E}\left(  m\right)  Y=Y\left(  m\right)  $

$A_{k}=\cap_{m_{p}\in\Omega_{k}}\mathcal{E}\left(  m_{p}\right)  ^{-1}\left(
\widehat{X}_{k}\left(  m_{p}\right)  \right)  \subset H_{J}$ for $k=1,2$. It
is not empty because it contains at least $\widehat{X}_{k}$.

$B_{k}$ the closed convex hull of $A_{k}$ in $H_{J}$

Then : $\forall Y\in B_{k},m\in\Omega_{k}:Y\left(  m\right)  =\widehat{X}%
_{k}\left(  m\right)  $

If $B_{1}\cap B_{2}\neq\varnothing$ there is at least one map, which can be
defined uniquely on M, belongs to $H_{J}$ and is equivalent to $\widehat
{X}_{1},\widehat{X}_{2}.$

\bigskip

This issue is of importance because many experiments aim at proving the
existence of a special behavior. We need, in addition, a test of the
hypothesis (denoted $H_{0})$ : there is a partition (and then the best
solution would be $\widehat{X}_{1},\widehat{X}_{2})$ against the hypothesis
(denoted $H_{1})$ there is no partition : there is a unique map $\widehat
{X}\in H_{J}$ for the domain $\Omega.$ The simplest test is to compare
$\sum_{m_{p}\in\Omega}\left\Vert X\left(  m_{p}\right)  -\varphi\left(
\Omega\right)  \left(  m_{p}\right)  \right\Vert $ to $\rho\left(  \Omega
_{1},\Omega_{2}\right)  .$ If $\varphi\left(  \Omega\right)  $ gives results
as good as $\widehat{X}_{1},\widehat{X}_{2}$ we can reject the
hypothesis.\ Notice that it accounts for the properties assumed for the maps
in $H_{J}.$ For instance if $H_{J}$ is comprised uniquely of continuous maps,
then $\varphi\left(  X\left(  \Omega\right)  \right)  $ is continuous, and
clearly distinct from the maps $\widehat{X}_{1},\widehat{X}_{2}$ continuous
only on $M_{1},M_{2}.$

It is quite obvious that the efficiency of this test decreases with N : the
smaller N, the greater the chance to accept $H_{0}$. Is there a way to control
the validity of an experiment ? The Theory of Tests, a branch of Statistics,
studies this kind of problems.

\bigskip

The problem is, given a sample of points $\Omega=\left(  m_{p}\right)
_{p=1}^{N}$ and the corresponding values $x=\left(  x_{p}\right)  _{p=1}^{N},$
decide if they obey to a simple ($X$, Hypothesis $H_{1})$ or a double
($X_{1},X_{2},$ Hypothesis $H_{0})$ distribution law.

The choice of the points $\left(  m_{p}\right)  _{p=1}^{N}$in a sample is
assumed to be random : all the points $m$ of M have the same probability to be
in $\Omega,$ but the size of $M_{1},M_{2}$ can be different, so it could give
a different chance for a point of $M_{1}$ or $M_{2}$ to be in the sample. Let
us say that :

$\Pr\left(  m\in M_{1}|H_{0}\right)  =1-\lambda,\Pr\left(  m\in M_{2}%
|H_{0}\right)  =\lambda,\Pr\left(  m\in M|H_{1}\right)  =1$

(all the probabilities are for a sample of a given size N)

Then the probability for any vector of $E$ to have a given value $x$ depends
only on the map $X$ : this is the number of points $m$ of M for which
$X\left(  m\right)  =x.$ For instance if there are two points $m$ with
$X(m)=x$ then $x$ has two times the probability to appear, and if $X$ is more
concentrated in an area of $E$, this area has more probability to appear. Let
us denote this value $\rho\left(  x\right)  \in\left[  0,1\right]  .$

Rigorously (Maths.869), with a measure $dx$ on $E$, $\mu$ on M, $\rho\left(
x\right)  dx$ is the pull-back of the measure $\mu$ on M. For any $\varpi$
belonging to the Borel algebra $\sigma E$ of E :

$\int_{\varpi}\rho\left(  x\right)  dx=\int_{\mathcal{E}\left(  m\right)
^{-1}\left(  \varpi\right)  }\mathcal{\mu}\left(  m\right)  \Leftrightarrow
\rho\left(  x\right)  dx=X^{\ast}\mu$

If $H_{1}$ is true, the probability $\Pr\left(  x|H_{1}\right)  =\rho\left(
x\right)  $ depends only on the value $x,$ that is of the map $X$.

If $H_{0}$ is true the probability depends on the maps and if $m\in M_{1}$ or
$m\in M_{2}$ $(M=M_{1}+M_{2})$

$\Pr\left(  x|H_{0}\wedge m\in M_{1}\right)  =\rho_{1}\left(  x\right)  $

$\Pr\left(  x|H_{0}\wedge m\in M_{2}\right)  =\rho_{2}\left(  x\right)  $

$\Rightarrow\Pr\left(  x|H_{0}\right)  =\left(  1-\lambda\right)  \rho
_{1}\left(  x\right)  +\lambda\rho_{2}\left(  x\right)  $

Moreover we have with some measure $dx$ on E :

$\int_{E}\rho\left(  x\right)  dx=\int_{E}\rho_{1}\left(  x\right)
dx=\int_{E}\rho_{2}\left(  x\right)  dx=1$

The likehood function is the probability of a given batch of data.\ It depends
on the hypothesis :

$L\left(  x|H_{0}\right)  =\Pr\left(  x_{1},x_{2},...x_{N}|H_{0}\right)  =%
{\displaystyle\prod\limits_{p=1}^{N}}
\left(  \left(  1-\lambda\right)  \rho_{1}\left(  x_{p}\right)  +\lambda
\rho_{2}\left(  x_{p}\right)  \right)  $

$L\left(  x|H_{1}\right)  =\Pr\left(  x_{1},x_{2},...x_{N}|H_{1}\right)  =%
{\displaystyle\prod\limits_{p=1}^{N}}
\rho\left(  x_{p}\right)  $

The Theory of Tests gives us some rules (see Kendall t.II). A critical region
is an area $w\subset E^{N}$ such that $H_{0}$ is rejected if $x\in w$.

One considers two risks :

- the risk of type I is to wrongly reject $H_{0}.$ It has the probability :
$\alpha=\Pr(x\in w|H_{0})$

- the risk of type II is to wrongly accept $H_{0}.$ It has the probability :
$1-\beta=\Pr(x\in E^{N}-w|H_{0})$ called the power of the test thus :

$\beta=\Pr(x\in w|H_{1})$

A simple rule, proved by Neyman and Pearson, says that the best critical
region $w$ is defined by :

$w=\left\{  x:\frac{L\left(  x|H_{0}\right)  }{L\left(  x|H_{1}\right)  }\leq
k\right\}  $

the scalar $k$ being defined by : $\alpha=\Pr(x\in w|H_{0}).$ So we are left
with a single parameter $\alpha,$ which can be seen as the rigor of the test.

The critical area $w\subset E^{N}$ is then :

$w=\left\{  x\in E^{N}:%
{\displaystyle\prod\limits_{p=1}^{N}}
\frac{\left(  \left(  1-\lambda\right)  \rho_{1}\left(  x_{p}\right)
+\lambda\rho_{2}\left(  x_{p}\right)  \right)  }{\rho\left(  x_{p}\right)
}\leq k\right\}  $

with :

$\alpha=\int_{w}%
{\displaystyle\prod\limits_{p=1}^{N}}
\left(  \left(  1-\lambda\right)  \rho_{1}\left(  \xi_{p}\right)  +\lambda
\rho_{2}\left(  \xi_{p}\right)  \right)  \left(  d\xi\right)  ^{N}$

It provides a reliable method to build a test, but requires to know, or to
estimate, $\rho,\rho_{1},\rho_{2},\lambda.$

\bigskip

In most of the cases encountered, actually one looks for an anomaly.

$H_{1}$ is unchanged, there is only one map $X,$defined over M. Then :
$\Pr\left(  x|H_{1}\right)  =\rho\left(  x\right)  $

$H_{0}$ becomes :

$M=M_{1}+M_{2}$

$\Pr\left(  m\in M_{1}|H_{0}\right)  =1-\lambda,\Pr\left(  m\in M_{2}%
|H_{0}\right)  =\lambda$

On $M_{1}$ the variable is $X$ :

$\Pr\left(  x_{p}|H_{0}\wedge m_{p}\in M_{1}\right)  =\rho\left(  x\right)
\Rightarrow\Pr\left(  x_{p}|H_{0}\right)  =\left(  1-\lambda\right)
\rho\left(  x\right)  $

On $M_{2}$ the variable becomes $X_{2}$

$\Pr\left(  x_{p}|H_{0}\wedge m_{p}\in M_{2}\right)  =\rho_{2}\left(
x\right)  \Rightarrow\Pr\left(  x_{p}|H_{0}\right)  =\lambda\rho_{2}\left(
x\right)  $

And $w$ is :

$w=\left\{  x\in E^{N}:%
{\displaystyle\prod\limits_{p=1}^{N}}
\frac{\left(  \left(  1-\lambda\right)  \rho\left(  x_{p}\right)  +\lambda
\rho_{2}\left(  x_{p}\right)  \right)  }{\rho\left(  x_{p}\right)  }\leq
k\right\}  $

$w=\left\{  x\in E^{N}:%
{\displaystyle\prod\limits_{p=1}^{N}}
\left(  1-\lambda+\lambda\frac{\rho_{2}\left(  x_{p}\right)  }{\rho\left(
x_{p}\right)  }\right)  \leq k\right\}  $

$\alpha=\int_{w}%
{\displaystyle\prod\limits_{p=1}^{N}}
\left(  \left(  1-\lambda\right)  \rho\left(  x_{p}\right)  +\lambda\rho
_{2}\left(  x_{p}\right)  \right)  \left(  dx\right)  ^{N}$

$\beta=\Pr(x\in w|H_{1})=\int_{w}\left(
{\displaystyle\prod\limits_{p=1}^{N}}
\rho\left(  x_{p}\right)  \right)  \left(  dx\right)  ^{N}$

If there is one observed value such that $\rho\left(  x_{p}\right)  =0$ then
$H_{0}$ should be accepted.\ But, because $\rho,\rho_{2}$ are not well known,
and the imprecision of the experiments, $H_{0}$ would be proven if
$\frac{L\left(  x|H_{0}\right)  }{L\left(  x|H_{1}\right)  }>k$ for a great
number of experiments. So we can say that $H_{0}$ is scientifically proven if :

$\forall\left(  x_{1},x_{2},...x_{N}\right)  :%
{\displaystyle\prod\limits_{p=1}^{N}}
\left(  \left(  1-\lambda\right)  +\lambda\frac{\rho_{2}\left(  x_{p}\right)
}{\rho\left(  x_{p}\right)  }\right)  >k$

By taking $x_{1}=x_{2}=...=x_{N}=x:$

$\forall x:\left(  1-\lambda\right)  +\lambda\frac{\rho_{2}\left(  x\right)
}{\rho\left(  x\right)  }>k^{1/N}$

$\frac{\rho_{2}\left(  x\right)  }{\rho\left(  x\right)  }>\left(
k^{1/N}+\lambda-1\right)  /\lambda$

When $N\rightarrow\infty:k^{1/N}\rightarrow1\Rightarrow\frac{\rho_{2}\left(
x\right)  }{\rho\left(  x\right)  }>1$

So a necessary condition to have a chance to say that a singularity has been
reliably proven is that : $\forall x:\frac{\rho_{2}\left(  x\right)  }%
{\rho\left(  x\right)  }>1.$

The function $\frac{\rho_{2}\left(  x\right)  }{\rho\left(  x\right)  }$ can
be called the Signal to Noise Ratio, by similarity with the Signal Theory.
Notice that we have used very few assumptions about the variables. And we can
state :

\begin{theorem}
\label{QMSingularity}In a system represented by variables $X$ which are maps
defined on a set M and valued in a vector space E, a necessary condition for a
singularity to be detected is that the Signal to Noise Ratio is greater than 1
for all values of the variables in E.
\end{theorem}

This result can be seen in another way : if a signal is acknowledged, then
necessarily it is such that $\frac{\rho_{2}\left(  x\right)  }{\rho\left(
x\right)  }>1.$ Any other signal would be interpreted as related to the
imprecision of the measure.\ So there is a threshold under which phenomena are
not acknowledged, and their value is necessarily above this threshold.\ The
singular phenomena are quantized. One application is the Planck's law (see
JC.Dutailly \textquotedblleft Mathematics in Physics\textquotedblright).

\newpage

\section{PROBABILITY}

One of the main purposes of the model is to know the state $X$, represented by
some vector $\psi\in H.$ The model is fully determinist, in that the values of
the variables $X$ are not assumed to depend on a specific event : there is no
probability law involved in its definition.\ However the value of $X$ which
will be acknowledged at the end of the experiment, when all the data have been
collected and analyzed, differs from its actual value. The discrepancy stems
from the usual imprecision of any measure, but also more fundamentally from
the fact that we estimate a vector in an infinite dimensional vector space
from a batch of data, which is necessarily finite. We will focus on this later
aspect, that is on the discrepancy between an observable $\Phi\left(
X\right)  $ and $X$.

In any practical physical experiment the estimation of $X$ requires the choice
of an observable. We have seen that the most efficient solution is to choose a
primary observable which, furthermore, provides the best statistical
estimator.\ However usually neither the map $\Phi$ nor the basis $\left(
e_{i}\right)  _{i\in I}$ are explicit, even if they do exist. So we can look
at the discrepancy $X-\Phi\left(  X\right)  $ from a different point of view :
for a given, fixed, value of the state $X$, what is the uncertainty which
stems from the choice of $\Phi$ among a large class of observables ? This sums
up to assess the risk linked to the choice of a specification for the
estimation of $X$.

\subsection{Primary observables}

Let us start with primary observables : the observable $\Phi$ is some
projection on a finite dimensional vector subspace of V.

The bases of the vector space $V_{0}$ (such that $O\subset V_{0})$ have the
same cardinality, so we can consider that the set I does not depend on a
choice of a basis (actually one can take $I=$ $%
\mathbb{N}
)$. The set $2^{I}$ is the largest $\sigma-$algebra on I. The set $\left(
I,2^{I}\right)  $ is measurable (Maths.802).

For any fixed $\psi\neq0\in H$ the function

$\widehat{\mu}_{\psi}:2^{I}\rightarrow%
\mathbb{R}
::\widehat{\mu}_{\psi}\left(  J\right)  =\frac{\left\langle \widehat{Y}%
_{J}\psi,\psi\right\rangle }{\left\Vert \psi\right\Vert ^{2}}=\frac{\left\Vert
\widehat{Y}_{J}\psi\right\Vert ^{2}}{\left\Vert \psi\right\Vert ^{2}}$

is a probability law on $\left(  I,2^{I}\right)  $ : it is positive, countably
additive and $\widehat{\mu}_{\psi}\left(  I\right)  =1$ (Maths.11.4.1).

If we see the choice of a finite subset $J\in2^{I}$ as an event in a
probabilist point of view, for a given $\psi\neq0\in H$ the quantity
$\widehat{Y}_{J}\left(  \psi\right)  $ is a random variable, with a
distribution law $\widehat{\mu}_{\psi}$

The operator $\widehat{Y}_{J}$ has two eigen values : 1 with eigen space
$\widehat{Y}_{J}\left(  H\right)  $ and 0 with eigen space $\widehat{Y}%
_{J^{c}}\left(  H\right)  $ . Whatever the primary observable, the value of
$\Phi\left(  X\right)  $ will be $Y_{J}\left(  X\right)  $ for some $J$, that
is an eigen vector of the operator $\Phi=Y_{J},$ and the probability to
observe $\Phi\left(  X\right)  $ , if the system is in the state $X$, is :

$\Pr\left(  \Phi\left(  X\right)  =Y_{J}\left(  X\right)  \right)  =\Pr\left(
J|\psi\right)  =\widehat{\mu}_{\psi}\left(  J\right)  =\frac{\left\Vert
\widehat{Y}_{J}\psi\right\Vert ^{2}}{\left\Vert \psi\right\Vert ^{2}}%
=\frac{\left\Vert \widehat{\Phi}\left(  \Upsilon\left(  X\right)  \right)
\right\Vert _{H}^{2}}{\left\Vert \Upsilon\left(  X\right)  \right\Vert
_{H}^{2}}$

This result still holds if another basis had been chosen : $\Phi\left(
X\right)  $ will be $Y_{J}\left(  X\right)  $ for some $J,$ expressed in the
new basis, but with a set J of same cardinality. And some specification must
always be chosen. So we have :

\bigskip

\begin{theorem}
\label{QMPRobPrimary}For any primary observable $\Phi$, the value $\Phi\left(
X\right)  $ which is measured is an eigen vector of the operator $\Phi,$ and
the probability to measure a value $\Phi\left(  X\right)  $ if the system is
in the state X is :

$\Pr\left(  \Phi\left(  X\right)  |X\right)  =\frac{\left\Vert \widehat{\Phi
}\left(  \Upsilon\left(  X\right)  \right)  \right\Vert _{H}^{2}}{\left\Vert
\Upsilon\left(  X\right)  \right\Vert _{H}^{2}}$
\end{theorem}

\subsection{Secondary observables}

For a secondary observable, as defined previously :

$\Phi=\sum_{p=1}^{n}\lambda_{p}Y_{J_{p}}$

$\widehat{\Phi}=\sum_{p=1}^{n}\lambda_{p}\widehat{\pi}_{H_{p}}$

The vectors decompose as :

$X=Y_{J^{c}}\left(  X\right)  +\sum_{p=1}^{n}X_{p}$

with $X_{p}=Y_{J_{p}}\left(  X\right)  =\sum_{i\in J_{p}}\left\langle \phi
_{i},\widehat{Y}_{Jp}\left(  \Upsilon\left(  X\right)  \right)  \right\rangle
_{H}e_{i}\in V_{p}$

$\Upsilon\left(  X\right)  =\psi=\psi_{c}+\sum_{p=1}^{n}\psi_{p}$ with
$\psi_{p}=\widehat{\pi}_{H_{p}}\left(  \psi\right)  ,\psi_{c}=\widehat{\pi
}_{H_{c}}\left(  \psi\right)  $

where $\psi_{p}$is an eigen vector of $\widehat{\Phi},X_{p}$ is an eigen
vector of $\Phi$ both for the eigen value $\lambda_{p}$

and

$\Phi\left(  X\right)  =\sum_{p=1}^{n}\lambda_{p}X_{p}$

$\widehat{\Phi}\left(  \psi\right)  =\sum_{p=1}^{n}\lambda_{p}\psi_{p}$

If, as above, we see the choice of a finite subset $J\in2^{I}$ as an event in
a probabilist point of view then the probability that $\Phi\left(  X\right)
=\lambda_{p}X_{p}$ if the system is in the state $X$, is given by $\Pr\left(
J_{p}|X\right)  =\frac{\left\Vert \widehat{Y}_{p}\psi\right\Vert ^{2}%
}{\left\Vert \psi\right\Vert ^{2}}=\frac{\left\Vert \psi_{p}\right\Vert ^{2}%
}{\left\Vert \psi\right\Vert ^{2}}$

And we have :

\bigskip

\begin{theorem}
For any secondary observable $\Phi,$ the value $\Phi\left(  X\right)  $ which
is observed if the system is in the state $X$ is a linear combination of eigen
vectors $X_{p}$ of $\Phi$ for the eigen value $\lambda_{p}$: $\Phi\left(
X\right)  =\sum_{p=1}^{n}\lambda_{p}X_{p}$

The probability that $\Phi\left(  X\right)  =\lambda_{p}X_{p}$ is:

$\Pr\left(  \Phi\left(  X\right)  =\lambda_{p}X_{p}|X\right)  =\frac
{\left\Vert \Upsilon\left(  X_{p}\right)  \right\Vert ^{2}}{\left\Vert
\Upsilon\left(  X\right)  \right\Vert ^{2}}$
\end{theorem}

\bigskip

Which can also be stated as : $\Phi\left(  X\right)  $ can take the values
$\lambda_{p}X_{p},$ each with the probability $\frac{\left\Vert \psi
_{p}\right\Vert ^{2}}{\left\Vert \psi\right\Vert ^{2}},$ then $\Phi\left(
X\right)  $ reads as an expected value. This is the usual way it is expressed
in QM.

The interest of these results comes from the fact that we do not need to
explicit any basis, or even the set I. And we do not involve any specific
property of the estimator of $X$, other than $\Phi$ is an observable. The
operator $\widehat{\Phi}$ sums up the probability law.

Of course this result can be seen in another way : as only $\Phi\left(
X\right)  $ can be accessed, one can say that the system takes only the states
$\Phi\left(  \lambda_{p}X_{p}\right)  ,$ with a probability $\frac{\left\Vert
\psi_{p}\right\Vert ^{2}}{\left\Vert \psi\right\Vert ^{2}}.$ This gives a
probabilistic behavior to the system ($X$ becoming a random variable) which is
not present in its definition, but is closer to the usual interpretation of QM.

This result can be illustrated by a simple example. Let us take a single
continuous variable $x$, which takes its values in $%
\mathbb{R}
.$ It is clear that any physical measure will at best give a rational number
$Y\left(  x\right)  \in%
\mathbb{Q}
$ up to some scale. There are only countably many rational numbers for
unaccountably many real scalars. So the probability to get $Y\left(  x\right)
\in%
\mathbb{Q}
$ should be zero. The simple fact of the measure gives the paradox that
rational numbers have an incommensurable weight, implying that each of them
has some small, but non null, probability to appear. In this case I can be
assimilated to $%
\mathbb{Q}
$ , the subsets J are any finite collection of rational numbers.

\subsection{Wave function}

The wave function is a central object in QM, but it has no general definition
and is deemed non physical (except in the Bohm's interpretation).\ Usually
this is a complex valued function, defined over the space of configuration of
the system : the set of all possible values of the variables representing the
system. If it is square integrable, then it belongs to a Hilbert space, and
can be assimilated to the vector representing the state. Because its arguments
comprise the coordinates of objects such as particles, it has a value at each
point, and the square of the module of the function is proportional to the
probability that the measure of the variable takes the values of the arguments
at this point. Its meaning is relatively clear for systems comprised of
particles, but less so for systems which include force fields, because the
space of configuration is not defined. But we will see now how it can be
precisely defined in our framework.

\bigskip

\begin{theorem}
\label{QMWavefunction}In a system modelled by N variables, collectively
denoted $X$, which are maps : $X:M\rightarrow F$ from a common measured set M
to a finite dimensional normed vector space $F$ and belonging to an open
subset of an infinite dimensional, separable, real Fr\'{e}chet vector space V,
such that the evaluation map : $\mathcal{E}\left(  m\right)  :V\rightarrow
F::\mathcal{E}\left(  m\right)  \left(  X\right)  =X\left(  m\right)  $ which
assigns at any X its value in a fixed point m in M is measurable : then for
any state $X$ of the system there is a function : $W:M\times F\rightarrow%
\mathbb{R}
$ such that $W\left(  m,y\right)  =\Pr\left(  \Phi\left(  X\right)  \left(
m\right)  =y|X\right)  $ is the probability that the measure of the value of
any primary observable $\Phi\left(  X\right)  $ at $m$ is $y$.
\end{theorem}

\begin{proof}
The conditions 1 apply, there is a Hilbert space H and an isometry
$\Upsilon:V\rightarrow H.$

To the primary observable $\Phi:V\rightarrow V_{J}$ is associated the
self-adjoint operator $\widehat{\Phi}=\Upsilon\circ\Phi\circ\Upsilon^{-1}$

We can apply the theorem \ref{QMPRobPrimary}: the probability to measure a
value $\Phi\left(  X\right)  =Y$ if the system is in the state $X$ is :

$\Pr\left(  \Phi\left(  X\right)  =Y|X\right)  =\frac{\left\Vert \widehat
{\Phi}\left(  \Upsilon\left(  Y\right)  \right)  \right\Vert _{H}^{2}%
}{\left\Vert \Upsilon\left(  X\right)  \right\Vert _{H}^{2}}=\pi\left(
Y\right)  $

Because only the maps belonging to $V_{J}$ are observed it provides a
probability law $\pi$ on the set $V_{J}$ : $\pi:V_{\sigma}\rightarrow\left[
0,1\right]  $ where $V_{\sigma}$ is the Borel algebra of $V_{J}$.

The evaluation map : $\mathcal{E}_{J}\left(  m\right)  :V_{J}\rightarrow
F::\mathcal{E}_{J}\left(  m\right)  \left(  Y\right)  =Y\left(  m\right)  $
assigns at any $Y\in V_{J}$ its value in the fixed point $m$ in M.

If $y\in F$ is a given vector of $F$, the set of maps in $V_{J}$ which gives
the value $y$ in $m$ is : $\varpi\left(  m,y\right)  =\mathcal{E}_{J}\left(
m\right)  ^{-1}\left(  y\right)  \subset V_{J}.$

The probability that the observable takes the value $y$ at $m$ $\Phi\left(
X\right)  (m)=y$ is

$\pi\left(  \varpi\left(  m,y\right)  \right)  =\pi\left(  \mathcal{E}%
_{J}\left(  m\right)  ^{-1}\left(  x\right)  \right)  $

$=\frac{1}{\left\Vert \Upsilon\left(  X\right)  \right\Vert _{H}^{2}}%
\int_{Y\in\varpi\left(  m,y\right)  }\left\Vert \widehat{\Phi}\left(
\Upsilon\left(  Y\right)  \right)  \right\Vert _{H}^{2}\pi\left(  Y\right)
=W\left(  m,y\right)  $
\end{proof}

\bigskip

If M is endowed with a positive measure $\mu$ and $X$ is a scalar function,
the space V of square integrable maps $\int_{\Omega}\left\vert X\left(
m\right)  \right\vert ^{2}\mu\left(  m\right)  <\infty$ is a separable Hilbert
space $H$, then the conditions 1 are met and $H$ can be identified with the
space of the states.

$W\left(  m,y\right)  =\frac{1}{\left\Vert X\right\Vert _{H}^{2}}\int
_{Y\in\varpi\left(  m,y\right)  }\left\vert Y\right\vert _{H}^{2}=\left(
\int_{\Omega}\left\vert X\right\vert ^{2}\mu\right)  ^{-1}\mu\left(
Y^{-1}\left(  m,y\right)  \right)  $

No structure, other than the existence of the measure $\mu$, is required on M.
But of course if the variables $X$ include derivatives M must be at least a
differentiable manifold.

$W$ can be identified with the square of the wave function of QM.

\newpage

\section{CHANGE\ OF\ VARIABLES}

In the conditions 1 we have noticed that, in the model, the variables could be
defined over different connected domains.\ Actually one can go further and
consider the change of variables, which leads to a theorem similar to the well
known Wigner's theorem. The problem appears in Physics in two different ways,
which reflect the interpretations of Scientific laws.

\subsection{Two ways to define the same state of a system}

\subsubsection{The first way : from a theoretical model}

In the first way the scientist has built a theoretical model, using known
concepts and their usual representation by mathematical objects. A change of
variables appears notably when :

i) The variables are the components of a geometric quantity (a vector, a
tensor,...) expressed in some basis. According to the general Principle of
Relativity, the state of the system shall not depend on the observers (those
measuring the coordinates).\ For instance it should not matter if the state of
a system is measured in different units. The data change, but according to
rules which depend on the mathematical representation which is used, and not
on the system itself. In a change of basis coordinates change but they
represent the same vectorial quantity. We will see another example with
interacting, indistinguishable systems.

ii) The variables are maps, depending on arguments which are themselves
coordinates of some event : $X_{k}=X_{k}\left(  \xi_{1},...\xi_{p_{k}}\right)
.$ Similarly these coordinates $\xi$ can change according to some rules, while
the variable $X_{k}$ represents the same event. A simple example that we will
develop later on is a simple function of the time $X_{k}\left(  t\right)  $
such that the time $t$ can be expressed in different units, or with different
origin : $X_{k}\left(  t\right)  $ and $X_{k}^{\prime}\left(  t\right)
=X_{k}\left(  t+\theta\right)  $ represent the same state.

By definition in both cases there is a continuous bijective map
$U:V\rightarrow V^{\prime}$ such that $X$ and $X^{\prime}=U(X)$ represent the
same state of the system. This is the way mathematicians see a change of
variables, and is usually called the passive way by physicists.

Any primary or secondary observable $\Phi$ is a linear map $\Phi\in L\left(
V;W\right)  $ into a finite dimensional vector subspace $W$. For the new
variable the observable is $\Phi^{\prime}\in L\left(  V;W^{\prime}\right)  .$
Both $W,W^{\prime}\subset V$ but $W^{\prime}$ is not necessarily identical to
$W$. However the assumption that $X^{\prime}=U(X)$ and $X$ represents the same
state of the system implies that for any measure of the state we have a
similar relation : $\Phi^{\prime}\circ U\left(  X\right)  =U\circ\Phi\left(
X\right)  \Leftrightarrow\Phi^{\prime}\circ U=U\circ\Phi$. This is actually
the true meaning of \textquotedblleft represent the same
state\textquotedblright. This means that actually one makes the measures
according to a fixed procedure, given by $\Phi,$ on variables which vary with
$U$. Because $U$ is a bijection on V : $\Phi^{\prime}=U\circ\Phi\circ U^{-1}$ .

\subsubsection{The second way : from experimental measures}

In the second way the scientist makes measures with a device that can be
adjusted according to different values of a parameter, say $\theta$ : the
simplest example is using different units, but often it is the orientation of
the device which can be changed. And the measures $Y\left(  \theta\right)  $
which are taken are related to the choice of parameter for the device. If the
results of experiments show that $Y\left(  \theta\right)  =Q\left(
\theta\right)  Y\left(  \theta_{0}\right)  $ with a bijective map $Q\left(
\theta\right)  $ and $\theta_{0}$ some fixed value of the parameter one can
assume that this experimental relation is a feature of the system itself.

Physicists distinguish a passive transformation, when only the device changes,
and an active transformation, when actually the experiment involves a physical
change on the system. In a passive transformation we come back to the first
way and it is legitimate to assume that we have actually the same state,
represented by different data, reflecting some mathematical change in their
expression, even if the observable, which is valued in a finite dimensional
space, does not account for all the possible values of the variables. In an
active transformation (for instance in the Stern-Gerlach experiment one
changes the orientation of a magnetic field to which the particles are
submitted) one can say that there is some map $U$ acting on the space V of the
states of the system, such that the measure is done by a unique procedure
$\widetilde{\Phi}$ on a state $X$ which is changed by a map $U\left(
\theta\right)  .$ So that the measures are $Y\left(  \theta\right)
=\widetilde{\Phi}\circ U\left(  \theta\right)  X$ and the relation $Y\left(
\theta\right)  =Q\left(  \theta\right)  Y\left(  \theta_{0}\right)  $ reads :
$\widetilde{\Phi}\circ U\left(  \theta\right)  \left(  X\right)  =U\left(
\theta\right)  \circ\widetilde{\Phi}\left(  X\right)  .$ So this is very
similar to the first case, where $\theta$ represents the choice of a frame.

\bigskip

In both cases there is the general idea that the state of the system is
represented by some fixed quantity, which can be measured in different
procedures, so that there is a relation, given by the way one goes from one
procedure to the others, between the measures of the state. In the first way
the conclusion comes from the mathematical definition in a theoretical model :
this is a simple mathematical deduction using the Principle of Relativity.\ In
the second way there is an assumption : that one can extend the experimental
facts, necessarily limited to a finite number of data, to the whole set of
possible values of the variable.

The Theorem 2 is based on the existence of a Fr\'{e}chet manifold structure on
the set of possible values of the maps $X$.\ The same manifold structure can
be defined by different, compatible, atlas.\ So the choice of other variables
can lead to the same manifold, and the fixed quantity that we identify with a
state is just a point on the manifold, and the change of variables is a change
of charts between compatible atlas.\ The variables must be related by
transition maps, that is continuous bijections, but additional conditions are
required, depending on the manifold structure considered.\ For instance for
differentiable manifolds the transition maps must be differentiable.\ We will
request that the transition maps preserve the positive kernel, which plays a
crucial role in Fr\'{e}chet manifolds.

\subsection{Fundamental theorem for a change of variables}

We will summarize these features in the following :

\begin{condition}
\textbf{\label{QMCond2}}

\textit{i) The same system is represented by the variables }$X=\left(
X_{1},...X_{N}\right)  $\textit{\ and }$X^{\prime}=\left(  X_{1}^{\prime
},...X_{N^{\prime}}^{\prime}\right)  $\textit{\ which belong to open subsets
}$\mathit{O}$\textit{,}$\mathit{O}^{\prime}$\textit{\ of the infinite
dimensional, separable, Fr\'{e}chet vector space V.}

\textit{ii) There is a continuous map }$U:V\rightarrow V,$\textit{\ bijective
on }$\left(  O,O^{\prime}\right)  ,$\textit{\ such that }$\mathit{X}%
$\textit{\ and }$X^{\prime}=U(X)$\textit{\ represent the same state of the
system}

iii) $U$ preserves the positive kernel on $V$\footnote{The positive kernel
plays a role similar to the probability of transition between states of the
Wigner's Theorem.}

iv) For any observable $\Phi$ of $X$, and $\Phi^{\prime}$ of $X^{\prime}$ :
$\Phi^{\prime}\circ U=U\circ\Phi$
\end{condition}

\bigskip

The map $U$ shall be considered as part of the model, as it is directly
related to the definition of the variables, and is assumed to be known. There
is no hypothesis that it is linear.

\bigskip

\begin{theorem}
\label{QMChangeVar}Whenever a change of variables on a system meets the
conditions 20 above,

i) there is a unitary, linear, bijective map $\widehat{U}\in%
\mathcal{L}%
\left(  H;H\right)  $ such that : $\forall X\in O:\widehat{U}\left(
\Upsilon\left(  X\right)  \right)  =\Upsilon\left(  U\left(  X\right)
\right)  $ where H is the Hilbert space and $\Upsilon$ is the linear map :
$\Upsilon:V\rightarrow H$ associated to $X,X^{\prime}$

ii) $U$ is necessarily a bijective linear map.

For any observables $\Phi,$ $\Phi^{\prime}$:

iii) $W^{\prime}=\Phi^{\prime}\left(  V\right)  $ is a finite dimensional
vector subspace of $V$, isomorphic to $W=\Phi\left(  V\right)  :W^{\prime
}=U\left(  W\right)  $

iv) the associated operators $\widehat{\Phi}=\Upsilon\circ\Phi\circ
\Upsilon^{-1},\widehat{\Phi}^{\prime}=\Upsilon\circ\Phi^{\prime}\circ
\Upsilon^{-1}$are such that : $\widehat{\Phi}^{\prime}=\widehat{U}%
\circ\widehat{\Phi}\circ\widehat{U}^{-1}$and $H_{\Phi^{\prime}}^{\prime
}=\widehat{\Phi}^{\prime}\left(  H\right)  $ is a vector subspace of H
isomorphic to $H_{\Phi}=\widehat{\Phi}\left(  H\right)  $
\end{theorem}

\begin{proof}
i) Let $V_{0}=O\cup O^{\prime}.$ This is an open set and we can apply the
theorem 2. There is a homeomorphism $\Xi:V_{0}\rightarrow H_{0}$ where $H_{0}
$ is an open subset of a Hilbert space H.\ For a basis $\left(  e_{i}\right)
_{i\in I}$ of Span$\left(  V_{0}\right)  $ there is an isometry $\Upsilon$
such that :

$\Upsilon:V_{0}\rightarrow H_{0}::\Upsilon\left(  Y\right)  =\sum_{i\in
I}\left\langle \phi_{i},\Upsilon\left(  Y\right)  \right\rangle _{H}%
\varepsilon_{i}$

such that :

$\forall i\in I:\varepsilon_{i}=\Upsilon\left(  e_{i}\right)  ;$

$\forall i,j\in I:\left\langle \phi_{i},\varepsilon_{j}\right\rangle
_{H}=\delta_{ij};$

ii) $\Upsilon$defines a positive kernel on $V_{0}:K_{V}\left(  Y_{1}%
,Y_{2}\right)  =\left\langle \Upsilon Y_{1},\Upsilon Y_{2}\right\rangle _{H}$

The sets $\left(  V_{0},\Upsilon,H\right)  $ and $\left(  V_{0},\Upsilon
U,H\right)  $ are two realizations triple of $K_{V}.$ Then there is an
isometry $\varphi$ such that :

$\Upsilon U=\varphi\circ\Upsilon$ (Maths.1200).

$\left\langle UX_{1},UX_{2}\right\rangle _{V}=\left\langle \Upsilon
UX_{1},\Upsilon UX_{2}\right\rangle _{H}=\left\langle \varphi\circ\Upsilon
X_{1},\varphi\circ\Upsilon X_{2}\right\rangle _{H}$

$=\left\langle \Upsilon X_{1},\Upsilon X_{2}\right\rangle _{H}=\left\langle
X_{1},X_{2}\right\rangle $

So $U$ preserves the scalar product on V

Let be : $\widehat{U}=\Upsilon\circ U\circ\Upsilon^{-1}$

$\left\langle \widehat{U}\psi_{1},\widehat{U}\psi_{2}\right\rangle
_{H}=\left\langle \Upsilon\circ U\circ\left(  \Upsilon^{-1}\psi_{1}\right)
,\Upsilon\circ U\circ\left(  \Upsilon^{-1}\psi_{2}\right)  \right\rangle _{H}$

$=\left\langle U\circ\left(  \Upsilon^{-1}\psi_{1}\right)  ,U\circ\left(
\Upsilon^{-1}\psi_{2}\right)  \right\rangle _{V}=\left\langle \left(
\Upsilon^{-1}\psi_{1}\right)  ,\left(  \Upsilon^{-1}\psi_{2}\right)
\right\rangle _{V}$

$=\left\langle \psi_{1},\psi_{2}\right\rangle _{H}$

So $\widehat{U}$ preserves the scalar product on $H$

iii) As seen in Theorem 2 starting from the basis $\left(  \varepsilon
_{i}\right)  _{i\in I}$ of $H$ one can define a Hermitian basis $\left(
\widetilde{\varepsilon}_{i}\right)  _{i\in I}$ of $H$, an orthonormal basis
$\left(  \widetilde{e}_{i}\right)  _{i\in I}$ of V for the scalar product
$K_{V}=\left\langle {}\right\rangle _{V}$ with $\widetilde{e}_{i}%
=\Upsilon^{-1}\left(  \widetilde{\varepsilon}_{i}\right)  $

U is defined for any vector of V, so for $\left(  \widetilde{e}_{i}\right)
_{i\in I}$ of V.\ 

Define : $\widehat{U}\left(  \widetilde{\varepsilon}_{i}\right)  =\widehat
{U}\left(  \Upsilon\left(  \widetilde{e}_{i}\right)  \right)  =\Upsilon\left(
U\left(  \widetilde{e}_{i}\right)  \right)  =\widetilde{\varepsilon}%
_{i}^{\prime}$

The set of vectors $\left(  \widetilde{\varepsilon}_{i}^{\prime}\right)
_{i\in I}$ is an orthonormal basis of $H$:

$\left\langle \widetilde{\varepsilon}_{i}^{\prime},\widetilde{\varepsilon}%
_{j}^{\prime}\right\rangle _{H}=\left\langle \widehat{U}\left(  \Upsilon
\left(  \widetilde{e}_{i}\right)  \right)  ,\widehat{U}\left(  \Upsilon\left(
\widetilde{e}_{j}\right)  \right)  \right\rangle _{H}=\left\langle
\widetilde{e}_{i},\widetilde{e}_{j}\right\rangle _{V}=\delta_{ij}$

The map : $\chi:\ell^{2}\left(  I\right)  \rightarrow H::\chi\left(  y\right)
=\sum_{i\in I}y_{i}\widetilde{\varepsilon}_{i}^{\prime}$ is an isomorphism
(same as in Theorem 2) and $\left(  \widetilde{\varepsilon}_{i}^{\prime
}\right)  _{i\in I}$ is a Hilbertian basis of $H$. So we can write :

$\forall\psi\in H:\psi=\sum_{i\in I}\psi^{i}\widetilde{\varepsilon}%
_{i},\widehat{U}\left(  \psi\right)  =\sum_{i\in I}\psi^{\prime i}%
\widetilde{\varepsilon}_{i}^{\prime}$

and : $\psi^{i}=\left\langle \widetilde{\varepsilon}_{i},\psi\right\rangle
_{H}=\left\langle \widehat{U}\left(  \widetilde{\varepsilon}_{i}\right)
,\widehat{U}\left(  \psi\right)  \right\rangle _{H}=\left\langle
\widetilde{\varepsilon}_{i}^{\prime},\sum_{j\in I}\psi^{\prime j}%
\widetilde{\varepsilon}_{j}^{\prime}\right\rangle _{H}=\psi^{\prime i}$

Thus the map $\widehat{U}$ \ reads : $\widehat{U}:H\rightarrow H::\widehat
{U}\left(  \sum_{i\in I}\psi^{i}\widetilde{\varepsilon}_{i}\right)
=\sum_{i\in I}\psi^{i}\widetilde{\varepsilon}_{i}^{\prime}$

It is linear, continuous and unitary : $\left\langle \widehat{U}\left(
\psi_{1}\right)  ,\widehat{U}\left(  \psi_{2}\right)  \right\rangle
=\left\langle \psi_{1},\psi_{2}\right\rangle $ and $\widehat{U}$ is invertible

$U=\Upsilon^{-1}\circ\widehat{U}\circ\Upsilon$ is linear and bijective

iv) For any primary or secondary observable $\Phi$ there is a self-adjoint,
Hilbert-Schmidt and trace class operator $\widehat{\Phi}$ on the associated
Hilbert space $H$ such that : $\widehat{\Phi}=\Upsilon\circ\Phi\circ
\Upsilon^{-1}.$ For the new variable the observable is $\Phi^{\prime}\in
L\left(  V;W^{\prime}\right)  $ and $W\prime\subset V$ is not necessarily
identical to W. It is associated to the operator : $\widehat{\Phi}^{\prime
}=\Upsilon\circ\Phi^{\prime}\circ\Upsilon^{-1}.$ W and W' are finite
dimensional vector subspaces of V.

\bigskip

$%
\begin{array}
[c]{ccccccccccccc}
&  &  &  &  &  &  &  &  &  &  &  & \\
&  & \Phi &  &  &  & U &  &  &  & \Phi^{\prime} &  & \\
W & \leftarrow & \leftarrow & \leftarrow & V & \rightarrow & \rightarrow &
\rightarrow & V & \rightarrow & \rightarrow & \rightarrow & W^{\prime}\\
\downarrow &  &  &  & \downarrow &  &  &  & \downarrow &  &  &  & \downarrow\\
\downarrow & \Upsilon &  &  & \downarrow & \Upsilon &  & \Upsilon & \downarrow
&  &  & \Upsilon & \downarrow\\
\downarrow &  & \widehat{\Phi} &  & \downarrow &  & \widehat{U} &  &
\downharpoonleft &  & \widehat{\Phi^{\prime}} &  & \downharpoonleft\\
H_{\Phi} & \leftarrow & \leftarrow & \leftarrow & H & \rightarrow &
\rightarrow & \rightarrow & H & \rightarrow & \rightarrow & \rightarrow &
H_{\Phi^{\prime}}\\
&  &  &  &  &  &  &  &  &  &  &  &
\end{array}
$

Because U is a bijection on V : $\Phi^{\prime}\circ U=U\circ\Phi
\Rightarrow\Phi^{\prime}=U\circ\Phi\circ U^{-1}$ and V is globally invariant
by U

$\Phi^{\prime}\left(  V\right)  =W^{\prime}=U\circ\Phi\circ U^{-1}\left(
V\right)  =U\circ\Phi\left(  V\right)  =U\left(  W\right)  $

thus W' is a vector subspace of V isomorphic to W

$\widehat{\Phi}^{\prime}=\Upsilon\circ\Phi^{\prime}\circ\Upsilon^{-1}%
=\Upsilon\circ U\circ\Phi\circ U^{-1}\circ\Upsilon^{-1}=\widehat{U}%
\circ\Upsilon\circ\Phi\circ\Upsilon^{-1}\circ\widehat{U}^{-1}=\widehat{U}%
\circ\widehat{\Phi}\circ\widehat{U}^{-1}$

Let us denote : $\widehat{\Phi}\left(  H\right)  =H_{\Phi},\widehat{\Phi
}^{\prime}\left(  H\right)  =H_{\Phi^{\prime}}$

$\widehat{U}\left(  H\right)  =H$ because it is a unitary map

$\widehat{\Phi}^{\prime}\left(  H\right)  =\widehat{U}\circ\widehat{\Phi}%
\circ\widehat{U}^{-1}\left(  H\right)  =\widehat{U}\circ\widehat{\Phi}\left(
H\right)  =\widehat{U}\left(  H_{\Phi}\right)  =H_{\Phi^{\prime}}$

thus $H_{\Phi^{\prime}}$ is a vector subspace of $H$ isomorphic to $H_{\Phi}$
\end{proof}

\bigskip

As a consequence the map U is necessarily linear, even if this was not assumed
in the conditions 20 : variables which are not linearly related (in the
conditions 20) cannot represent the same state.

As $\widehat{U}$ is unitary, it cannot be self adjoint or trace class (except
if $U=Id$). So it differs from an observable.

\subsubsection{Change of units}

A special case of this theorem is the choice of units to measure the
variables. A change of units is a map : $X_{k}^{\prime}=\alpha_{k}X_{k}$ with
fixed scalars $\left(  \alpha_{k}\right)  _{k=1}^{N}.$ As we must have :

$\left\langle U\left(  X_{1}\right)  ,U\left(  X_{2}\right)  \right\rangle
_{V}=\left\langle X_{1},X_{2}\right\rangle _{V}=\sum_{k=1}^{N}\alpha_{k}%
^{2}\left\langle X_{1},X_{2}\right\rangle _{V}=\left\langle X_{1}%
,X_{2}\right\rangle _{V}\Rightarrow\sum_{k=1}^{N}\alpha_{k}^{2}=1$

which implies for any single variable $X_{k}:\alpha_{k}=1.$ So the variables
in the model should be dimensionless quantities. This is in agreement with the
elementary rule that any formal theory should not depend on the units which
are used.\ 

More generally whenever one has a law which relates quantities which are not
expressed in the same units, there should be some fundamental constant
involved, to absorb the discrepancy between the units. For instance some
Physicals laws involve an exponential, such as the wave equation for a plane
wave :

$\psi=\exp i\left(  \left\langle \overrightarrow{k},\overrightarrow
{r}\right\rangle -\varpi t\right)  $

They require that the argument in the exponential is dimensionless, and
because $\overrightarrow{r}$ is a length and $t$ a time we should have a
fundamental constant with the dimension of a speed (in this case $c$).

But also it implies that there should be some \textquotedblleft universal
system of units\textquotedblright\ (based on a single quantity) in which all
quantities of the theory can be measured. In Physics this is the Planck's
system which relate the units of different quantities through the values of
the fundamental constants c, G (gravity), R (Boltzmann constant), $\hbar,$ and
the charge of the electron (see Wikipedia for more).

\bigskip

Usually the variables are defined with respect to some frame, then the rules
for a change of frame have a special importance and are a defining feature of
the model.\ When the rules involve a group, the previous theorem can help to
precise the nature of the abstract Hilbert space H and from there the choice
of the maps $X$.

\subsection{Group representation}

\subsubsection{Summary of representation of groups}

The theory of group representation is a key tool in Physics. We will remind
some basic results here, see Maths.23 for a comprehensive study of this topic.

The left action of a group G on a set E is a map :$\lambda:G\times
E\rightarrow E::\lambda\left(  g,x\right)  $ such that $\lambda\left(
gg^{\prime},x\right)  =\lambda\left(  g,\lambda\left(  g^{\prime},x\right)
\right)  ,\lambda\left(  1,x\right)  =x.\ $And similarly for a right action
$\rho\left(  x,g\right)  .$

The representation of a group G is a couple $(E,f)$ of a vector space E and a
continuous map $f:G\rightarrow G%
\mathcal{L}%
\left(  E;E\right)  $ (the set of linear invertible maps from E to E) such
that :

$\forall g,g^{\prime}\in G:f\left(  g\cdot g^{\prime}\right)  =f\left(
g\right)  \circ f\left(  g^{\prime}\right)  ;f\left(  1\right)  =Id\Rightarrow
f\left(  g^{-1}\right)  =f\left(  g\right)  ^{-1}$

A representation is \textbf{faithful} if $f$ is bijective.

A vector subspace F is \textbf{invariant} if $\forall u\in F,g\in G:f\left(
g\right)  u\in F$

A representation is \textbf{irreducible} if there is no other invariant
subspace than $E,0$.

A representation is not unique : from a given representation one can build
many others. The sum of the representations $\left(  E_{1},f_{1}\right)
,\left(  E_{2},f_{2}\right)  $ is $\left(  E_{1}\oplus E_{2},f_{1}%
+f_{2}\right)  .$

A representation is \textbf{unitary} if there is a scalar product on $E$ and
$f\left(  g\right)  $ is unitary : $\forall u,v\in F,g\in G:\left\langle
f\left(  g\right)  u,f\left(  g\right)  v\right\rangle =\left\langle
u,v\right\rangle $

If two groups G,G' are isomorphic by $\phi$, then a representation $\left(
E,f\right)  $ of G provides a representation of G':

$\phi:G^{\prime}\rightarrow G::\forall g,g^{\prime}\in G^{\prime}:\phi\left(
g\cdot g^{\prime}\right)  =\phi\left(  g\right)  \cdot\phi\left(  g^{\prime
}\right)  ;\phi\left(  1_{G^{\prime}}\right)  =1_{G}\Rightarrow\phi\left(
g^{-1}\right)  =\phi\left(  g\right)  ^{-1}$

$f:G\rightarrow G%
\mathcal{L}%
\left(  E;E\right)  $

Define $f^{\prime}:G^{\prime}\rightarrow G%
\mathcal{L}%
\left(  E;E\right)  ::f^{\prime}\left(  g^{\prime}\right)  =f\left(
\phi\left(  g^{\prime}\right)  \right)  $

$f^{\prime}\left(  g_{1}^{\prime}\cdot g_{2}^{\prime}\right)  =f\left(
\phi\left(  g_{1}^{\prime}\cdot g_{2}^{\prime}\right)  \right)  =f\left(
\phi\left(  g_{2}^{\prime}\right)  \right)  \circ f\left(  \phi\left(
g_{1}^{\prime}\right)  \right)  =f^{\prime}\left(  g_{1}^{\prime}\right)
\circ f^{\prime}\left(  g_{2}^{\prime}\right)  $

A \textbf{Lie group} is a group endowed with the structure of a manifold.\ On
the tangent space $T_{1}G$ at its unity (that we will denote 1) there is an
algebraic structure of \textbf{Lie algebra}, that we will also denote
generally $T_{1}G,$ endowed with a bracket $\left[  {}\right]  $ which is a
bilinear antisymmetric map on $T_{1}G.$

If $G$ is a Lie group with Lie algebra $T_{1}G$ and $\left(  E,f\right)  $ a
representation of $G$, then $\left(  E,f^{\prime}\left(  1\right)  \right)  $
is a representation of the Lie algebra $T_{1}G:$

$f^{\prime}\left(  1\right)  \in%
\mathcal{L}%
\left(  T_{1}G;%
\mathcal{L}%
\left(  E;E\right)  \right)  $

$\forall X,Y\in T_{1}G:f^{\prime}\left(  1\right)  \left(  \left[  X,Y\right]
\right)  =f^{\prime}\left(  1\right)  \left(  X\right)  \circ f^{\prime
}\left(  1\right)  \left(  Y\right)  -f^{\prime}\left(  1\right)  \left(
Y\right)  \circ f^{\prime}\left(  1\right)  \left(  X\right)  $

The converse, from the Lie algebra to the group, holds if $G$ is simply
connected, otherwise a representation of the Lie algebra provides usually
multiple valued representations of the group (we will see important examples later).

Any Lie group G has the \textbf{adjoint representation} $\left(
T_{1}G,Ad\right)  $ over its Lie algebra.

Any irreducible representation of a commutative (abelian) group is unidimensional.

Any unitary representation of a compact or finite group is reducible in the
sum of orthogonal, finite dimensional, irreducible unitary representations.

Any representation of a group on a finite dimensional vector space becomes a
representation on a set of matrices by choosing a basis. The representations
of the common groups of matrices are tabulated. In the standard representation
$\left(  K^{n},\imath\right)  $ of a group G of $n\times n$ matrices on a
field K the map $\imath$\ is the usual action of matrices on column vectors in
the space $K^{n}.$ If G is a Lie group then the standard representation of its
Lie algebra is the representation $\left(  K^{n},\imath\right)  $ by matrices,
deduced by derivation.

Two representations $\left(  E,f\right)  ,\left(  F,\rho\right)  $ of the same
group G are \textbf{equivalent} if there is an isomorphism : $\phi
:E\rightarrow F$ such that :

$\forall g\in G:f\left(  g\right)  =\phi^{-1}\circ\rho\left(  g\right)
\circ\phi$

Then from a basis $\left(  e_{i}\right)  _{i\in I}$ of $E$ one deduces a basis
$|e_{i}>$ of $F$ by : $|e_{i}>=\phi\left(  e_{i}\right)  .$ Because $\phi$ is
an isomorphism $|e_{i}>$ is a basis of $F$.\ Moreover the matrix of the action
of $G$ is in this basis the same as for $\left(  E,f\right)  :$

$\rho\left(  g\right)  $%
$\vert$%
$e_{i}>=$ $\sum_{j\in J}\left[  \rho\left(  g\right)  \right]  _{j}^{i}%
|e_{j}>=\rho\left(  g\right)  \phi\left(  e_{i}\right)  =\phi\circ f\left(
g\right)  \left(  e_{i}\right)  $

$=\phi\left(  \sum_{j\in I}\left[  f\left(  g\right)  \right]  _{i}^{j}%
e_{j}\right)  =\sum_{p\in I}\left[  f\left(  g\right)  \right]  _{i}^{j}%
\phi\left(  e_{j}\right)  =\sum_{p\in I}\left[  f\left(  g\right)  \right]
_{i}^{j}|e_{j}>$

$\left[  \rho\left(  g\right)  \right]  =\left[  f\left(  g\right)  \right]  $

If $K$ is a subgroup of $G$, and $\left(  E,f\right)  $ a representation of
$G$, then $\left(  E,f\right)  $ is a subrepresentation of $K$.

The vector subspaces $F$ of $E$ which are invariant by $K$ provide
representations $\left(  F,f\right)  $ of $K$.

\subsubsection{Change of variable parametrized by a group}

This is the usual case in Physics. The second point of view that we have
noticed above is clear when $U$ is defined by a group. The system is
represented by fixed variables, and the measures are taken according to
procedures which change with $g$ and we have :

$\Phi\left(  g\right)  \left(  X\right)  =U\left(  g\right)  \circ\Phi\left(
1\right)  \left(  X\right)  $

$\Phi\in L\left(  V;W\right)  $ and $U(g)$ is a bijection so $X$ and
$\Phi\left(  1\right)  \left(  X\right)  $ are in bijective correspondence and
$X$ must belong to $W\subset V$ : we reduce the definition of the states at
what can be observed. And to assume that this is true for any observable leads
to redefine $X$ as in the first way, but this requires and additional assumption.

\bigskip

\begin{theorem}
\label{QMGroup}If the conditions 20 are met, and $(V,U)$ is a representation
of the group $G$, then:

i) $\left(  H,\widehat{U}\right)  $ is a unitary representation of the group
$G $ with $\widehat{U}\left(  g\right)  =\Upsilon\circ U\left(  g\right)
\circ\Upsilon^{-1}$

ii) For any observable $\Phi\in L\left(  V;W\right)  $ the vector space
$W\subset V$ is invariant by $U$ and $\left(  W,U\right)  $ is a
representation of $G$, and for the associated operator $\widehat{\Phi
}=\widehat{U}\left(  g\right)  \circ\widehat{\Phi}\circ\widehat{U}\left(
g\right)  ^{-1}\in L\left(  H;H_{\Phi}\right)  ,$ $\left(  H_{\Phi}%
,\widehat{U}\right)  $ is a finite dimensional unitary representation of the
group $G$.

If $G$ is a Lie group, and $U$ continuous, then :

iii) $U$ is smooth, $\widehat{U}$ is differentiable and $\left(  \widehat
{U}^{\prime}\left(  1\right)  ,H\right)  $ is an anti-symmetric representation
of the Lie algebra $T_{1}G$ of $G$

iv) For any observable $\Phi\in L\left(  V;W\right)  $ $\left(  H_{\Phi
},\widehat{U}^{\prime}\left(  1\right)  \right)  $ is an anti-symmetric
representation of the Lie algebra $T_{1}G$ of $G$

If $(F,f)$ is a unitary representation of $G$, equivalent to $\left(  H_{\Phi
},\widehat{U}\right)  ,$ and $\Phi$ a primary or secondary observable, then :

v) The results of measures of $\Phi$ for two values $1,g$ and the same state
of the system are related by :

$\Phi\circ U\left(  1\right)  \left(  X\right)  =\sum_{j\in J}X^{j}\left(
1\right)  e_{j},\Phi\circ U\left(  g\right)  \left(  X\right)  =\sum_{j\in
J}X^{j}\left(  g\right)  e_{j}$ for some basis $\left(  e_{i}\right)  _{i\in
I}$ of V

$X^{j}\left(  g\right)  =\sum_{k\in J}\left[  f\left(  g\right)  \right]
_{k}^{j}X^{k}\left(  1\right)  $ where $\left[  f\left(  g\right)  \right]  $
is the matrix of $f(g)$ in orthonormal bases of F

vi) If moreover $G$ is a Lie group and $U,f$ continuous, then the action
$U^{\prime}\left(  1\right)  \left(  \kappa_{a}\right)  $ of $U^{\prime}(1)$
for vectors $\kappa_{a}$ of $T_{1}G$ are expressed by the same matrices
$\left[  K_{a}\right]  $ of the action $f^{\prime}\left(  1\right)  \left(
\kappa_{a}\right)  $ :

$f^{\prime}\left(  1\right)  \left(  \kappa_{a}\right)  \left(  f_{j}\right)
=\sum_{k\in J}\left[  K_{a}\right]  _{j}^{k}f_{k}\rightarrow U^{\prime
}(1)\left(  \kappa_{a}\right)  \left(  e_{j}\right)  =\sum_{k\in J}\left[
K_{a}\right]  _{j}^{k}e_{k}$

and similarly for the observable $\Phi:\Phi\circ U^{\prime}\left(  1\right)
\left(  \kappa_{a}\right)  \left(  e_{j}\right)  =\sum_{k\in J}\left[
K_{a}\right]  _{j}^{k}e_{k}$
\end{theorem}

\begin{proof}
i) The map : $U:G\rightarrow G%
\mathcal{L}%
\left(  V;V\right)  $ is such that : $U\left(  g\cdot g^{\prime}\right)
=U\left(  g\right)  \circ U\left(  g^{\prime}\right)  ;U\left(  1\right)  =Id$
where $G$ is a group and 1 is the unit in $G$.

Then $U(g)$ is necessarily invertible, because $U\left(  g^{-1}\right)
=U\left(  g\right)  ^{-1}$

$\widehat{U}:G\rightarrow%
\mathcal{L}%
\left(  H;H\right)  ::\widehat{U}=\Upsilon\circ U\circ\Upsilon^{-1}$ is such
that :

$\widehat{U}\left(  g\cdot g^{\prime}\right)  =\Upsilon\circ U\left(  g\cdot
g^{\prime}\right)  \circ\Upsilon^{-1}=\Upsilon\circ U\left(  g\right)  \circ
U\left(  g^{\prime}\right)  \circ\Upsilon^{-1}=\Upsilon\circ U\left(
g\right)  \circ\Upsilon^{-1}\circ\Upsilon\circ U\left(  g^{\prime}\right)
\circ\Upsilon^{-1}=\widehat{U}\left(  g\right)  \circ\widehat{U}\left(
g^{\prime}\right)  $

$\widehat{U}\left(  1\right)  =\Upsilon\circ U\left(  1\right)  \circ
\Upsilon^{-1}=Id$

So $\left(  H,\widehat{U}\right)  $ is a unitary representation of the group
$G $ ($\widehat{U}\left(  g\right)  $ is bijective, thus invertible).

ii) For any observable : $\Phi\circ U\left(  g\right)  =U\left(  g\right)
\circ\Phi,\widehat{\Phi}=\widehat{U}\left(  g\right)  \circ\widehat{\Phi}%
\circ\widehat{U}\left(  g\right)  ^{-1}$

Let us take $Y\in W=\Phi\left(  V\right)  :\exists X\in V:Y=\Phi\left(
X\right)  $

$U\left(  g\right)  Y=U\left(  g\right)  \left(  \Phi\left(  X\right)
\right)  =\Phi\left(  U\left(  g\right)  X\right)  \in\Phi\left(  V\right)  $

And similarly

$\widehat{Y}\in\widehat{\Phi}\left(  H\right)  :\exists\psi\in H:\widehat
{Y}=\widehat{\Phi}\left(  \psi\right)  $

$\widehat{U}\left(  g\right)  \widehat{Y}=\widehat{U}\left(  g\right)  \left(
\widehat{\Phi}\left(  \psi\right)  \right)  =\widehat{\Phi}\left(  \widehat
{U}\left(  g\right)  \psi\right)  \in\widehat{\Phi}\left(  H\right)  $

thus $W,H_{\Phi}=\widehat{\Phi}\left(  H\right)  $ are invariant by
$U$,$\widehat{U}$

The scalar product on $H$ holds on the finite dimensional subspace
$\widehat{\Phi}\left(  H\right)  ,$ which is a Hilbert space.

iii) If $G$ is a Lie group and the map $U:G\rightarrow%
\mathcal{L}%
\left(  V;V\right)  $ continuous, then it is smooth (Maths.1789), $\widehat
{U}$ is differentiable and $\left(  \widehat{U}^{\prime}\left(  1\right)
,H\right)  $ is an anti-symmetric representation of the Lie algebra $T_{1}G$
of $G$ :

$\forall\kappa\in T_{1}G:\left(  \widehat{U}^{\prime}\left(  1\right)
\kappa\right)  ^{\ast}=-\left(  \widehat{U}^{\prime}\left(  1\right)
\kappa\right)  $

$\widehat{U}\left(  \exp\kappa\right)  =\exp\widehat{U}^{\prime}\left(
1\right)  \kappa$ where the first exponential is taken on $T_{1}G$ and the
second on
$\mathcal{L}$%
(H;H) (Maths.1886).

iv) $\Phi$ is a primary or secondary observable, and so is $\Phi\circ U\left(
g\right)  ,$ then $\widehat{\Phi}\circ\widehat{U}\left(  g\right)
=\widehat{U}\left(  g\right)  \circ\widehat{\Phi}$ is a self-adjoint, compact
operator, and by the Riesz theorem (Math.1142) its spectrum is either finite
or is a countable sequence converging to 0 (which may or not be an eigen
value) and, except possibly for 0, is identical to the set $\left(
\lambda_{p}\left(  g\right)  \right)  _{p\in%
\mathbb{N}
}$ of its eigen values (Maths.1020). For each distinct eigen value the eigen
spaces $H_{p}\left(  g\right)  $ are orthogonal and H is the direct sum
$H=\oplus_{p\in%
\mathbb{N}
}H_{p}\left(  g\right)  $. For each non null eigen value $\lambda_{p}\left(
g\right)  $ the eigen space $H_{p}\left(  g\right)  $ is finite dimensional.
For a primary observable the eigen values are either 1 or 0.

Because $H_{\Phi}$ is finite dimensional, for each value of g there is an
orthonormal basis $\left(  \widetilde{\varepsilon}_{i}\left(  g\right)
\right)  _{i\in J}$ of $H_{\Phi}$ comprised of a finite number of vectors
which are eigen vectors of $\widehat{\Phi}\circ\widehat{U}\left(  g\right)
:\widehat{\Phi}\circ\widehat{U}\left(  g\right)  \left(  \widetilde
{\varepsilon}_{j}\left(  g\right)  \right)  =\lambda_{j}\left(  g\right)
\widetilde{\varepsilon}_{j}\left(  g\right)  $

Any vector of $H_{\Phi}$ reads :

$\psi=\sum_{j\in J}\psi^{j}\left(  g\right)  \widetilde{\varepsilon}%
_{j}\left(  g\right)  $ and

$\widehat{\Phi}\circ\widehat{U}\left(  g\right)  =\sum_{p\in%
\mathbb{N}
}\lambda_{p}\left(  g\right)  \widehat{\pi}_{H_{p}\left(  g\right)  }$with the
orthogonal projection $\widehat{\pi}_{H_{p}\left(  g\right)  }$ on
$H_{p}\left(  g\right)  .$

And, because any measure belongs to $H_{\Phi}$ it is a linear combination of
eigen vectors

$\Phi\circ U\left(  g\right)  \left(  X\right)  =\Upsilon^{-1}\circ
\widehat{\Phi}\circ\widehat{U}\left(  g\right)  \circ\Upsilon\left(  X\right)
=\Upsilon^{-1}\left(  \sum_{j\in J}\lambda_{j}\left(  g\right)  \psi
^{j}\left(  g\right)  \widetilde{\varepsilon}_{j}\left(  g\right)  \right)  $

$=\sum_{j\in J}\lambda_{j}\left(  g\right)  \psi^{j}\Upsilon^{-1}\left(
\widetilde{\varepsilon}_{j}\left(  g\right)  \right)  =\sum_{j\in J}%
\lambda_{j}\left(  g\right)  \psi^{j}e_{j}\left(  g\right)  $

for some basis $\left(  e_{i}\right)  _{i\in I}$ of V : $e_{j}\left(
g\right)  =\Upsilon^{-1}\left(  \widetilde{\varepsilon}_{j}\left(  g\right)
\right)  $ and $\Phi\circ U\left(  g\right)  \left(  e_{j}\left(  g\right)
\right)  =\lambda_{j}e_{j}\left(  g\right)  $

That we can write :

$\Phi\circ U\left(  g\right)  \left(  X\right)  =\sum_{j\in J}\lambda_{j}%
\psi^{j}\left(  g\right)  e_{j}\left(  g\right)  =\sum_{j\in J}X^{j}\left(
g\right)  e_{j}\left(  g\right)  =U\left(  g\right)  \circ\Phi\left(
X\right)  $

$\Phi\left(  X\right)  =U\left(  g^{-1}\right)  \left(  \sum_{j\in J}%
X^{j}\left(  g\right)  e_{j}\left(  g\right)  \right)  $

v) If the representations $\left(  H_{\Phi},\widehat{U}\right)  ,\left(
F,f\right)  $ are equivalent (which happens if they have the same finite
dimension) there is an isomorphism $\phi:H_{\Phi}\rightarrow F$ which can be
defined by taking an orthonormal basis $\left(  \widetilde{\varepsilon}%
_{i}\left(  g_{0}\right)  \right)  _{i\in J},\left(  f_{j}\left(
g_{0}\right)  \right)  _{j\in J}$ in each vector space, for some fixed
$g_{0}\in G$ that we can take $g_{0}=1:$ $\phi\left(  \sum_{i\in J}\psi
^{j}\widetilde{\varepsilon}_{j}\left(  1\right)  \right)  =\sum_{i\in J}%
\psi^{j}f_{j}\left(  1\right)  \Leftrightarrow\phi\left(  \widetilde
{\varepsilon}_{j}\left(  1\right)  \right)  =f_{j}\left(  1\right)  $

To a change of $g$ corresponds a change of orthonormal basis, both in
$H_{\Phi}$ and $F$, given by the known unitary map $f(g)$ : $f_{j}\left(
g\right)  =f\left(  g\right)  \left(  f_{j}\left(  1\right)  \right)
=\sum_{k\in J}\left[  f\left(  g\right)  \right]  _{j}^{k}f_{k}\left(
1\right)  $ and thus we have the same matrix for $\widehat{U}\left(  g\right)
:$

$\widetilde{\varepsilon}_{j}\left(  g\right)  =\widehat{U}\left(  g\right)
\left(  \widetilde{\varepsilon}_{j}\left(  1\right)  \right)  =\phi^{-1}\circ
f\left(  g\right)  \circ\phi\left(  \widetilde{\varepsilon}_{j}\left(
1\right)  \right)  =\phi^{-1}\circ f\left(  g\right)  \left(  f_{j}\left(
1\right)  \right)  =\sum_{k\in J}\left[  f\left(  g\right)  \right]  _{j}%
^{k}\widetilde{\varepsilon}_{k}\left(  1\right)  $\newline

$%
\begin{array}
[c]{ccccccccccccc}
&  & U\left(  g\right)  &  &  &  & \Phi &  &  &  &  &  & \\
V & \rightarrow & \rightarrow & \rightarrow & V & \rightarrow & \rightarrow &
\rightarrow & W &  &  &  & \\
\downarrow &  &  &  & \downarrow &  &  &  & \downarrow &  &  &  & \\
\downarrow & \Upsilon &  & \Upsilon & \downarrow &  &  & \Upsilon & \downarrow
&  &  &  & \\
\downarrow &  & \widehat{U}\left(  g\right)  &  & \downharpoonleft &  &
\widehat{\Phi} &  & \downarrow &  & \widehat{U}\left(  g\right)  &  & \\
H & \rightarrow & \rightarrow & \rightarrow & H & \rightarrow & \rightarrow &
\rightarrow & H_{\Phi} & \rightarrow & \rightarrow & \rightarrow & H_{\Phi}\\
&  &  &  &  &  &  &  & \downarrow &  &  &  & \downarrow\\
&  &  &  &  &  &  & \phi & \downarrow &  &  & \phi & \downarrow\\
&  &  &  &  &  &  &  & \downarrow &  & f\left(  g\right)  &  & \downarrow\\
&  &  &  &  &  &  &  & F & \rightarrow & \rightarrow & \rightarrow & F
\end{array}
$\newline

$\widetilde{\varepsilon}_{j}\left(  g\right)  =\widehat{U}\left(  g\right)
\left(  \widetilde{\varepsilon}_{j}\left(  1\right)  \right)  =\sum_{k\in
J}\left[  f\left(  g\right)  \right]  _{j}^{k}\widetilde{\varepsilon}%
_{k}\left(  1\right)  $

$e_{j}\left(  g\right)  =\Upsilon^{-1}\left(  \widetilde{\varepsilon}%
_{j}\left(  g\right)  \right)  =\Upsilon^{-1}\left(  \sum_{k\in J}\left[
f\left(  g\right)  \right]  _{j}^{k}\widetilde{\varepsilon}_{k}\left(
1\right)  \right)  $

$=\sum_{k\in J}\left[  f\left(  g\right)  \right]  _{j}^{k}\Upsilon
^{-1}\left(  \widetilde{\varepsilon}_{k}\left(  1\right)  \right)  =\sum_{k\in
J}\left[  f\left(  g\right)  \right]  _{j}^{k}e_{k}\left(  1\right)  $

$e_{j}\left(  g\right)  =\Upsilon^{-1}\circ\widehat{U}\left(  g\right)
\circ\Upsilon\left(  e_{j}\left(  1\right)  \right)  =U\left(  g\right)
\left(  e_{j}\left(  1\right)  \right)  $

Thus the matrix of $U(g)$ to go from $1$ to $g$ is $\left[  f\left(  g\right)
\right]  $

$\Phi\left(  X\right)  =U\left(  g^{-1}\right)  \left(  \sum_{j\in J}%
X^{j}\left(  g\right)  e_{j}\left(  g\right)  \right)  $

$\Phi\circ U\left(  g\right)  \left(  X\right)  =\sum_{j\in J}X^{j}\left(
g\right)  e_{j}\left(  g\right)  =\sum_{j\in J}X^{j}\left(  g\right)
\sum_{k\in J}\left[  f\left(  g^{-1}\right)  \right]  _{j}^{k}e_{k}\left(
1\right)  $

$\Phi\circ U\left(  g_{0}\right)  \left(  X\right)  =\sum_{k\in J}X^{k}\left(
1\right)  e_{k}\left(  1\right)  \Rightarrow\sum_{j\in J}X^{j}\left(
g\right)  \left[  f\left(  g^{-1}\right)  \right]  _{j}^{k}=X^{k}\left(
1\right)  $

$X^{j}\left(  g\right)  =\sum_{k\in J}\left[  f\left(  g\right)  \right]
_{j}^{k}X^{j}\left(  1\right)  $

The measures $\Phi\circ U\left(  g\right)  \left(  X\right)  $ transform with
the known matrix $f\left(  g\right)  .$

vi) $\left(  H_{\Phi},\widehat{U}^{\prime}(1)\right)  ,\left(  F,f^{\prime
}\left(  1\right)  \right)  $ are equivalent, anti-symmetric (or
anti-hermitian for complex vector spaces) representations of the Lie algebra
$T_{1}G.$ If $\left(  \kappa_{a}\right)  _{a=1}^{m}$ is a basis of $T_{1}G$
then $f^{\prime}\left(  1\right)  $, which is a linear map, is defined by the
values of $f^{\prime}(1)\left(  \kappa_{a}\right)  \in L\left(  F;F\right)  .$

$%
\begin{array}
[c]{cccccc}
&  &  & \widehat{U}^{\prime}\left(  1\right)  \left(  \kappa\right)  &  & \\
& H_{\Phi} & \rightarrow & \rightarrow & \rightarrow & H_{\Phi}\\
& \downarrow &  &  &  & \downarrow\\
\phi & \downarrow &  &  & \phi & \downarrow\\
& \downarrow &  & f^{\prime}\left(  1\right)  \left(  \kappa\right)  &  &
\downarrow\\
& F & \rightarrow & \rightarrow & \rightarrow & F
\end{array}
$\newline

$\widehat{U}^{\prime}\left(  1\right)  \left(  \kappa\right)  \left(
\psi\right)  =\phi^{-1}\circ f^{\prime}\left(  1\right)  \left(
\kappa\right)  \circ\phi\left(  \psi\right)  $

If we know the values of the action of $f^{\prime}\left(  1\right)  \left(
\kappa_{a}\right)  $ on any orthonormal basis $\left(  f_{j}\right)  _{j\in
J}$ of F :

$f^{\prime}\left(  1\right)  \left(  \kappa_{a}\right)  \left(  f_{j}\right)
=\sum_{k\in J}\left[  K_{a}\right]  _{j}^{k}f_{k}$

we have the value of $\widehat{U}^{\prime}\left(  1\right)  \left(  \kappa
_{a}\right)  $ for the corresponding orthonormal basis $\left(  \widehat
{\varepsilon}_{j}\right)  _{j\in J}$ of $H_{\Phi}$

$\widehat{U}^{\prime}\left(  1\right)  \left(  \kappa_{a}\right)  \left(
\widehat{\varepsilon}_{j}\right)  =\widehat{U}^{\prime}\left(  1\right)
\left(  \kappa_{a}\right)  \phi^{-1}\left(  f_{j}\right)  =\phi^{-1}\circ
f^{\prime}\left(  1\right)  \left(  \kappa_{a}\right)  \left(  f_{j}\right)  $

$=\phi^{-1}\left(  \sum_{k\in J}\left[  K_{a}\right]  _{j}^{k}f_{k}\right)
=\sum_{k\in J}\left[  K_{a}\right]  _{j}^{k}\widehat{\varepsilon}_{k}$

So $\widehat{U}^{\prime}\left(  1\right)  $ is represented in an orthonormal
basis of $H_{\Phi}$ by the same matrices $\left[  K_{a}\right]  $

And similarly :

$\widehat{U}\left(  g\right)  =\Upsilon\circ U\left(  g\right)  \circ
\Upsilon^{-1}\Rightarrow\widehat{U}^{\prime}\left(  1\right)  \left(
\kappa\right)  =\Upsilon\circ U^{\prime}\left(  1\right)  \left(
\kappa\right)  \circ\Upsilon^{-1}$

$U^{\prime}(1)\left(  \kappa_{a}\right)  \left(  e_{j}\right)  =\Upsilon\circ
U^{\prime}\left(  1\right)  \left(  \kappa_{a}\right)  \circ\Upsilon
^{-1}\left(  e_{j}\right)  =\Upsilon\circ U^{\prime}\left(  1\right)  \left(
\kappa_{a}\right)  \left(  \widehat{\varepsilon}_{j}\right)  =\Upsilon\left(
\sum_{k\in J}\left[  K_{a}\right]  _{j}^{k}\widehat{\varepsilon}_{k}\right)
=\sum_{k\in J}\left[  K_{a}\right]  _{j}^{k}e_{k}$

vii) Because $\Phi\circ U\left(  g\right)  =U\left(  g\right)  \circ
\Phi\Rightarrow\Phi\circ U^{\prime}\left(  1\right)  \left(  \kappa
_{a}\right)  =U^{\prime}\left(  1\right)  \left(  \kappa_{a}\right)  \circ
\Phi:$

$\Phi\circ U^{\prime}\left(  1\right)  \left(  \kappa_{a}\right)  \left(
e_{j}\right)  =\sum_{k\in J}\left[  K_{a}\right]  _{j}^{k}\Phi\left(
e_{k}\right)  $
\end{proof}

\bigskip

This result is specially important in Physics. Any unitary representation of a
compact or finite group is reducible in the sum of orthogonal, finite
dimensional, irreducible unitary representations. As a consequence the space
$V$ of the variables $X$ has the same structure. If, as it can be assumed, the
state of the system stays in the same irreducible representation, it can
belong only to some specific finite dimensional spaces, defined through the
representation or an equivalent representation of $G$. $X$ depends only on a
finite number of parameters, This is the starting point of quantization.

Notice that the nature of the space $E$ does not matter, only the matrices
$\left[  f\left(  g\right)  \right]  ,\left[  K\right]  .$

Usually in Physics the changes are not parametrized by the group, but by a
vector of the Lie algebra (for instance rotations are not parametrized by a
matrix but by a vector representing the rotation), which gives a special
interest to the two last results.

The usual geometric representations, based on frames defined through a point
and a set of vectors, such as in Galilean Geometry and Special Relativity,
have been generalized by the formalism of fiber bundles, which encompasses
also General Relativity, and is the foundation of gauge theories. Gauge
theories use abundantly group transformations, so they are a domain of choice
to implement the previous results.

\subsubsection{Fourier transform}

If $G$ is an abelian group we have more. Irreducible representations of
abelian groups are unidimensional, and any unitary representation of an
abelian group is the sum of projections on unidimensional vector subspaces
which, for infinite dimensional representations, takes the form of spectral
integrals. More precisely, there is a bijective correspondence between the
unitary representation of an abelian group $G$ and the spectral measures on
the Pontryagin dual $\widehat{G},$ which is the space of continuous maps :
$\vartheta:G\rightarrow T$ where $T$ is the set of complex numbers of module 1
(Maths.1932). This can be made less abstract if $G$ is a topological, locally
compact group.\ Then it has a Haar measure $\mu$ and the representation
$\left(  H,\widehat{U}\right)  $ is equivalent to $\left(  L^{2}\left(  G,\mu,%
\mathbb{C}
\right)  ,%
\mathcal{F}%
\right)  $ that is to the Fourier transform $%
\mathcal{F}%
$\ on complex valued, square integrable, functions on $G$ (Maths.2421).

If $\varphi\in L^{2}\left(  G,\mu,%
\mathbb{C}
\right)  \cap L^{1}\left(  G,\mu,%
\mathbb{C}
\right)  :$

$%
\mathcal{F}%
\left(  \varphi\right)  \left(  \vartheta\right)  =\int_{G}\varphi\left(
g\right)  \overline{\vartheta\left(  g\right)  }\mu\left(  g\right)  $

$%
\mathcal{F}%
^{\ast}\left(  h\right)  \left(  g\right)  =\int_{\widehat{G}}h\left(
\vartheta\right)  \vartheta\left(  g\right)  \nu\left(  \vartheta\right)  $
for a unique Haar measure $\nu$ on $\widehat{G}$ and $%
\mathcal{F}%
^{\ast}=%
\mathcal{F}%
^{-1}$

If $G$ is a compact group then we have Fourier series on a space of periodic
functions, and if $G$ is a non compact, finite dimensional Lie group, $G$ is
isomorphic to some vector space $E$ and we have the usual Fourier transform on
functions on $E$.

These cases are important from a practical point of view as it is possible to
replace the abstract Hilbert space $H$ by more familiar spaces of functions,
and usually one can assume that the space $V$ is itself some Hilbert space.
The previous tools (observables,...) are then directly available.

The most usual application is about periodic phenomena : whenever a system is
inclosed in some box, it can be usually assumed that they are periodic (and
null out of the box). Then the representation is naturally through Fourier
series and we have convenient Hilbert bases.

\subsubsection{One parameter groups}

An important case, related to the previous one, is when the variables $X$
depend on a scalar real argument, and the model is such that $X\left(
t\right)  ,X^{\prime}(t^{\prime})=X(t+\theta),$ with any fixed $\theta,$
represent the same state. The associated operator is parametrized by a scalar
and we have a map :

$\widehat{U}:%
\mathbb{R}
_{+}\rightarrow G%
\mathcal{L}%
\left(  H,H\right)  $ such that :

$\widehat{U}\left(  t+t^{\prime}\right)  =\widehat{U}\left(  t\right)
\circ\widehat{U}\left(  t^{\prime}\right)  $

$\widehat{U}\left(  0\right)  =Id$

Then we have a one parameter semi-group. If moreover the map $\widehat{U}$ is
strongly continuous (that is $\lim_{\theta\rightarrow0}\left\Vert \widehat
{U}\left(  \theta\right)  -Id\right\Vert =0$ ), it can be extended to $%
\mathbb{R}
.\left(  \widehat{U},H\right)  $ is a unitary representation of the abelian
group $\left(
\mathbb{R}
,+\right)  .$ We have a one parameter group, and because $\widehat{U}$ is a
continuous Lie group morphism it is differentiable with respect to $\theta$ (Maths.1784).

Any strongly continuous one parameter group of operators on a Banach vector
space admits an infinitesimal generator $S\in%
\mathcal{L}%
\left(  H;H\right)  $ such that : $\widehat{U}\left(  t\right)  =\sum
_{n=0}^{\infty}\frac{t^{n}}{n!}S^{n}=\exp tS$ (Maths.1033). By derivation with
respect to $t$ we get : $\frac{d}{ds}\widehat{U}\left(  s\right)
|_{t=s}=\left(  \exp tS\right)  \circ S\Rightarrow S=\frac{d}{ds}\widehat
{U}\left(  s\right)  |_{t=0}$

Because $\widehat{U}\left(  t\right)  $ is unitary S is anti-hermitian :

$\left\langle \widehat{U}\left(  t\right)  \psi,\widehat{U}\left(  t\right)
\psi^{\prime}\right\rangle _{H}=\left\langle \psi,\psi^{\prime}\right\rangle
_{H}$

$\Rightarrow\left\langle \frac{d}{dt}\widehat{U}\left(  t\right)
\psi,\widehat{U}\left(  t\right)  \psi^{\prime}\right\rangle _{H}+\left\langle
\widehat{U}\left(  t\right)  \psi,\frac{d}{dt}\widehat{U}\left(  t\right)
\psi^{\prime}\right\rangle _{H}=0\Rightarrow S=-S^{\ast}$

$S$ is normal and has a spectral resolution $P$ :

$S=\int_{Sp\left(  S\right)  }sP\left(  s\right)  $

$S$ is anti-hermitian so its eigen-values are pure imaginary : $\lambda
=-\overline{\lambda}.$ $\widehat{U}\left(  t\right)  $ is not compact and $S$
is not compact, usually its spectrum is continuous, so it is not associated to
any observable.

We will see in the next Chapter a striking application of this case.

\subsection{Extension to manifolds}

Several extensions of the theorem 2 can be considered.\ One problem that we
will meet in the next chapters is the following.\ In a model variables $X$ are
maps defined on a manifold $M$, valued in a fixed vector space, and belong to
a space $V$ of maps with the required properties. But a variable $Y $ is
defined through $X$ : $Y\left(  m\right)  =f\left(  X\left(  m\right)
\right)  $ and belongs to a manifold $N\left(  X\right)  $ depending on
$X$.\ So the conditions 1 do not apply.

To address this kind of problem we need to adapt our point of view. We have
seen the full mathematical definition of a manifold in the first section. A
manifold $M$ is a class of equivalence : the same point $m$ of $M$ can be
defined by several charts, maps $\varphi:E\rightarrow M$ from a vector space
$E$ to $M$, with different coordinates : $m=\varphi_{a}\left(  \xi_{a}\right)
=\varphi_{b}\left(  \xi_{b}\right)  $ so that it defines classes of
equivalence between sets of coordinates : $\xi_{a}\sim\xi_{b}\Leftrightarrow
\varphi_{a}\left(  \xi_{a}\right)  =\varphi_{b}\left(  \xi_{b}\right)
.\ $These classes of equivalence are made clear by the transitions maps
$\chi_{ba}:E\rightarrow E$, which are bijective : $\xi_{a}\sim\xi
_{b}\Leftrightarrow\xi_{b}=\chi_{ba}\left(  \xi_{a}\right)  .$ And these
transitions maps are the key characteristic of the manifold. To a point $m$ of
$M$ corresponds a class of equivalence of coordinates.

So let us consider a system represented by a model which meets the following :

\begin{condition}
\label{QMCond3Manifold}

The model is comprised of :

i) A finite number of variables, collectively denoted $X$, which are maps
valued in a vector space $E$ and meeting the conditions 1 : they belong to an
open subset $O$ of a separable, infinite dimensional Fr\'{e}chet space $V$.

ii) A variable $Y$, valued in a set $F$, defined by a map : $f:O\rightarrow
F::Y=f\left(  X\right)  $

iii) A collection of linear continuous bijective maps $\mathfrak{U=}\left(
U_{a}\in G%
\mathcal{L}%
\left(  V;V\right)  \right)  _{a\in A},$ comprising the identity, closed under
composition : $\forall a,b\in A:U_{a}\circ U_{b}\in\mathfrak{U}$

iv) On $V$ and $F$ the equivalence relation :

$R:X\sim X^{\prime}\Leftrightarrow\exists a\in A:X^{\prime}=U_{a}\left(
X\right)  :f\left(  X\right)  =f\left(  X^{\prime}\right)  $
\end{condition}

Denote the set $N=\left\{  Y=f\left(  X\right)  ,X\in O\right\}  .$ The
quotient set : $N/R$ is comprised of classes of equivalence of points Y which
can be defined by related coordinates. This is a manifold, which can be
discrete and comprising only a finite number of points. One can also see the
classes of equivalence of $N/R$ as representing states of the system, defined
equivalently by the variable $X,X^{\prime}=U_{a}\left(  X\right)  .$

Notice that $f$ is unique, no condition is required on $E$ other than to be a
vector space, and nothing on $F$. Usually the maps $U_{a}$ are defined by :
$U_{a}\left(  X\right)  =\chi_{a}\circ X$ where the maps $\chi_{a}\in
GL\left(  E;E\right)  $ are bijective on $E$ (not $F$ or $V$) but only the
continuity of $U_{a}$ can be defined.

We have the following result :

\begin{theorem}
\label{QMMAnifold}For a system represented by a model meeting the conditions
23 :

i) V can be embedded as an open of a Hilbert space $H$ with a linear isometry
$\Upsilon:V\rightarrow H,$ to each $U_{a}$ is associated the unitary operator
$\widehat{U}_{a}=\Upsilon\circ U_{a}\circ\Upsilon^{-1}$ on H, each class of
equivalence $\left[  V\right]  _{y}$ of R on V is associated to a class of
equivalence $\left[  H\right]  _{y}$ in H of :

$\widehat{R}:\psi\sim\psi^{\prime}\Leftrightarrow\exists a\in A:\psi^{\prime
}=\widehat{U}_{a}\left(  \psi\right)  .$ $\left[  V\right]  _{y}$ is a
partition of V and $\left[  H\right]  _{y}$ of H.

ii) If $\left(  V,U\right)  $ is a representation of a Lie group $G$, then
$\left(  H,\widehat{U}\right)  $ is a unitary representation of $G$ and each
$\left[  H\right]  _{y}$is invariant by the action of G.
\end{theorem}

\begin{proof}
i) R defines a partition of V, we can label each class of equivalence by the
value of Y, and pick one element $X_{y}$ in each class :

$\left[  V\right]  _{y}=\left\{  X\in O:f\left(  X\right)  \sim f\left(
X_{y}\right)  =y\right\}  \equiv\left\{  X\in O:\exists a\in A:X=U_{a}\left(
X_{y}\right)  \right\}  $

$\equiv\left\{  X\in O:X=U_{a}\left(  X_{y}\right)  ,a\in A\right\}  $

The variables $X$ meet the conditions 1, $O$ can be embedded as an open of a
Hilbert space $H$ and there is linear isomorphism : $\Upsilon:V\rightarrow H$

In $\left[  V\right]  _{y}$ the variables $X,X^{\prime}=U_{a}\left(  X\right)
$ define the same state and we can implement the theorem \ref{QMChangeVar}%
.\ $\widehat{U}_{a}=\Upsilon\circ U_{a}\circ\Upsilon^{-1}$ is an unitary
operator on $H$

$\forall X\in\left[  V\right]  _{y}:\widehat{U}_{a}\circ\Upsilon\left(
X_{y}\right)  =\Upsilon\circ U_{a}\left(  X_{y}\right)  =\Upsilon\left(
X\right)  $

The set $\left[  H\right]  _{y}=\Upsilon\left(  \left[  V\right]  _{y}\right)
=\left\{  \psi\in H:\psi=\widehat{U}_{a}\left(  \Upsilon\left(  X_{y}\right)
\right)  ,a\in A\right\}  $ is the class of equivalence of :

$\widehat{R}:\psi\sim\psi^{\prime}\Leftrightarrow\exists a\in A:\psi^{\prime
}=\widehat{U}_{a}\left(  \psi\right)  $

R defines a partition of $V$ : $V=\cup_{y}\left[  V\right]  _{y}$ and
$\widehat{R}$ defines a partition of $H$ : $H=\cup_{y}\left[  H\right]  _{y}$

ii) If $\left(  V,U\right)  $ is a representation of a Lie group $G$ then
$\left[  V\right]  _{y}$ is the orbit of $X_{y},$ $\left(  H,\widehat
{U}\right)  $ is a unitary representation of $G$

Each $\left[  H\right]  _{y}$ is invariant by $G$. The vector subspace
$\left[  F\right]  _{y}$ spanned by $\left[  H\right]  _{y}$ is invariant by
$G$, so $\left(  \left[  F\right]  _{y},\widehat{U}\right)  $ is a
representation of G.
\end{proof}

\bigskip

As a consequence of the last result : if $U$ is a compact group, then the
representation $\left(  H,\widehat{U}\right)  $ is the sum of irreducible,
orthogonal, finite dimensional representations.\ For each value of $Y$ the
subset $\left[  H\right]  _{y}$is invariant by the action of $G$, so it must
belong to one of the irreducible representations, as well as $\left[
F\right]  _{y}$.\ The maps $X$, for a given value of $Y$, belong to a finite
dimensional vector space, and depend on a finite number of parameters. This is
the usual meaning of the quantization of $X$.

\newpage

\section{THE\ EVOLUTION OF\ THE\ SYSTEM}

In many models involving maps, the variables $X_{k}$ are functions of the time
$t$, which represents the evolution of the system. So this is a privileged
argument of the functions. So far we have not made any additional assumption
about the model : the open $\Omega$ of the Hilbert space contains all the
possible values but, due to the laws to which it is subject, only some
solutions will emerge, depending on the initial conditions. They are fixed by
the value $X(0)$ of the variables at some origin $0$ of time. They are
specific to each realization of the system, but we should expect that the
model and the laws provide a general solution, that is a map : $X\left(
0\right)  \rightarrow X$ which determines $X$ for each specific occurrence of
$X(0)$. It will happen if the laws are determinist. One says that the problem
is well posed if for any initial conditions there is a unique solution $X$,
and that $X$ depends continuously on $X(0)$. We will give a more precise
meaning of determinism by enlarging the conditions 1 as follows :

\begin{condition}
\textbf{\label{QMCon4EvolutionGen}} : \textit{The model representing the
system meets the conditions 1. Moreover :}

i) $V$ is an \textit{infinite dimensional separable Fr\'{e}chet space
}$\mathit{V}$\textit{\ of maps : }$X=\left(  X_{k}\right)  _{k=1}%
^{N}::R\rightarrow E$ \textit{where }$R$\textit{\ is an open subset of }$%
\mathbb{R}
$\textit{\ and }$E$\textit{\ a normed vector space}

ii) $\forall t\in R$ \textit{the evaluation map} : $\mathcal{E}\left(
t\right)  :V\rightarrow E:\mathcal{E}\left(  t\right)  X=X\left(  t\right)  $
\textit{is continuous}

\textit{The laws for the evolution of the system are such that the variables
}$\left(  X_{k}\right)  _{k=1}^{N}$\textit{, which define the possible states
considered for the system (that we call the admissible states) meet the
conditions :}

iii) \textit{The initial state of the system, defined at }$t=0\in R$\textit{,
belongs to an open subset }$A$\textit{\ of }$E$

iv) \textit{For any solutions }$X,X^{\prime}$\textit{\ belonging to }$O$
\textit{if the set }$\varpi=\left\{  t,X(t)=X(t^{\prime})\right\}
$\textit{\ has a non null Lebesgue measure then }$X=X^{\prime}$\textit{.}
\end{condition}

The last condition iv) means that the system is semi determinist : to the same
initial conditions can correspond several different solutions, but if two
solutions are equal on some interval then they are equal almost everywhere.

The condition ii) is rather technical and should be usually met. Practically
it involves some relation between the semi-norms on $V$ and the norm on $E$
(this is why we need a norm on $E$) : when two variables $X,X^{\prime}$ are
close in $V$, then their values $X\left(  t\right)  ,X^{\prime}\left(
t\right)  $ must be close for almost all $t$. More precisely, because
$\mathcal{E}\left(  t\right)  $ is linear, the continuity can be checked at
$X=$ $0$ and reads:

$\forall t\in R,\forall X\in O:\forall\varepsilon>0,\exists\eta:d\left(
X,0\right)  _{V}<\eta\Rightarrow\left\Vert X\left(  t\right)  \right\Vert
_{E}<\varepsilon$ where $d$ is the metric on $V$

In all usual cases (such as $L^{p}$ spaces or spaces of differentiable
functions) $d\left(  X,0\right)  _{V}\rightarrow0\Rightarrow\forall t\in
R:\left\Vert X\left(  t\right)  \right\Vert _{E}\rightarrow0$ and the
condition ii) is met, but this is not a general result.

Notice that :

- the variables $X$ can depend on any other arguments besides $t$ as previously

- $E$ can be infinite dimensional but must be normed

- no continuity condition is imposed on $X$.

\subsection{Fundamental theorems for the evolution of a system}

If the model meets the conditions 25 then it meets the conditions 1 : there is
a separable, infinite dimensional, Hilbert space H, defined up to isomorphism,
such that the states (admissible or not) $\mathcal{S}$ belonging to O can be
embedded as an open subset $\Omega\subset H$ which contains 0 and a convex
subset. Moreover to any basis of V is associated a bijective linear map
$\Upsilon:V\rightarrow H.$

\begin{theorem}
\label{QMEvolutGen}If the conditions 25 are met, then there are :

i) a Hilbert space $F$, an open subset $\widetilde{A}\subset F$

ii) a map : $\Theta:R\rightarrow%
\mathcal{L}%
\left(  F;F\right)  $ such that $\Theta\left(  t\right)  $ is unitary and, for
the admissible states $X\in O\subset V:$

$X\left(  0\right)  \in\widetilde{A}\subset F$

$\forall t:X\left(  t\right)  =\Theta\left(  t\right)  \left(  X\left(
0\right)  \right)  \in F$

iii) for each value of $t$ an isometry : $\widehat{\mathcal{E}}\left(
t\right)  \in%
\mathcal{L}%
\left(  H;F\right)  $ such that for the admissible states $X\in O\subset V:$

$\forall X\in O:\widehat{\mathcal{E}}\left(  t\right)  \Upsilon\left(
X\right)  =X\left(  t\right)  $

where $H$ is the Hilbert space and $\Upsilon$ is the linear chart associated
to X and any basis of V
\end{theorem}

\begin{proof}
i) Define the equivalence relation on V :

$\mathcal{R}:X\sim X^{\prime}\Leftrightarrow X(t)=X^{\prime}(t)$ for almost
every $t\in R$

and take the quotient space $V/\mathcal{R},$ then the set of admissible states
is a set $\widetilde{O}$ such that :

$\widetilde{O}\in O\subset V$

$\forall X\in\widetilde{O}:X\left(  0\right)  \in A$

$\forall X,X^{\prime}\in\widetilde{O},\forall t\in R:X(t)=X^{\prime
}(t)\Rightarrow X=X^{\prime}$

ii) Define :

$\forall t\in R:\widetilde{F}\left(  t\right)  =\left\{  X\left(  t\right)
,X\in\widetilde{O}\right\}  $ thus $\widetilde{F}\left(  0\right)  =A$

A is a subset of E.\ There are families of independent vectors belonging to A,
and a largest family $\left(  f_{j}\right)  _{j\in J}$ of independent vectors.
It generates a vector space $F(0)$ which is a vector subspace of E, containing
A$.$

$\forall u\in F\left(  0\right)  :\exists\left(  x_{j}\right)  _{j\in J}\in%
\mathbb{R}
_{0}^{J}:u=\sum_{j\in J}x_{j}f_{j}$

The map :

$\widetilde{\Theta}\left(  t\right)  :\widetilde{F}\left(  0\right)
\rightarrow\widetilde{F}\left(  t\right)  ::\widetilde{\Theta}\left(
t\right)  u=\mathcal{E}\left(  t\right)  \circ\mathcal{E}\left(  0\right)
^{-1}u$

is bijective and continuous

The set $F\left(  t\right)  =\widetilde{\Theta}\left(  t\right)  F\left(
0\right)  \subset E$ is well defined by linearity :

$\widetilde{\Theta}\left(  t\right)  \left(  \sum_{j\in J}x_{j}f_{j}\right)
=\sum_{j\in J}x_{j}\widetilde{\Theta}\left(  t\right)  \left(  f_{j}\right)  $

The map : $\widetilde{\Theta}\left(  t\right)  :F\left(  0\right)  \rightarrow
F\left(  t\right)  $ is linear, bijective, continuous on an open subset A,
thus continuous, and the spaces $F(t)$ are isomorphic, vector subspaces of E,
containing $\widetilde{F}\left(  t\right)  .$

Define : $\left(  \varphi_{j}\right)  _{j\in J}$ the largest family of
independent vectors of

$\left\{  \widetilde{\Theta}\left(  t\right)  \left(  f_{j}\right)  ,t\in
R\right\}  $. This is a family of independent vectors of E, which generates a
subspace $\widetilde{F}$ of E, containing each of the $F(t)$ and thus each of
the $\widetilde{F}\left(  t\right)  .$ Moreover each of the $\varphi_{j}$ is
the image of a unique vector $f_{j}$ for some $t_{j}\in R.$

The map $\widetilde{\Theta}\left(  t\right)  $ is then a continuous linear map
$\widetilde{\Theta}\left(  t\right)  \in%
\mathcal{L}%
\left(  \widetilde{F};\widetilde{F}\right)  $

iii) The conditions of proposition 1 are met for O and V, so there are a
Hilbert space $H$ and a linear map : $\Upsilon:O\rightarrow\Omega$

Each of the $\varphi_{j}$ is the image of a unique vector $f_{j}$ for some
$t\in R,$and thus there is a uniquely defined family $\left(  X_{j}\right)
_{j\in J}$ of $\widetilde{O}$ such that $X_{j}\left(  t_{j}\right)
=\varphi_{j}$.

Define on $\widetilde{F}$ the bilinear symmetric definite positive form with
coefficients :

$\left\langle \varphi_{j},\varphi_{k}\right\rangle _{\widetilde{F}}%
=K_{V}\left(  \mathcal{E}\left(  t_{j}\right)  ^{-1}\varphi_{j},\mathcal{E}%
\left(  t_{k}\right)  ^{-1}\varphi_{k}\right)  $

$=\left\langle \Upsilon\mathcal{E}\left(  t_{j}\right)  ^{-1}\varphi
_{j},\Upsilon\mathcal{E}\left(  t_{k}\right)  ^{-1}\varphi_{k}\right\rangle
_{H}=\left\langle X_{j},X_{k}\right\rangle _{H}$

By the Gram-Schmidt procedure we can build an orthonormal basis $\left(
\widetilde{\varphi}_{j}\right)  _{j\in J}$ of $\widetilde{F}$ : $\widetilde
{F}=Span\left(  \widetilde{\varphi}_{j}\right)  _{j\in J}$ and the Hilbert
vector space : $F=\left\{  \sum_{j\in J}\widetilde{x}_{j}\widetilde{\varphi
}_{j},\left(  \widetilde{x}_{j}\right)  _{j\in J}\in\ell^{2}\left(  J\right)
\right\}  $ which is a vector space containing $\widetilde{F}$ (but is not
necessarily contained in E).

iv) The map : $\widetilde{\Theta}\left(  t\right)  \in%
\mathcal{L}%
\left(  \widetilde{F};\widetilde{F}\right)  $ is a linear homomorphism,
$\widetilde{F} $ is dense in F, thus $\widetilde{\Theta}\left(  t\right)  $
can be extended to a continuous operator $\Theta\left(  t\right)  \in%
\mathcal{L}%
\left(  F;F\right)  $ (Math.1003).

$\widetilde{\Theta}\left(  t\right)  $ is unitary on $\widetilde{F}$ :
$\left\langle u,v\right\rangle _{\widetilde{F}}=K_{V}\left(  \mathcal{E}%
\left(  0\right)  ^{-1}u,\mathcal{E}\left(  0\right)  ^{-1}v\right)  $ so
$\Theta\left(  t\right)  $ is unitary on F.

iv) Define the map :

$\widehat{\mathcal{E}}\left(  t\right)  :\Omega\rightarrow F::\widehat
{\mathcal{E}}\left(  t\right)  \Upsilon\left(  X\right)  =X\left(  t\right)  $

where $\Omega\subset H$ is the open associated to V and O.

For $X\in\widetilde{O}:$

$\widehat{\mathcal{E}}\left(  t\right)  \Upsilon\left(  X\right)  =X\left(
t\right)  =\widetilde{\Theta}\left(  t\right)  X=\mathcal{E}\left(  t\right)
\circ\mathcal{E}\left(  0\right)  ^{-1}X$

$\widehat{\mathcal{E}}\left(  t\right)  =\mathcal{E}\left(  t\right)
\circ\mathcal{E}\left(  0\right)  ^{-1}\circ\Upsilon^{-1}$

$\widehat{\mathcal{E}}\left(  t\right)  $ is linear, continuous, bijective on
$\Omega,$ it is an isometry :

$\left\langle \widehat{\mathcal{E}}\left(  t\right)  \psi,\widehat
{\mathcal{E}}\left(  t\right)  \psi^{\prime}\right\rangle _{F}=\left\langle
X\left(  t\right)  ,X^{\prime}\left(  t\right)  \right\rangle _{F}%
=\left\langle \Upsilon X,\Upsilon X^{\prime}\right\rangle _{H}=\left\langle
\psi,\psi^{\prime}\right\rangle _{H}$

v) $A=\widetilde{F}\left(  0\right)  $ is an open subset of $F(0)$, which is
itself an open vector subspace of F.\ Thus A can be embedded as an open subset
$\widetilde{A}$ of F.
\end{proof}

\bigskip

When $X$ depends on other arguments $\xi$, the result reads :

$\forall t,\forall\xi:X\left(  t,\xi\right)  =\Theta\left(  t\right)  \left(
X\left(  0,\xi\right)  \right)  \in F$

Indeed the basic feature which is used is :

$\forall X,X^{\prime}\in\widetilde{O},\forall t\in R:X(t)=X^{\prime
}(t)\Rightarrow X=X^{\prime}$

which means : $\forall t,\forall\xi:X(t,\xi)=X^{\prime}(t,\xi)\Leftrightarrow
X=X^{\prime}$

As a consequence the model is determinist, up to the equivalence between maps
almost everywhere equal. But the operator $\Theta\left(  t\right)  $ depends
on t and not necessarily continuously, so the problem is not necessarily well
posed. Notice that each solution $X(t)$ belong to E, but the Hilbert space F
can be larger than E. Moreover the result holds if the conditions apply to
some variables only.

But we have a stronger result.

\begin{theorem}
\label{QMEvolutSchrod}If t\textit{he model representing the system meets the
conditions 1 and moreover :}

i) V is an \textit{infinite dimensional separable Fr\'{e}chet space V of maps
: }$X=\left(  X_{k}\right)  _{k=1}^{N}::R\rightarrow E$ \textit{where E is a
normed vector space}

ii) $\forall t\in%
\mathbb{R}
$ \textit{the evaluation map} : $\mathcal{E}\left(  t\right)  :V\rightarrow
E:\mathcal{E}\left(  t\right)  X=X\left(  t\right)  $ \textit{is continuous}

iii) the variables $X_{k}^{\prime}\left(  t\right)  =X_{k}\left(
t+\theta\right)  $ and $X_{k}\left(  t\right)  $ represent the same state of
the system, for any $t^{\prime}=t+\theta$ with a fixed $\theta\in%
\mathbb{R}
$

then :

i) there is a continuous map $S\in%
\mathcal{L}%
\left(  V;V\right)  $ such that :

$\mathcal{E}\left(  t\right)  =\mathcal{E}\left(  0\right)  \circ\exp tS$

$\forall t\in%
\mathbb{R}
:X\left(  t\right)  =\left(  \exp tS\circ X\right)  \left(  0\right)  =\left(
\sum_{n=0}^{\infty}\frac{t^{n}}{n!}S^{n}X\right)  \left(  0\right)  $

and the operator $\widehat{S}=\Upsilon\circ S\circ\Upsilon^{-1}$ associated to
S is anti-hermitian

ii) there are a Hilbert space F, an open $\widetilde{A}\subset F,$ a
continuous anti-hermitian map $\widetilde{S}\in%
\mathcal{L}%
\left(  F;F\right)  $ such that :

$\forall X\in O\subset V:X\left(  0\right)  \in\widetilde{A}\subset F$

$\forall t:X\left(  t\right)  =\left(  \exp t\widetilde{S}\right)  \left(
X\left(  0\right)  \right)  \in F$

iii) The maps X are smooth and :

$\frac{d}{ds}X\left(  s\right)  |_{s=t}=\widetilde{S}X\left(  t\right)  $
\end{theorem}

\begin{proof}
i) We have a change of variables U depending on a parameter $\theta\in%
\mathbb{R}
$ which reads with the evaluation map : $\mathcal{E}:%
\mathbb{R}
\times V\rightarrow F::\mathcal{E}\left(  t\right)  X=X\left(  t\right)  $ :

$\forall t,\theta\in%
\mathbb{R}
:\mathcal{E}\left(  t\right)  \left(  U\left(  \theta\right)  X\right)
=\mathcal{E}\left(  t+\theta\right)  \left(  X\right)  \Leftrightarrow
\mathcal{E}\left(  t\right)  U\left(  \theta\right)  =\mathcal{E}\left(
t+\theta\right)  =\mathcal{E}\left(  \theta\right)  U\left(  t\right)  $:

U defines a one parameter group of linear operators:

$U\left(  \theta+\theta^{\prime}\right)  X\left(  t\right)  =X\left(
t+\theta+\theta^{\prime}\right)  =U\left(  \theta\right)  \circ U\left(
\theta^{\prime}\right)  X\left(  t\right)  $

$U\left(  0\right)  X\left(  t\right)  =X\left(  t\right)  $

It is obviously continuous at $\theta=0$ so it is continuous.

ii) The conditions 1 are met, so there are a Hilbert space $H$, a linear chart
$\Upsilon,$ and $\widehat{U}:%
\mathbb{R}
\rightarrow%
\mathcal{L}%
\left(  H;H\right)  $ such that $\widehat{U}\left(  \theta\right)  $ is
linear, bijective, unitary :

$\forall X\in O:\widehat{U}\left(  \theta\right)  \left(  \Upsilon\left(
X\right)  \right)  =\Upsilon\left(  U\left(  \theta\right)  \left(  X\right)
\right)  $

$\widehat{U}\left(  \theta+\theta^{\prime}\right)  =\Upsilon\circ U\left(
\theta+\theta^{\prime}\right)  \circ\Upsilon^{-1}=\Upsilon\circ U\left(
\theta\right)  \circ U\left(  \theta^{\prime}\right)  \circ\Upsilon
^{-1}=\Upsilon\circ U\left(  \theta\right)  \circ\Upsilon^{-1}\circ
\Upsilon\circ U\left(  \theta^{\prime}\right)  \circ\Upsilon^{-1}=\widehat
{U}\left(  \theta\right)  \circ\widehat{U}\left(  \theta^{\prime}\right)  $

$\widehat{U}\left(  0\right)  =\Upsilon\circ U\left(  0\right)  \circ
\Upsilon^{-1}=Id$

The map : $\widehat{U}:%
\mathbb{R}
\rightarrow%
\mathcal{L}%
\left(  H;H\right)  $ is uniformly continuous with respect to $\theta,$ it
defines a one parameter group of unitary operators. So there is an
anti-hermitian operator $\widehat{S}$ with spectral resolution P such that :

$\widehat{U}\left(  \theta\right)  =\sum_{n=0}^{\infty}\frac{\theta^{n}}%
{n!}\widehat{S}^{n}=\exp\theta\widehat{S}$

$\frac{d}{ds}\widehat{U}\left(  s\right)  |_{\theta=s}=\left(  \exp
\theta\widehat{S}\right)  \circ\widehat{S}$

$\widehat{S}=\int_{Sp\left(  S\right)  }sP\left(  s\right)  $

$\left\Vert \widehat{U}\left(  \theta\right)  \right\Vert =1\leq\exp\left\Vert
\theta\widehat{S}\right\Vert $

iii) $S=\Upsilon^{-1}\circ\widehat{S}\circ\Upsilon$ is a continuous map on the
largest vector subspace $V_{0}$ of V which contains O, which is a normed
vector space with the norm induced by the positive kernel.

$\left\Vert S\right\Vert \leq\left\Vert \Upsilon^{-1}\right\Vert \left\Vert
\widehat{S}\right\Vert \left\Vert \Upsilon\right\Vert =\left\Vert \widehat
{S}\right\Vert $ because $\Upsilon$ is an isometry.

So the series $\sum_{n=0}^{\infty}\frac{\theta^{n}}{n!}S^{n}$ converges in
$V_{0}$ and :

$U\left(  \theta\right)  =\Upsilon^{-1}\circ\widehat{U}\left(  \theta\right)
\circ\Upsilon=\sum_{n=0}^{\infty}\frac{\theta^{n}}{n!}S^{n}=\exp\theta S$

$\forall\theta,t\in%
\mathbb{R}
:U\left(  \theta\right)  X\left(  t\right)  =X\left(  t+\theta\right)
=\left(  \exp\theta S\right)  X\left(  t\right)  $

$\mathcal{E}\left(  t\right)  \exp\theta S=\mathcal{E}\left(  t+\theta\right)
$

Exchange $\theta,t$ and take $\theta=0:$

$\mathcal{E}\left(  \theta\right)  \exp tS=\mathcal{E}\left(  t+\theta\right)
$

$\mathcal{E}\left(  0\right)  \exp tS=\mathcal{E}\left(  t\right)  \in%
\mathcal{L}%
\left(  V;E\right)  $

which reads :

$\forall t\in%
\mathbb{R}
:U\left(  t\right)  X\left(  0\right)  =X\left(  t\right)  =\left(  \exp
tS\right)  X\left(  0\right)  $

$\left(  U,V_{0}\right)  $ is a continuous representation of $\left(
\mathbb{R}
,+\right)  ,$ U is smooth and $X$ is smooth :

$\frac{d}{ds}U\left(  s\right)  X\left(  0\right)  |_{s=t}=\frac{d}%
{ds}X\left(  s\right)  |_{s=t}=SX\left(  t\right)  $

$\Leftrightarrow\frac{d}{ds}\mathcal{E}\left(  s\right)  |_{s=t}%
=S\mathcal{E}\left(  t\right)  $

The same result holds whatever the size of O in V, so S is defined over V.

iv) The set : $F\left(  t\right)  =\left\{  X\left(  t\right)  ,X\in
V\right\}  $ is a vector subspace of E.

Each map is fully defined by its value at one point :

$\forall t\in%
\mathbb{R}
:X\left(  t\right)  =\left(  \exp tS\circ X\right)  \left(  0\right)  $

$X\left(  t\right)  =X^{\prime}\left(  t\right)  \Rightarrow\forall
\theta:X\left(  t+\theta\right)  =X^{\prime}\left(  t+\theta\right)
\Leftrightarrow X=X^{\prime}$

So the conditions 25 are met.

$\Theta\left(  t\right)  :F\left(  0\right)  \rightarrow F\left(  t\right)
::\Theta\left(  t\right)  u=\mathcal{E}\left(  t\right)  \circ\mathcal{E}%
\left(  0\right)  ^{-1}u=\mathcal{E}\left(  0\right)  \circ\exp tS\circ
\mathcal{E}\left(  0\right)  ^{-1}u$

The map $\Theta\left(  \theta\right)  :F\rightarrow F$ defines a one parameter
group, so it has an infinitesimal generator $\widetilde{S}\in%
\mathcal{L}%
\left(  F;F\right)  :\Theta\left(  \theta\right)  =\exp\theta\widetilde{S}$
and because $\Theta\left(  \theta\right)  $ is unitary $\widetilde{S}$ is anti-hermitian.

$\frac{d}{ds}\Theta\left(  s\right)  X\left(  0\right)  |_{s=t}=\frac{d}%
{ds}X\left(  s\right)  |_{s=t}=\widetilde{S}X\left(  t\right)  $
\end{proof}

\bigskip

As a consequence such a model is necessarily determinist, and the system is
represented by smooth maps whose evolution is given by a unique operator. It
is clear that the conditions 25 are then met, so this case is actually a
special case of the previous one. Notice that, even if $X$ was not assumed to
be continuous, smoothness is a necessary result. This result can seem
surprising, but actually the basic assumption about a translation in time
means that the laws of evolution are smooth, and as a consequence the
variables depend smoothly on the time. And conversely this implies that,
whenever there is some discontinuity in the evolution of the system, the
conditions above cannot hold : time has a specific meaning, related to a
change in the environment.

\subsection{Comments}

The conditions above depend deeply on how the time is understood in the model.
We have roughly two cases :

A) $t$ is a parameter used only to identify a temporal location. In Galilean
Geometry the time is independent from the spatial coordinates for any observer
and one can consider a change of coordinates such as : $t^{\prime}=t+\theta$
with any constant $\theta.$ The variables $X,X^{\prime}$ such that $X^{\prime
}\left(  t^{\prime}\right)  =X\left(  t+\theta\right)  $ represent the same
system. Similarly in Relativist Geometry the universe can be modelled as a
manifold, and a change of coordinates with affine parameters, $\xi^{\prime
}=\xi+\theta$ with a fixed 4 vector $\theta,$ is a change of charts. The
components of any quantity defined on the tensorial tangent bundle change
according to the jacobian $\left[  \frac{\partial\xi^{\prime}}{\partial\xi
}\right]  $which is the identity, so the corresponding variables represent the
same system. Then we are usually in the conditions of the Theorem
\ref{QMEvolutSchrod}, and this is the basis of the Schr\"{o}dinger equation.

B) $t$ is a parameter used to measure the duration of a phenomenon, usually
the time elapsed since some specific event, and it is clear that the origin of
time matters and the variables $X,X^{\prime}$ such that $X^{\prime}\left(
t^{\prime}\right)  =X\left(  t+\theta\right)  $ do not represent the same
system. This is the case in more specific models, such as in Engineering. The
proposition \ref{QMEvolutSchrod} does not hold, but the proposition
\ref{QMEvolutGen} holds if the model is determinist.

\bigskip

The conditions 25 require at least that all the variables which are deemed
significant are accounted for. As it as been discussed in the previous
chapter, usually probabilist laws appear because some of them are missing. The
Theorem \ref{QMEvolutGen} precises this issue : by denoting the missing
variables $Y$, one needs to enlarge the vector space E, and similarly F. The
map $\Theta\left(  t\right)  $ still exists, but it encompasses the couples
$\left(  X\left(  t\right)  ,Y\left(  t\right)  \right)  .$ The dispersion of
the observed values of $X(t)$ are then imputed to the distribution of the
unknown values $Y(t)$.

\subsection{Observables}

When a system is studied through its evolution, the observables can be
considered from two different points of view :

- in the movie way : the estimation of the parameters is done at the end of
the period considered, from a batch of data corresponding to several times
(which are not necessarily the same for all variables). So this is the map X
which is estimated through an observable $X\rightarrow\Phi\left(  X\right)  $.

- in the picture way : the estimation is done at different times (the same for
all the variables which are measured). So there are the values $X(t)$ which
are estimated. Then the estimation of $X(t)$ is given by $\varphi\left(
X\left(  t\right)  \right)  =\varphi\left(  \mathcal{E}\left(  t\right)
X\right)  ,$ with $\mathcal{\varphi}$ a linear map from E to a finite
dimensional vector space, which usually does not depend on t (the
specification stays the same).

In the best scenario the two methods should give the same result, which reads :

$\varphi\left(  \mathcal{E}\left(  t\right)  X\right)  =\mathcal{E}\left(
t\right)  \left(  \Phi X\right)  \Leftrightarrow\varphi=\mathcal{E}\left(
t\right)  \circ\Phi\circ\mathcal{E}\left(  t\right)  ^{-1}$

But usually, when it is possible, the first way gives a better statistical estimation.

\subsection{Phases Transitions}

There is a large class of problems which involve transitions in the evolution
of a system. They do not involve the maps $X$, which belong to the same family
as above, but the values $X(t)$ which are taken over a period of time in some
vector space E. There are distinct subsets of E, that we will call
\textbf{phases} (to avoid any confusion with states which involves the map
$X$), between which the state of the system goes during its evolution, such as
the transition solid / gas or between magnetic states. The questions which
arise are then : what are the conditions, about the initial conditions or the
maps X, for the occurrence of such an event ? Can we forecast the time at
which such event takes place ?

\bigskip

Staying in the general model meeting the conditions 25, the first issue is the
definition of the phases. The general idea is that they are significantly
different states, and it can be formalized by : the set $\left\{  X(t),t\in
R,X\in O\right\}  $ is disconnected, it comprises two disjoint subsets
$E_{1},E_{2}$ closed in $E.$

If the maps $X:R\rightarrow F$ are continuous and $R$ is an interval of $%
\mathbb{R}
$ (as we will assume) then the image $X(R)$ is connected, the maps $X$ cannot
be continuous, and we cannot be in the conditions of proposition
\ref{QMEvolutSchrod} (a fact which is interesting in itself), but we can be in
the case of proposition \ref{QMEvolutGen}. This is a difficult but also very
common issue : in the real life such discontinuous evolutions are the rule.
However, as we have seen, in the physical world discontinuities happen only at
isolated points : the existence of a singularity is what makes interesting a
change of phase. If the transition points are isolated, there is an open
subset of $R$ which contains each of them, a finite number of them in each
compact subset of $R$, and at most a countable number of transition points. A
given map $X$ is then continuous (with respect to $t$) except in a set of
points $\left(  \theta_{\alpha}\right)  _{\alpha\in A},A\subset%
\mathbb{N}
.$ If $X(0)\in E_{1}$ then the odd transition points $\theta_{2\alpha+1}$ mark
a transition $E_{1}\rightarrow E_{2}$ and the opposite for the even points
$\theta_{2\alpha}.$

If the conditions 25 are met then $\Theta$ is continuous except in $\left(
\theta_{\alpha}\right)  _{\alpha\in A},$ the transition points do not depend
on the initial state $X(0)$, but the phase on each segment does. Then it is
legitimate to assume that there is some probability law which rules the
occurrence of a transition. We will consider two cases.

\bigskip

The simplest assumption is that the probability of the occurrence of a
transition at any time $t$ is constant.\ Then it depends only on the cumulated
lengths of the periods $T_{1}=\sum_{\alpha=0}\left[  \theta_{2\alpha}%
,\theta_{2\alpha+1}\right]  ,T_{2}=\sum_{\alpha=0}\left[  \theta_{2\alpha
+1},\theta_{2\alpha+2}\right]  $ respectively.

Let us assume that $X\left(  0\right)  \in E_{1}$ then the changes
$E_{1}\rightarrow E_{2}$ occur for $t=\theta_{2\alpha+1},$ the probability of
transitions read :

$\Pr\left(  X\left(  t+\varepsilon\right)  \in E_{2}|X\left(  t\right)  \in
E_{1}\right)  =\Pr\left(  \exists\alpha\in%
\mathbb{N}
:t+\varepsilon\in\left[  \theta_{2\alpha+1},\theta_{2\alpha+2}\right]
\right)  $

$=T_{2}/\left(  T_{1}+T_{2}\right)  $

$\Pr\left(  X\left(  t+\varepsilon\right)  \in E_{1}|X\left(  t\right)  \in
E_{2}\right)  =\Pr\left(  \exists\alpha\in%
\mathbb{N}
:t+\varepsilon\in\left[  \theta_{2\alpha},\theta_{2\alpha+1}\right]  \right)
$

$=T_{1}/\left(  T_{1}+T_{2}\right)  $

$\Pr\left(  X\left(  t\right)  \in E_{1}\right)  =T_{1}/\left[  R\right]
;\Pr\left(  X\left(  t\right)  \in E_{2}\right)  =T_{2}/\left[  R\right]  $

The probability of a transition at $t$ is : $T_{2}/\left(  T_{1}+T_{2}\right)
\times T_{1}/\left(  T_{1}+T_{2}\right)  +T_{1}/\left(  T_{1}+T_{2}\right)
\times T_{2}/\left(  T_{1}+T_{2}\right)  =2T_{1}T_{2}/\left(  T_{1}%
+T_{2}\right)  ^{2}$. It does not depend of the initial phase, and depends
only on $\Theta.$ This probability law can be checked from a batch of data
about the values of $T_{1},T_{2}$ for each observed transition.

\bigskip

However usually the probability of a transition depends on the values of the
variables. The phases are themselves characterized by the value of $X(t)$, so
a sensible assumption is that the probability of a transition increases with
the proximity of the other phase . Using the Hilbert space structure of F it
is possible to address practically this case.

If $E_{1},E_{2}$ are \textit{closed convex subsets} of F, which is a Hilbert
space, there is a unique map : $\pi_{1}:F\rightarrow E_{1}$. The vector
$\pi_{1}\left(  x\right)  $ is the unique $y\in E_{1}$ such that $\left\Vert
x-y\right\Vert _{F}$ is minimum. The map $\pi_{1}$ is continuous and $\pi
_{1}^{2}=\pi_{1}$. And similarly for $E_{2}.$

The quantity $r=\left\Vert X\left(  t\right)  -\pi_{1}\left(  X\left(
t\right)  \right)  \right\Vert _{F}+\left\Vert X\left(  t\right)  -\pi
_{2}\left(  X\left(  t\right)  \right)  \right\Vert _{F}=$ the distance to the
other subset than where $X(t)$ lies, so one can assume that the probability of
a transition at $t$ is : $f\left(  r\right)  $ where $f:$ $%
\mathbb{R}
\rightarrow\left[  0,1\right]  $ is a probability density. The probability of
a transition depends only on the state at $t$, but one cannot assume that the
transitions points $\theta_{\alpha}$ do not depend on X.

The result holds if $E_{1},E_{2}$ are closed \textit{vector subspaces} of F
such that $E_{1}\cap E_{2}=\left\{  0\right\}  .$ Then

$X\left(  t\right)  =\pi_{1}\left(  X\left(  t\right)  \right)  +\pi
_{2}\left(  X\left(  t\right)  \right)  $

and $\left\Vert X\left(  t\right)  \right\Vert ^{2}=\left\Vert \pi_{1}\left(
X\left(  t\right)  \right)  \right\Vert ^{2}+\left\Vert \pi_{2}\left(
X\left(  t\right)  \right)  \right\Vert ^{2}$

$\frac{\left\Vert \pi_{1}\left(  X\left(  t\right)  \right)  \right\Vert ^{2}%
}{\left\Vert X\left(  t\right)  \right\Vert ^{2}}$ can be interpreted as the
probability that the system at $t$ is in the phase $E_{1}$.

\bigskip

One important application is forecasting a transition for a given map $X$.
From the measure of $X(t)$ one can compute for each $t$ the quantity
$r(t)=\left\Vert X\left(  t\right)  -\pi_{1}\left(  X\left(  t\right)
\right)  \right\Vert _{F}+\left\Vert X\left(  t\right)  -\pi_{2}\left(
X\left(  t\right)  \right)  \right\Vert _{F}$ and, if we know $f$, we have the
probability of a transition at $t$. The practical problem is then to estimate
$f$ from the measure of $r$ over a past period $[0,T]$. A very simple, non
parametric, estimator can be built when $X$ are maps depending only of $t$
(see J.C.Dutailly \textit{Estimation of the probability of transitions between
phases}). It can be used to forecast the occurrence of events such as earth quakes.

\newpage

\section{INTERACTING\ SYSTEMS}

\subsection{Representation of interacting systems}

In the propositions above no assumption has been done about the interaction
with exterior variables. If the values of some variables are given (for
instance to study the impact of external factors with the system) then they
shall be fully integrated into the set of variables, at the same footing as
the others.

A special case occurs when one considers two systems $S_{1},S_{2}$, which are
similarly represented, meaning that that we have the same kind of variables,
defined as identical mathematical objects and related significance. To account
for the interactions between the two systems the models are of the form :

\bigskip%

\begin{tabular}
[c]{lllllllllll}%
$\ulcorner$ & $S_{1}$ & $\urcorner$ &  &  &  &  &  & $\ulcorner$ & $S_{2}$ &
$\urcorner$\\
$X_{1}$ &  & $Z_{1}$ &  &  &  &  &  & $X_{2}$ &  & $Z_{2}$\\
$V_{1}$ & \multicolumn{1}{c}{$\times$} & $W_{1}$ &  &  &  &  &  & $V_{2}$ &
\multicolumn{1}{c}{$\times$} & $W_{2}$\\
& $\downarrow\Upsilon_{1}$ &  &  &  &  &  &  &  & $\downarrow\Upsilon_{2}$ &
\\
& $\psi_{1}$ &  &  &  &  &  &  &  & $\psi_{2}$ & \\
& $H_{1}$ &  &  &  &  &  &  &  & $H_{2}$ & \\
&  &  &  & $\ulcorner$ & \multicolumn{1}{c}{$S_{1+2}$} & $\urcorner$ &  &  &
& \\
&  &  &  & $X_{1}$ & \multicolumn{1}{c}{} & $X_{2}$ &  &  &  & \\
&  &  &  & $V_{1}$ & \multicolumn{1}{c}{$\times$} & $V_{2}$ &  &  &  & \\
&  &  &  &  & \multicolumn{1}{c}{} &  &  &  &  & \\
&  &  &  & $\psi_{1}$ & \multicolumn{1}{c}{} & $\psi_{2}$ &  &  &  & \\
&  &  &  & $H_{1}$ & \multicolumn{1}{c}{$\times$} & $H_{2}$ &  &  &  &
\end{tabular}

\bigskip

$X_{1},X_{2}$ are the variables (as above $X$ denotes collectively a set of
variables) characteristic of the systems $S_{1},S_{2},$and $Z_{1},Z_{2}$ are
variables representing the interactions. Usually these variables are difficult
to measure and to handle.\ One can consider the system $S_{1+2}$ with the
direct product $X_{1}\times X_{2}$ , but doing so we obviously miss the
interactions $Z_{1},Z_{2}$.

We see now how it is possible to build a simpler model which keeps the
features of $S_{1},S_{2}$ and accounts for their interactions.

\bigskip

We consider the models without interactions (so with only $X_{1},X_{2})$ and
we assume that they meet the conditions 1. For each model $S_{k},k=1,2$ there are

a linear map : $\Upsilon_{k}:V_{k}\rightarrow H_{k}::\Upsilon_{k}\left(
X_{k}\right)  =\psi_{k}=\sum_{i\in I_{k}}\left\langle \phi_{ki},\psi
_{k}\right\rangle e_{ki}$

a positive kernel : $K_{k}:V_{k}\times V_{k}\rightarrow%
\mathbb{R}
$

Let us denote S the new model.\ Its variables will be collectively denoted Y,
valued in a Fr\'{e}chet vector space V'. There will be another Hilbert space
H', and a linear map $\Upsilon^{\prime}:V^{\prime}\rightarrow H^{\prime}%
$\ similarly defined. As we have the choice of the model, we will impose some
properties to Y and V' in order to underline both that they come from
$S_{1},S_{2}$ and that they are interacting.

\bigskip

\begin{condition}
\label{QMTensor}

\textit{i) The variable }$\mathit{Y}$\textit{\ can be deduced from the value
of }$X_{1},X_{2}$\textit{\ : there must be a bilinear map : }$\Phi:V_{1}\times
V_{2}\rightarrow V^{\prime}$

\textit{ii) }$\Phi$\textit{\ must be such that whenever the systems }%
$S_{1},S_{2}$\textit{\ are in the states }$\psi_{1},\psi_{2}$\textit{\ then S
is in the state }$\psi^{\prime}$\textit{\ and}

$\Upsilon^{\prime-1}\left(  \psi^{\prime}\right)  =\Phi\left(  \Upsilon
_{1}^{-1}\left(  \psi_{1}\right)  ,\Upsilon_{2}^{-1}\left(  \psi_{2}\right)
\right)  $

\textit{iii) The positive kernel is a defining feature of the models, so we
want a positive kernel K' of }$\left(  V^{\prime},\Upsilon^{\prime}\right)
$\textit{\ such that :}

$\forall X_{1},X_{1}^{\prime}\in V_{1},\forall X_{2},X_{2}^{\prime}\in V_{2}:
$

$K^{\prime}\left(  \Phi\left(  X_{1},X_{2}\right)  ,\Phi\left(  X_{1}^{\prime
},X_{2}^{\prime}\right)  \right)  =K_{1}\left(  X_{1},X_{1}^{\prime}\right)
\times K_{2}\left(  X_{2},X_{2}^{\prime}\right)  $
\end{condition}

\bigskip

We will prove the following :

\begin{theorem}
Whenever two systems $S_{1},S_{2}$ interact, there is a model S encompassing
the two systems and meeting the conditions \ref{QMTensor} above. It is
obtained by taking the tensor product of the variables specific to
$S_{1},S_{2}.$ Then the Hilbert space of S is the tensorial product of the
Hilbert spaces associated to each system.
\end{theorem}

\begin{proof}
First let us see the consequences of the conditions if they are met.

The map : $\varphi:H_{1}\times H_{2}\rightarrow H^{\prime}::\varphi\left(
\psi_{1},\psi_{2}\right)  =\Phi\left(  \Upsilon_{1}^{-1}\left(  \psi
_{1}\right)  ,\Upsilon_{2}^{-1}\left(  \psi_{2}\right)  \right)  $ is
bilinear. So, by the universal property of the tensorial product, there is a
unique map $\widehat{\varphi}:H_{1}\otimes H_{2}\rightarrow H^{\prime}$ such
that : $\varphi=\widehat{\varphi}\circ\imath$ where $\imath:H_{1}\times
H_{2}\rightarrow H_{1}\otimes H_{2}$ is the tensorial product (Maths.369).

The condition iii) reads :

$\left\langle \Upsilon_{1}\left(  X_{1}\right)  ,\Upsilon_{1}\left(
X_{1}^{\prime}\right)  \right\rangle _{H_{1}}\times\left\langle \Upsilon
_{2}\left(  X_{2}\right)  ,\Upsilon_{2}\left(  X_{2}^{\prime}\right)
\right\rangle _{H_{2}}$

$=\left\langle \left(  \Upsilon^{\prime}\circ\Phi\left(  \Upsilon_{1}\left(
X_{1}\right)  ,\Upsilon_{2}\left(  X_{2}\right)  \right)  ,\Upsilon^{\prime
}\circ\Phi\left(  \Upsilon_{1}\left(  X_{1}^{\prime}\right)  ,\Upsilon
_{2}\left(  X_{2}^{\prime}\right)  \right)  \right)  \right\rangle
_{H^{\prime}}$

$\left\langle \psi_{1},\psi_{1}^{\prime}\right\rangle _{H_{1}}\times
\left\langle \psi_{2},\psi_{2}^{\prime}\right\rangle _{H_{2}}=\left\langle
\varphi\left(  \psi_{1},\psi_{2}\right)  ,\varphi\left(  \psi_{1}^{\prime
},\psi_{2}^{\prime}\right)  \right\rangle _{H^{\prime}}$

$=\left\langle \widehat{\varphi}\left(  \psi_{1}\otimes\psi_{2}\right)
,\widehat{\varphi}\left(  \psi_{1}^{\prime}\otimes\psi_{2}^{\prime}\right)
\right\rangle _{H^{\prime}}$

The scalar products on $H_{1},H_{2}$ extend in a scalar product on
$H_{1}\otimes H_{2},$ endowing the latter with the structure of a Hilbert
space with :

$\left\langle \left(  \psi_{1}\otimes\psi_{2}\right)  ,\left(  \psi
_{1}^{\prime}\otimes\psi_{2}^{\prime}\right)  \right\rangle _{H_{1}\otimes
H_{2}}=\left\langle \psi_{1},\psi_{1}^{\prime}\right\rangle _{H_{1}%
}\left\langle \psi_{2},\psi_{2}^{\prime}\right\rangle _{H_{2}}$

and then the reproducing kernel is the product of the reproducing kernels (Maths.1208).

So we must have : $\left\langle \widehat{\varphi}\left(  \psi_{1}\otimes
\psi_{2}\right)  ,\widehat{\varphi}\left(  \psi_{1}^{\prime}\otimes\psi
_{2}^{\prime}\right)  \right\rangle _{H^{\prime}}=\left\langle \psi_{1}%
\otimes\psi_{2},\psi_{1}^{\prime}\otimes\psi_{2}^{\prime}\right\rangle
_{H_{1}\otimes H_{2}}$ and $\widehat{\varphi}$ must be an isometry :
$H_{1}\otimes H_{2}\rightarrow H^{\prime}$

So by taking $H^{\prime}=H_{1}\otimes H_{2}$ and $V^{\prime}=V_{1}\otimes
V_{2}$ we meet the conditions.
\end{proof}

\bigskip

The conditions above are a bit abstract, but are logical and legitimate in the
view of the Hilbert spaces. They lead to a natural solution, which is not
unique and makes sense only if the systems are defined by similar variables.
The measure of the tensor S can be addressed as before, the observables being
linear maps defined in the tensorial products $V_{1}\otimes V_{2},$
$H_{1}\otimes H_{2}$ and valued in finite dimensional vector subspaces of
these tensor products.

\subsection{Comments}

A key point in this representation is the difference between the simple direct
product : $V_{1}\times V_{2}$ and the tensorial product $V_{1}\otimes V_{2},$
an issue about which there is much confusion.

The knowledge of the states $\left(  X_{1},X_{2}\right)  $ of both systems
requires two vectors of I components each, that is $2\times I$ scalars, and
the knowledge of the state S requires a vector of $I%
{{}^2}%
$ components. So the measure of S requires more data, and brings more
information, because it encompasses all the interactions. Moreover \textit{a
tensor is not necessarily the tensorial product of vectors} (if it is so it is
said to be \textbf{decomposable}), it is the sum of such tensors. There is no
canonical map : $V_{1}\otimes V_{2}\rightarrow V_{1}\times V_{2}.$ So
\textit{there is no simple and unique way to associate two vectors} $\left(
X_{1},X_{2}\right)  $ \textit{to one tensor S}. This seems paradoxical, as one
could imagine that both systems can always be studied, and their states
measured, even if they are interacting. But the simple fact that we consider
interactions means that the measure of the state of one of the system shall
account for the conditions in which the measure is done, so it shall precise
the value of the state of the other system and of the interactions
$Z_{1},Z_{2}$.

If a model is arbitrary, its use must be consistent : if the scientist assumes
that there are interactions, they must be present somewhere in the model, as
variables for the computations as well as data to be collected. They can be
dealt with in two ways. Either we opt for the two systems model, and we have
to introduce the variables $Z_{1},Z_{2}$ representing the interactions, then
we have two separate models as in the first section. The study of their
interactions can be a topic of the models, but this is done in another picture
and requires additional hypotheses about the laws of the interactions. Or, if
we intend to account for both systems and their interactions in a single
model, we need a representation which supports more information that can bring
$V_{1}\times V_{2}.$ The tensorial product is one way to enrich the model,
this is the most economical and, as far as one follows the guidelines
i),ii),iii) above, the only one. The complication in introducing general
tensors is the price that we have to pay to account for the interactions. This
representation does not, in any way, imply anything about \textit{how} the
systems interact, or even if they interact at all (in this case S is always
decomposable). As usual the choice is up to the scientist, based upon how he
envisions the problem at hand. But he has to live with his choice.

This issue is at the root of the paradoxes of entanglement.\ With many
variants it is an experiment which involves two objects, which interact at the
beginning, then are kept separated and non interacting, and eventually one
measures the state of one of the two objects, from which the state of the
other can be deduced with some probability. If we have two objects which
interact at some point, with a significant result because it defines a new
state, and we compare their states, then we must either incorporate the
interactions, or consider that they constitute a single system and use the
tensorial product. The fact that the objects cease to interact at some point
does not matter : they are considered together if we compare their states. The
interactions must be accounted for, one way or another and, when an evolution
is considered, this is the map which represents the whole of the evolution
which is significant, not its value at some time.

A common interpretation of this representation is to single out decomposable
tensors $\Psi=\psi_{1}\otimes\psi_{2}$ , called \textquotedblleft pure
states\textquotedblright, so that actual states would be a superposition of
pure states (a concept popularized by the famous Schr\"{o}dinger's cat). It is
clear that in an interacting system the pure states are an abstraction, which
actually would represent two non interacting systems, so their superposition
is an artificial construction. It can be convenient in simple cases, where the
states of each system can be clearly identified, or in complicated models to
represent quantities which are defined over the whole system as we will see
later. But it does not imply any mysterious feature, notably any probabilist
behavior, for the real systems. A state of the two interacting systems is
represented by a single tensor, and a tensor is not necessarily decomposable,
but it is a sum of decomposable tensors.

\subsection{Homogeneous systems}

The previous result can be extended to N (a number that we will assumed to be
fixed) similar systems (that we will call \textbf{microsystems}), represented
by the same model, interacting together. For each microsystem, identified by a
label $s$, the Hilbert space H and the linear map $\Upsilon$ are the same, the
state S of the total system can be represented as a vector belonging to the
tensorial product $\mathbf{V}_{N}=\otimes_{s=1}^{N}V,$ associated to a tensor
$\Psi$ belonging to the tensorial product $\mathbf{H}_{N}=\otimes_{s=1}^{N}H.$
The linear maps $\Upsilon\in%
\mathcal{L}%
\left(  V;H\right)  $ can be uniquely extended as maps $\Upsilon_{N}\in%
\mathcal{L}%
\left(  \mathbf{V}_{N};\mathbf{H}_{N}\right)  $ such that (Maths.423) :

$\Upsilon_{N}\left(  X_{1}\otimes...\otimes X_{N}\right)  =\Upsilon\left(
X_{1}\right)  \otimes...\otimes\Upsilon\left(  X_{N}\right)  $

The state of the system is then totally defined by the value of tensors
$S,\Psi$, with $I^{N}$ components.

We have general properties on these tensorial products (Maths.1208).

If $\left(  \widetilde{\varepsilon}_{i}\right)  _{i\in I}$ is a Hilbertian
basis of H then $E_{i_{1}...i_{N}}=\widetilde{\varepsilon}_{i_{1}}%
\otimes...\otimes\widetilde{\varepsilon}_{i_{N}}$ is a Hilbertian basis of
$\otimes_{s=1}^{N}H.$ The scalar product is defined by linear extension of

$\left\langle \Psi,\Psi^{\prime}\right\rangle _{\mathbf{H}_{N}}=\left\langle
\psi_{1},\psi_{1}^{\prime}\right\rangle _{H}\times...\times\left\langle
\psi_{N},\psi_{N}^{\prime}\right\rangle _{H}$

for decomposable tensors : $\Psi=\psi_{1}\otimes...\otimes\psi_{N}%
,\Psi^{\prime}=\psi_{1}^{\prime}\otimes...\otimes\psi_{N}^{\prime}.$

The subspaces $\otimes_{s=1}^{p}H\otimes\widetilde{\varepsilon}_{i}%
\otimes_{s=p+2}^{N}H$ are orthogonal and $\otimes_{s=1}^{N}H\simeq\ell
^{2}\left(  I^{N}\right)  $

Any operator on H can be extended on $\otimes_{s=1}^{N}H$ with similar
properties : a self adjoint, unitary or compact operator extends uniquely as a
self adjoint, unitary or compact operator (Maths.1211).

\bigskip

In the general case the label matters : the state $S=X_{1}\otimes...\otimes
X_{N}$ is deemed different from $S=X_{\sigma\left(  1\right)  }\otimes
...\otimes X_{\sigma\left(  N\right)  }$ where $\left(  X_{\sigma\left(
p\right)  }\right)  _{p=1}^{N}$ is a permutation of $\left(  X_{s}\right)
_{s=1}^{N}$. If the microsystems have all the same behavior they are, for the
observer, indistinguishable. Usually the behavior is related to a parameter
analogous to a size, so in such cases the microsystems are assumed to have the
same size. We will say that these interacting systems are homogeneous :

\begin{definition}
\label{QMDefHomSyst}A \textbf{homogeneous system} is a system comprised of a
fixed number N of microsystems, represented in the same model, such that any
permutation of the N microsystems gives the same state of the total system.
\end{definition}

We have the following result :

\begin{proposition}
\label{QMThHomSyst}The states $\Psi$ of homogeneous systems belong to an open
subset of a subspace \textbf{h} of the Hilbert space $\otimes_{s=1}^{N}H $ ,
defined by :

i) a class of conjugacy $\mathfrak{S}\left(  \lambda\right)  $ of the group of
permutations $\mathfrak{S}\left(  N\right)  ,$defined itself by a
decomposition of N in $p$ parts :

$\lambda=\left\{  0\leq n_{p}\leq...\leq n_{1}\leq N,n_{1}+...n_{p}=N\right\}
.$

ii) $p$ distinct vectors $\left(  \widetilde{\varepsilon}_{j}\right)
_{j=1}^{p}$ of a Hermitian basis of $H$ which together define a subspace
$H_{J}$

iii) The space \textbf{h} of tensors representing the states of the system is
then :

either the symmetric tensors belonging to : $\odot_{n_{1}}H_{J}\otimes
\odot_{n_{2}}H_{J}...\otimes\odot_{n_{p}}H_{J}$

or the antisymmetric tensors belonging to : $\wedge_{n_{1}}H_{J}\otimes
\wedge_{n_{2}}H_{J}...\otimes\wedge_{n_{p}}H_{J}$
\end{proposition}

\begin{proof}
i) In the representation of the general system the microsystems are identified
by some label s = 1 ...\ N. An exchange of labels $U(\sigma)$ is a change of
variables, represented by an action of the group of permutations
$\mathfrak{S}\left(  N\right)  $: U is defined uniquely by linear extension of
$U(\sigma)\left(  X_{1}\otimes...\otimes X_{N}\right)  =X_{\sigma\left(
1\right)  }\otimes...\otimes X_{\sigma\left(  N\right)  }$ on decomposable tensors.

We can implement the Theorem \ref{QMGroup} proven previously. The tensors
$\psi$ representing the states of the system belong to a Hilbert space
$\mathbf{H}_{N}\subset\otimes_{s=1}^{N}H$ such that $\left(  \mathbf{H}%
_{N},\widehat{U}\right)  $ is a unitary representation of $\mathfrak{S}\left(
N\right)  $ . Which implies that $\mathbf{H}_{N}$ is invariant by $\widehat
{U}$ . The action of $\widehat{U}$ on $\otimes_{s=1}^{N}H$ is defined uniquely
by linear extension of

$\widehat{U}(\sigma)\left(  \psi_{1}\otimes...\otimes\psi_{N}\right)
=\psi_{\sigma\left(  1\right)  }\otimes...\otimes\psi_{\sigma\left(  N\right)
}$ on decomposable tensors.

$\Psi\in\otimes_{s=1}^{N}H$ reads in a Hilbert basis $\left(  \widetilde
{\varepsilon}_{i}\right)  _{i\in I}$ of $H:$

$\Psi=\sum_{i_{1}...i_{N}\in I}\Psi^{i_{1}...i_{N}}\widetilde{\varepsilon
}_{i_{1}}\otimes...\widetilde{\varepsilon}_{i_{N}}$ and :

$\widehat{U}(\sigma)\Psi=\sum_{i_{1}...i_{N}\in I}\Psi^{i_{1}...i_{N}}%
\widehat{U}(\sigma)\left(  \widetilde{\varepsilon}_{i_{1}}\otimes
...\widetilde{\varepsilon}_{i_{N}}\right)  =\sum_{i_{1}...i_{N}\in I}%
\Psi^{i_{1}...i_{N}}\widetilde{\varepsilon}_{\sigma\left(  i_{1}\right)
}\otimes...\widetilde{\varepsilon}_{\sigma\left(  i_{N}\right)  }$

$=\sum_{i_{1}...i_{N}\in I}\Psi^{\sigma\left(  i_{1}\right)  ...\sigma\left(
i_{N}\right)  }\widetilde{\varepsilon}_{i_{1}}\otimes...\widetilde
{\varepsilon}_{i_{N}}$

$\left\langle \widehat{U}(\sigma)\Psi,\widehat{U}(\sigma)\Psi^{\prime
}\right\rangle =\left\langle \Psi,\Psi^{\prime}\right\rangle $

$\Leftrightarrow\sum_{i_{1}...i_{N}\in I}\Psi^{\sigma\left(  i_{1}\right)
...\sigma\left(  i_{N}\right)  }\Psi^{\prime\sigma\left(  i_{1}\right)
...\sigma\left(  i_{N}\right)  }=\sum_{i_{1}...i_{N}\in I}\Psi^{i_{1}...i_{N}%
}\Psi^{\prime i_{1}...i_{N}}$

The only vector subspaces of $\otimes_{s=1}^{N}H$ which are invariant by
$\widehat{U}$ and on which $\widehat{U}$ is unitary are spaces of symmetric or
antisymmetric tensors :

symmetric : $\Psi^{\sigma\left(  i_{1}\right)  ...\sigma\left(  i_{N}\right)
}=\Psi^{i_{1}...i_{N}}$

antisymmetric : $\Psi^{\sigma\left(  i_{1}\right)  ...\sigma\left(
i_{N}\right)  }=\epsilon\left(  \sigma\right)  \Psi^{i_{1}...i_{N}}$

ii) $\mathfrak{S}\left(  N\right)  $ is a finite, compact group. Its unitary
representations are the sum of orthogonal, finite dimensional, unitary,
irreducible representations (Maths.1948). Let $\mathbf{h}\subset\otimes
_{s=1}^{N}H$ be an irreducible, finite dimensional, representation of
$\widehat{U}.$ Then $\forall\sigma\in\mathfrak{S}\left(  N\right)
:\widehat{U}(\sigma)\mathbf{h\subset h}$

iii) Let J a finite subset of I with $card(J)\geq N$, $H_{J}$ the associated
Hilbert space, $\widehat{Y}_{J}:H\rightarrow H_{J}$ the projection, and
$\widehat{Y}_{J_{N}}=\otimes_{N}\widehat{Y}_{J}$ be the extension of
$\widehat{Y}_{J}$ to $\otimes_{s=1}^{N}H$ :

$\widehat{Y}_{J_{N}}\left(  \sum_{i_{1}...i_{N}\in I}\Psi^{i_{1}...i_{N}%
}\widetilde{\varepsilon}_{i_{1}}\otimes...\widetilde{\varepsilon}_{i_{N}%
}\right)  =\sum_{i_{1}...i_{N}\in J}\Psi^{i_{1}...i_{N}}\widetilde
{\varepsilon}_{i_{1}}\otimes...\widetilde{\varepsilon}_{i_{N}}$

Then :

$\forall\sigma\in\mathfrak{S}\left(  N\right)  :\widehat{U}(\sigma)\widehat
{Y}_{J_{N}}\left(  \sum_{i_{1}...i_{N}\in I}\Psi^{i_{1}...i_{N}}%
\widetilde{\varepsilon}_{i_{1}}\otimes...\widetilde{\varepsilon}_{i_{N}%
}\right)  $

$=\sum_{i_{1}...i_{N}\in J}\Psi^{\sigma\left(  i_{1}\right)  ...\sigma\left(
i_{N}\right)  }\widetilde{\varepsilon}_{i_{1}}\otimes...\widetilde
{\varepsilon}_{i_{N}}=\widehat{Y}_{J_{N}}\widehat{U}(\sigma)\Psi$

So if $\mathbf{h}$ is invariant by $\widehat{U}$ then $\widehat{Y}_{J_{N}%
}\mathbf{h}$ is invariant by $\widehat{U}.$ If $\left(  \mathbf{h,}\widehat
{U}\right)  $ is an irreducible representation then the only invariant
subspace are 0 and $\mathbf{h}$ itself, so necessarily $\mathbf{h\subset
}\widehat{Y}_{J_{N}}\left(  \otimes_{s=1}^{N}H\right)  $ for $card(J)=N.$
Which implies : $\mathbf{h\subset}\otimes_{N}H_{J}$ with $H_{J}=\widehat
{Y}_{J}H$ and $card(J)=N.$

iv) There is a partition of $\mathfrak{S}\left(  N\right)  $ in conjugacy
classes $\mathfrak{S}\left(  \lambda\right)  $ which are subgroups defined by
a decomposition of N in p parts :

$\lambda=\left\{  0\leq n_{p}\leq...\leq n_{1}\leq N,n_{1}+...n_{p}=N\right\}
.$ Notice that there is an order on the sets $\left\{  \lambda\right\}  .$
Each element of a conjugacy class is then defined by a repartition of the
integers $\left\{  1,2,...N\right\}  $ in $p$ subsets of $n_{k}$ items (this
is a Young Tableau) (Maths. 5.2.2). A class of conjugacy is an abelian
subgroup of $\mathfrak{S}\left(  N\right)  $ : its irreducible representations
are unidimensional.

The irreducible representations of $\mathfrak{S}\left(  N\right)  $ are then
defined by a class of conjugacy, and the choice of a vector.

\textbf{h} is a Hilbert space, thus it has a Hilbertian basis, composed of
decomposable tensors which are of the kind $\widetilde{\varepsilon}_{j_{1}%
}\otimes...\otimes\widetilde{\varepsilon}_{j_{N}}$ where $\widetilde
{\varepsilon}_{j_{k}}$ are chosen among the vectors of a Hermitian basis
$\left(  \widetilde{\varepsilon}_{j}\right)  _{j\in J}$ of $H_{J}$

If $\widetilde{\varepsilon}_{j_{1}}\otimes...\otimes\widetilde{\varepsilon
}_{j_{N}}\in H,\forall\sigma\in\mathfrak{S}\left(  N\right)  :\widehat
{U}(\sigma)\widetilde{\varepsilon}_{j_{1}}\otimes...\otimes\widetilde
{\varepsilon}_{j_{N}}=\widetilde{\varepsilon}_{j_{\sigma\left(  1\right)  }%
}\otimes...\otimes\widetilde{\varepsilon}_{j_{\sigma\left(  N\right)  }}%
\in\mathbf{h}$

and because the representation is irreducible the basis of \textbf{h }is
necessarily composed from a set of $p\leq N$ vectors $\widetilde{\varepsilon
}_{j}$ by action of $\widehat{U}(\sigma)$

Conversely : for any Hermitian basis $\left(  \widetilde{\varepsilon}%
_{i}\right)  _{i\in I}$ of H, any subset J of cardinality N of I, any
conjugacy class $\lambda,$ any family of vectors $\left(  \widetilde
{\varepsilon}_{j_{k}}\right)  _{k=1}^{p}$ chosen in $\left(  \widetilde
{\varepsilon}_{i}\right)  _{i\in J}$, the action of $\widehat{U}$ on the
tensor :

$\Psi_{\lambda}=\otimes_{n_{1}}\widetilde{\varepsilon}_{j_{1}}\otimes_{n_{2}%
}\widetilde{\varepsilon}_{j_{2}}...\otimes_{n_{p}}\widetilde{\varepsilon
}_{j_{p}},j_{1}\leq j_{2}..\leq j_{p}$

gives the same tensor if $\sigma\in\mathfrak{S}\left(  \lambda\right)
:\widehat{U}\left(  \sigma\right)  \Psi_{\lambda}=\Psi_{\lambda}$

gives a different tensor if $\sigma\in\mathfrak{S}\left(  \lambda^{c}\right)
$ the conjugacy class complementary to $\mathfrak{S}\left(  \lambda\right)
:\mathfrak{S}\left(  \lambda^{c}\right)  =\complement_{\mathfrak{S}\left(
N\right)  }^{\mathfrak{S}\left(  \lambda\right)  }$

so it provides an irreducible representation by :

$\forall\Psi\in\mathbf{h:}\Psi=\sum_{\sigma\in\mathfrak{S}\left(  \lambda
^{c}\right)  }\Psi^{\sigma}\widehat{U}\left(  \sigma\right)  \left(
\otimes_{n_{1}}\widetilde{\varepsilon}_{j_{1}}\otimes_{n_{2}}\widetilde
{\varepsilon}_{j_{2}}...\otimes_{n_{p}}\widetilde{\varepsilon}_{j_{p}}\right)
$

where the components $\Psi^{\sigma}$ are labeled by the vectors of a basis of
\textbf{h. }The dimension of \textbf{h} his given by the cardinality of
$\mathfrak{S}\left(  \lambda^{c}\right)  $ that is : $\frac{N!}{n_{1}%
!...n_{p}!}.$ All the vector spaces \textbf{h} of the same conjugacy class
(but different vectors $\widetilde{\varepsilon}_{i})$ have the same dimension,
thus they are isomorphic.

v) A basis of \textbf{h }is comprised of tensorial products of N vectors of a
Hilbert basis of H. So we can give the components of the tensors of \textbf{h}
with respect to\textbf{\ }$\otimes_{s=1}^{N}H.$ We have two non equivalent
representation :

By symmetric tensors : \textbf{h} is then\textbf{\ }isomorphic to
$\odot_{n_{1}}H_{J}\otimes\odot_{n_{2}}H_{J}...\otimes\odot_{n_{p}}H_{J}$
where the symmetric tensorial product $\odot$ and the space of n order
symmetric tensor on $H_{J}$ is $\odot_{n}H_{J}$

By antisymmetric tensors : \textbf{h} is then\textbf{\ }isomorphic to
$\wedge_{n_{1}}H_{J}\otimes\wedge_{n_{2}}H_{J}...\otimes\wedge_{n_{p}}H_{J}$
and the space of n order antisymmetric tensor on $H_{J}$ is $\wedge_{n}H_{J}$

The result extends to $V_{N}$ by : $S=\Upsilon_{N}^{-1}\left(  \Psi\right)  $
\end{proof}

\subsection{Remarks}

i) For each choice of a class of conjugacy, and each choice of the vectors
$\left(  \widetilde{\varepsilon}_{j}\right)  _{j=1}^{p}$ which defines
$H_{J},$ we have a different irreducible representation with vector space
\textbf{h. }Different classes of conjugacy gives non equivalent
representations. But different choices of the Hermitian basis $\left(
\widetilde{\varepsilon}_{j}\right)  _{j\in I}$ and the subset J of I, for a
given class of conjugacy, give equivalent representations, and they can be
arbitrary. So, for a given system, the set of states is characterized by a
subset J of N elements in any basis of H, and by a class of conjugacy.

A change of the state of the system can occur either inside the same vector
space \textbf{h}, or between irreducible representations: \textbf{h}%
$\rightarrow$\textbf{h'}.\ As we will see in the next chapters usually the
irreducible representation is fixed by other variables (such that energy) and
a change of irreducible representation implies a discontinuous process. The
states of the total system are quantized by the interactions.

ii) $\otimes_{n_{1}}\widetilde{\varepsilon}_{j_{1}}\otimes_{n_{2}}%
\widetilde{\varepsilon}_{j_{2}}...\otimes_{n_{p}}\widetilde{\varepsilon
}_{j_{p}}$ can be seen as representing a configuration where $n_{k}$
microsystems are in the same state $\widetilde{\varepsilon}_{j_{k}}.$The class
of conjugacy, characterized by the integers $n_{p},$ correspond to the
distribution of the microsystems between fixed states.

iii) If O is a convex subset then S belongs to a convex subset, and the basis
can be chosen such that $\forall\Psi\in\mathbf{h}$ is a linear combination
$\left(  y_{k}\right)  _{k=1}^{q}$ of the generating tensors with $y_{k}%
\in\left[  0,1\right]  ,\sum_{k=1}^{q}y_{k}=1.$ S can then be identified to
the expected value of a random variable which would take one of the value
$\otimes_{n_{1}}X_{1}\otimes_{n_{2}}X_{2}...\otimes_{n_{p}}X_{p},$ which
corresponds to $n_{k}$ microsystems having the state $X_{k}.$ As exposed above
the identification with a probabilist model is formal : there is no random
behavior assumed for the physical system.

iv) In the probabilist picture one can assume that each microsystem behaves
independently, and has a probability $\pi_{j}$ to be in the state represented
by $\widetilde{\varepsilon}_{j}$ and $\sum_{j=1}^{N}\pi_{j}=1.$ Then the
probability that we have $\left(  n_{k}\right)  _{k=1}^{p}$ microstates in the
states $\left(  \widetilde{\varepsilon}_{k}\right)  _{k=1}^{p}$ is $\frac
{N!}{n_{1}!...n_{p}!}\left(  \pi_{j_{1}}\right)  ^{n_{1}}...\left(  \pi
_{j_{p}}\right)  ^{n_{p}}.$

v) The set of symmetric tensor $\odot_{n}H_{J}$ is a closed vector subspace of
$\otimes_{n}H_{J},$ this is a Hilbert space, $\dim\otimes_{n}H_{J}%
=C_{p+n-1}^{p-1}$ with Hilbertian basis $\frac{1}{\sqrt{n!}}\odot_{j\in
J}\widetilde{\varepsilon}_{j}=\frac{1}{\sqrt{n!}}S_{n}\left(  \otimes_{j\in
J}\widetilde{\varepsilon}_{j}\right)  $ where the symmetrizer is :

$S_{n}\left(  \sum_{\left(  i_{1}...i_{n}\right)  }\psi^{i_{1}..i_{n}%
}\widetilde{\varepsilon}_{i_{1}}\otimes..\otimes\widetilde{\varepsilon}%
_{i_{n}}\right)  =\sum_{\left(  i_{1}...i_{n}\right)  }\psi^{i_{1}..i_{n}}%
\sum_{\sigma\in\mathfrak{S}\left(  n\right)  }\widetilde{\varepsilon}%
_{\sigma\left(  1\right)  }\otimes....\widetilde{\varepsilon}_{\sigma\left(
k\right)  }$

A tensor is symmetric iff : $\Psi\in\odot_{n}H_{J}\Leftrightarrow S_{n}\left(
\Psi\right)  =n!\Psi$ (Maths. 7.2.1,13.5.2).

The set of antisymmetric tensor $\wedge_{n}H_{J}$ is a closed vector subspace
of $\otimes_{n}H_{J},$ this is a Hilbert space, $\dim\wedge_{n}H_{J}=C_{p}%
^{n}$ with Hilbertian basis $\frac{1}{\sqrt{n!}}\wedge_{j\in J}\widetilde
{\varepsilon}_{j}=\frac{1}{\sqrt{n!}}A_{n}\left(  \otimes_{j\in J}%
\widetilde{\varepsilon}_{j}\right)  $ with the antisymmetrizer :

$A_{n}\left(  \sum_{\left(  i_{1}...i_{n}\right)  }\psi^{i_{1}..i_{n}%
}\widetilde{\varepsilon}_{i_{1}}\otimes..\otimes\widetilde{\varepsilon}%
_{i_{n}}\right)  =\sum_{\left(  i_{1}...i_{n}\right)  }\psi^{i_{1}..i_{n}}%
\sum_{\sigma\in\mathfrak{S}\left(  n\right)  }\epsilon\left(  \sigma\right)
\widetilde{\varepsilon}_{\sigma\left(  1\right)  }\otimes....\widetilde
{\varepsilon}_{\sigma\left(  k\right)  }$

A tensor is antisymmetric iff : $\Psi\in\wedge_{n}H_{J}\Leftrightarrow
A_{n}\left(  \Psi\right)  =n!\Psi$ (Maths. 7.2.2,13.5.2)

v) for $\theta\in\mathfrak{S}\left(  N\right)  :\widehat{U}(\theta)\Psi$ is
usually different from $\Psi$

\subsection{Global observables of homogeneous systems}

The previous definitions of observables can be extended to homogeneous
systems.\ An observable is defined on the total system, this is a map :
$\Phi:\mathbf{V}_{N}\rightarrow W$ where W is a finite dimensional vector
subspace of $\mathbf{V}_{N}$, but not necessarily a tensorial vector product
of spaces. To $\Phi$ is associated the self-adjoint operator $\widehat{\Phi
}=\Upsilon\circ\Phi\circ\Upsilon^{-1}$ and $H_{\Phi}=\widehat{\Phi}\left(
\otimes_{s=1}^{N}H\right)  \subset\otimes_{s=1}^{N}H.$

\begin{theorem}
Any observable of a homogeneous system is of the form :

$\Phi:\mathbf{V}_{N}\rightarrow W$ where W is generated by vectors
$\Phi_{\lambda}$ associated to each class of conjugacy of $\mathfrak{S}\left(
N\right)  $

The value of $\Phi\left(  X_{1}\otimes...\otimes X_{N}\right)  =\varphi\left(
X_{1},...,X_{N}\right)  \Phi_{\lambda}$ where $\varphi$ is a scalar linear
symmetric map, if the system is in a state corresponding to $\lambda$
\end{theorem}

\begin{proof}
The space W must be invariant by U and $H_{\Phi}$ invariant by $\widehat{U}.$
If the system is in a state belonging to \textbf{h }for a class of conjugacy
$\lambda,$ then $H_{\Phi}=\widehat{\Phi}\mathbf{h}$ and $\left(  \widehat
{\Phi}\mathbf{h,}\widehat{U}\right)  $ is an irreducible representation of the
abelian subgroup $\mathfrak{S}\left(  \lambda\right)  $ corresponding to
$\lambda.$ It is necessarily unidimensional and $\Phi\left(  X_{1}%
\otimes...\otimes X_{N}\right)  $ is proportional to a unique vector.\ The
observable being a linear map, the function $\varphi$ is a linear map of the
components of the tensor.
\end{proof}

There is no way to estimate the state of each microsystem. From a practical
point of view, this is a vector $\gamma=\widehat{\Phi}\left(  \otimes_{n_{1}%
}\widetilde{\varepsilon}_{j_{1}}\otimes_{n_{2}}\widetilde{\varepsilon}_{j_{2}%
}...\otimes_{n_{p}}\widetilde{\varepsilon}_{j_{p}}\right)  $ which is
measured, and from it $\lambda,\left(  \widetilde{\varepsilon}_{j_{k}}\right)
_{k=1}^{p}$ are estimated.

In the probabilist picture the expected value of $\gamma$ is :

$\left\langle \gamma\right\rangle =z\left(  \pi_{1},...,\pi_{N}\right)  $

with

$z\left(  \pi_{1},...,\pi_{N}\right)  $

$=\sum_{\lambda}\frac{N!}{n_{1}!...n_{p}!}\sum_{1\leq j_{1}\leq..\leq
j_{p}\leq N}\left(  \pi_{j_{1}}\right)  ^{n_{1}}...\left(  \pi_{j_{p}}\right)
^{n_{p}}\widehat{\Phi}\left(  \otimes_{n_{1}}\varepsilon_{j_{1}}%
...\otimes_{n_{p}}\varepsilon_{j_{p}}\right)  $

We have a classic statistical problem : estimate the $\pi_{i}$ from a
statistic given by the measure of $\gamma$. If the statistic $\widehat{\Phi}$
is sufficient, meaning that $\pi_{i}$ depends only on $\gamma,$ as F is finite
dimensional whatever the number of microsystems, the Pitman-Koopman-Darmois
theorem tells us that the probability law is exponential, then an estimation
by the maximum likehood gives the principle of Maximum Entropy with entropy :

$E=-\sum_{j=1}^{N}\pi_{j}\ln\pi_{j}$

In the usual interpretation of the probabilist picture, it is assumed that the
state of each microsystem can be measured independently. Then the entropy
$E=-\sum_{j=1}^{N}\pi_{j}\ln\pi_{j}$ can be seen as a measure of the
heterogeneity of the system. And, contrary to a usual idea, the interactions
between the micro-systems do not lead to the homogenization of their states,
but to their quantization : the states are organized according to the classes
of conjugacy.

\subsection{Evolution of homogeneous systems}

The evolution of homogeneous systems raises many interesting issues. The
assumptions are a combination of the previous conditions.

\begin{theorem}
\label{QMThEvHomSyst}For a model representing the evolution of a homogeneous
system comprised of a fixed number N of microsystems s = 1 ...N which are
represented by the same model, with variables $\left(  X_{s}\right)
_{s=1}^{N} $ such that, for each microsystem :

i) the variables $X_{s}$ are maps : $X_{s}::R\rightarrow E$ where R is an open
subset of $%
\mathbb{R}
$ and E a normed vector space, belonging to an open subset O of an infinite
dimensional Fr\'{e}chet space V

ii) $\forall t\in R$ the evaluation map : $\mathcal{E}\left(  t\right)
:O\rightarrow E:\mathcal{E}\left(  t\right)  X_{s}=X_{s}\left(  t\right)  $ is continuous

iii) $\forall t\in R:X_{s}\left(  t\right)  =X_{s}^{\prime}\left(  t\right)
\Rightarrow X_{s}=X_{s}^{\prime}$

There is a map : $S:R\rightarrow\otimes_{N}F$ such that $S(t)$ represents the
state of the system at t. $S(t)$ takes its value in a vector space $f(t)$ such
that $\left(  \mathbf{f}\left(  t\right)  ,\widehat{U}_{F}\right)  $ , where
$\widehat{U}_{F}$ is the permutation on $\otimes_{N}F,$ is an irreducible
representation of $\mathfrak{S}\left(  N\right)  $
\end{theorem}

The crucial point is that the homogeneity is understood as the microsystems
follow the same laws, but at a given time they do not have necessarily the
same state.

\begin{proof}
i) Implement the Theorem 2 for each microsystem : there is a common Hilbert
space $H$ associated to V and a continuous linear map $\Upsilon:V\rightarrow
H::\psi_{s}=\Upsilon\left(  X_{s}\right)  $

ii) Implement the Theorem \ref{QMThHomSyst} on the homogeneous system, that is
for the whole of its evolution. The state of the system is associated to a
tensor $\Psi\in\mathbf{h}$ where \textbf{h }is defined by a Hilbertian basis
$\left(  \widetilde{\varepsilon}_{i}\right)  _{i\in I}$ of H, a finite subset
J of I, a conjugacy class $\lambda$ and a family of p vectors $\left(
\widetilde{\varepsilon}_{j_{k}}\right)  _{k=1}^{p}$ belonging to $\left(
\widetilde{\varepsilon}_{i}\right)  _{i\in J}.$ The vector space \textbf{h}
stays the same whatever t.

iii) Implement the Theorem \ref{QMEvolutGen} on the evolution of each
microsystem : there is a common Hilbert space F, a map : $\widehat
{\mathcal{E}}:R\rightarrow%
\mathcal{L}%
\left(  H;F\right)  $ such that : $\forall X_{s}\in O:\widehat{\mathcal{E}%
}\left(  t\right)  \Upsilon\left(  X_{s}\right)  =X_{s}\left(  t\right)  $ and
$\forall t\in R,$ $\widehat{\mathcal{E}}\left(  t\right)  $ is an isometry

Define $\forall i\in I:\varphi_{i}:R\rightarrow F::\varphi_{i}\left(
t\right)  =\widehat{\mathcal{E}}\left(  t\right)  \widetilde{\varepsilon}_{i}$

iv) $\widehat{\mathcal{E}}\left(  t\right)  $ can be uniquely extended in a
continuous linear map :

$\widehat{\mathcal{E}}_{N}\left(  t\right)  :\otimes_{N}H\rightarrow
\otimes_{N}F$ such that : $\widehat{\mathcal{E}}_{N}\left(  t\right)  \left(
\otimes_{N}\psi_{s}\right)  =\otimes_{N}X_{s}\left(  t\right)  $

$\widehat{\mathcal{E}}_{N}\left(  t\right)  \left(  \otimes_{s=1}%
^{N}\widetilde{\varepsilon}_{i_{s}}\right)  =\otimes_{s=1}^{N}\varphi_{i_{s}%
}\left(  t\right)  $

$\widehat{\mathcal{E}}_{N}\left(  t\right)  $ is an isometry, so $\forall t\in
R:$ $\left\{  \otimes_{s=1}^{N}\varphi_{i_{s}}\left(  t\right)  ,i_{s}\in
I\right\}  $ is a Hilbertian basis of $\otimes_{N}F$

v) Define as the state of the system at t : $S\left(  t\right)  =\widehat
{\mathcal{E}}_{N}\left(  t\right)  \left(  \Psi\right)  \in\otimes_{N}F$

Define : $\forall\sigma\in\mathfrak{S}\left(  N\right)  :\widehat{U}%
_{F}\left(  \sigma\right)  \in%
\mathcal{L}%
\left(  \otimes_{N}F;\otimes_{N}F\right)  $ by linear extension of :
$\widehat{U}_{F}\left(  \sigma\right)  \left(  \otimes_{s=1}^{N}f_{s}\right)
=\otimes_{s=1}^{N}f_{\sigma\left(  s\right)  }$

$\widehat{U}_{F}\left(  \sigma\right)  \left(  \otimes_{s=1}^{N}\varphi
_{i_{s}}\left(  t\right)  \right)  =\otimes_{s=1}^{N}\varphi_{\sigma\left(
i_{s}\right)  }\left(  t\right)  =\widehat{\mathcal{E}}_{N}\left(  t\right)
\widehat{U}\left(  \sigma\right)  \left(  \otimes_{s=1}^{N}\widetilde
{\varepsilon}_{i_{s}}\right)  $

$\forall\Psi\in\mathbf{h:}\Psi=\sum_{\sigma\in\mathfrak{S}\left(  \lambda
^{c}\right)  }\Psi^{\sigma}\widehat{U}\left(  \sigma\right)  \left(
\otimes_{n_{1}}\widetilde{\varepsilon}_{j_{1}}\otimes_{n_{2}}\widetilde
{\varepsilon}_{j_{2}}...\otimes_{n_{p}}\widetilde{\varepsilon}_{j_{p}}\right)
$

$S\left(  t\right)  =\sum_{\sigma\in\mathfrak{S}\left(  \lambda^{c}\right)
}\Psi^{\sigma}\widehat{\mathcal{E}}_{N}\left(  t\right)  \circ\widehat
{U}\left(  \sigma\right)  \left(  \otimes_{n_{1}}\widetilde{\varepsilon
}_{j_{1}}\otimes_{n_{2}}\widetilde{\varepsilon}_{j_{2}}...\otimes_{n_{p}%
}\widetilde{\varepsilon}_{j_{p}}\right)  $

$S\left(  t\right)  =\sum_{\sigma\in\mathfrak{S}\left(  \lambda^{c}\right)
}\Psi^{\sigma}\widehat{U}_{F}\left(  \sigma\right)  \otimes_{n_{1}}%
\varphi_{j_{1}}\left(  t\right)  \otimes_{n_{2}}\varphi_{j_{2}}\left(
t\right)  ...\otimes_{n_{p}}\varphi_{j_{p}}\left(  t\right)  $

$\forall\theta\in\mathfrak{S}\left(  \lambda\right)  :\widehat{U}_{F}\left(
\theta\right)  \left(  \otimes_{n_{1}}\varphi_{j_{1}}\left(  t\right)
\otimes_{n_{2}}\varphi_{j_{2}}\left(  t\right)  ...\otimes_{n_{p}}%
\varphi_{j_{p}}\left(  t\right)  \right)  $

$=\otimes_{n_{1}}\varphi_{j_{1}}\left(  t\right)  \otimes_{n_{2}}%
\varphi_{j_{2}}\left(  t\right)  ...\otimes_{n_{p}}\varphi_{j_{p}}\left(
t\right)  $

$\forall\theta\in\mathfrak{S}\left(  \lambda^{c}\right)  :\widehat{U}%
_{F}\left(  \theta\right)  \left(  \otimes_{n_{1}}\varphi_{j_{1}}\left(
t\right)  \otimes_{n_{2}}\varphi_{j_{2}}\left(  t\right)  ...\otimes_{n_{p}%
}\varphi_{j_{p}}\left(  t\right)  \right)  $

$\neq\left(  \otimes_{n_{1}}\varphi_{j_{1}}\left(  t\right)  \otimes_{n_{2}%
}\varphi_{j_{2}}\left(  t\right)  ...\otimes_{n_{p}}\varphi_{j_{p}}\left(
t\right)  \right)  $

and the tensors are linearly independent

So $\left\{  \widehat{U}_{F}\left(  \sigma\right)  \left(  \otimes_{n_{1}%
}\varphi_{j_{1}}\left(  t\right)  \otimes_{n_{2}}\varphi_{j_{2}}\left(
t\right)  ...\otimes_{n_{p}}\varphi_{j_{p}}\left(  t\right)  \right)
,\sigma\in\mathfrak{S}\left(  \lambda^{c}\right)  \right\}  $ is an
orthonormal basis of

$\mathbf{f}\left(  t\right)  =Span\left\{  \widehat{U}_{F}\left(
\sigma\right)  \left(  \otimes_{n_{1}}\varphi_{j_{1}}\left(  t\right)
\otimes_{n_{2}}\varphi_{j_{2}}\left(  t\right)  ...\otimes_{n_{p}}%
\varphi_{j_{p}}\left(  t\right)  \right)  ,\sigma\in\mathfrak{S}\left(
\lambda^{c}\right)  \right\}  $

$\mathbf{f}\left(  t\right)  =\widehat{\mathcal{E}}_{N}\left(  t\right)
\left(  \mathbf{h}\right)  $

Let $\widetilde{f}\left(  t\right)  \subset\mathbf{f}\left(  t\right)  $ be
any subspace globally invariant by $\left\{  \widehat{U}_{F}\left(
\theta\right)  ,\theta\in\mathfrak{S}\left(  N\right)  \right\}  :\widehat
{U}_{F}\left(  \theta\right)  \widetilde{f}\left(  t\right)  \in\widetilde
{f}\left(  t\right)  $

$\widehat{\mathcal{E}}_{N}\left(  t\right)  $ is an isometry, thus a bijective map

$\widetilde{h}=\widehat{\mathcal{E}}_{N}\left(  t\right)  ^{-1}\widetilde
{f}\left(  t\right)  \Leftrightarrow\widetilde{f}\left(  t\right)
=\widehat{\mathcal{E}}_{N}\left(  t\right)  \widetilde{h}$

$\widehat{U}_{F}\left(  \theta\right)  \widehat{\mathcal{E}}_{N}\left(
t\right)  \widetilde{h}\in\widehat{\mathcal{E}}_{N}\left(  t\right)
\widetilde{h}$

$\forall\Psi\in\mathbf{h:}\widehat{U}_{F}\left(  \theta\right)  \widehat
{\mathcal{E}}_{N}\left(  t\right)  \Psi=\widehat{\mathcal{E}}_{N}\left(
t\right)  \widehat{U}\left(  \theta\right)  \Psi$

$\Rightarrow\widehat{\mathcal{E}}_{N}\left(  t\right)  \widehat{U}\left(
\theta\right)  \widetilde{h}\in\widehat{\mathcal{E}}_{N}\left(  t\right)
\widetilde{h}$

$\Rightarrow\widehat{U}\left(  \theta\right)  \widetilde{h}\in\widetilde{h}$

So $\left(  \mathbf{f}\left(  t\right)  ,\widehat{U}_{F}\right)  $ is an
irreducible representation of $\mathfrak{S}\left(  N\right)  $
\end{proof}

For each $t$ the space $\mathbf{f}\left(  t\right)  $ is defined by a
Hilbertian basis $\left(  f_{i}\right)  _{i\in I}$ of F, a finite subset J of
I, a conjugacy class $\lambda\left(  t\right)  $ and a family of $p$ vectors
$\left(  f_{j_{k}}\left(  t\right)  \right)  _{k=1}^{p}$ belonging to $\left(
f_{i}\right)  _{i\in J}.$ The set J is arbitrary but defined by \textbf{h}, so
it does not depend on $t$\textbf{. }For a given class of conjugacy different
families of vectors $\left(  f_{j_{k}}\left(  t\right)  \right)  _{k=1}^{p}$
generate equivalent representations and isomorphic spaces, by symmetrization
or antisymmetrization.\ So for a given system one can pick up a fixed ordered
family $\left(  f_{j}\right)  _{j=1}^{N}$ of vectors in $\left(  f_{i}\right)
_{i\in I}$ such that for each class of conjugacy $\lambda=\left\{  0\leq
n_{p}\leq...\leq n_{1}\leq N,n_{1}+...n_{p}=N\right\}  $ there is a unique
vector space $\mathbf{f}_{\lambda}$ defined by $\otimes_{n_{1}}f_{1}%
\otimes_{n_{2}}f_{2}...\otimes_{n_{p}}f_{p}.$ Then if $S\left(  t\right)
\in\mathbf{f}_{\lambda}:$

$S\left(  t\right)  =\sum_{\sigma\in\mathfrak{S}\left(  \lambda^{c}\right)
}S^{\sigma}\left(  t\right)  \widehat{U}_{F}\left(  \sigma\right)  \left(
\otimes_{n_{1}}f_{1}\otimes_{n_{2}}f_{2}...\otimes_{n_{p}}f_{p}\right)  $

and at all time $S\left(  t\right)  \in\otimes_{N}F_{J}.$

The vector spaces $\mathbf{f}_{\lambda}$ are orthogonal. With the orthogonal
projection $\pi_{\lambda}$ on $\mathbf{f}_{\lambda}:$

$\forall t\in R:S\left(  t\right)  =\sum_{\lambda}\pi_{\lambda}S\left(
t\right)  $

$\left\Vert S\left(  t\right)  \right\Vert ^{2}=\sum_{\lambda}\left\Vert
\pi_{\lambda}S\left(  t\right)  \right\Vert ^{2}$

The distance between S(t) and a given $\mathbf{f}_{\lambda}$ is well defined
and :

$\left\Vert S\left(  t\right)  -\pi_{\lambda}S\left(  t\right)  \right\Vert
^{2}=\left\Vert S\left(  t\right)  \right\Vert ^{2}-\left\Vert \pi_{\lambda
}S\left(  Ut\right)  \right\Vert ^{2}$

\bigskip

Whenever S, and thus $\Theta,$ is continuous, the space $\mathbf{f}_{\lambda}
$ stays the same. As we have seen previously one can assume that, in all
practical cases, $\Theta$ is continuous but for a countable set $\left\{
t_{k},k=1,2..\right\}  $ of isolated points. Then the different spaces
$\mathbf{f}_{\lambda}$ can be seen as phases, each of them associated with a
class of conjugacy $\lambda$. And there are as many possible phases as classes
of conjugacy. So, in a probabilist picture, one can assume that the
probability for the system to be in a phase $\lambda:\Pr\left(  S\left(
t\right)  \in\mathbf{f}_{\lambda}\right)  $ is a function of $\frac{\left\Vert
\pi_{\lambda}S\left(  t\right)  \right\Vert ^{2}}{\left\Vert S\left(
t\right)  \right\Vert ^{2}}.$ It can be estimated as seen previously from data
on a past period, with the knowledge of both $\lambda$ and $\frac{\left\Vert
\pi_{\lambda}S\left(  t\right)  \right\Vert ^{2}}{\left\Vert S\left(
t\right)  \right\Vert ^{2}}.$

\newpage

\section{CORRESPONDENCE\ WITH\ QUANTUM MECHANICS}

It is useful to compare the results proven in the present paper to the axioms
of QM as they are usually expressed.

\subsection{Hilbert space}

QM : 1. \textit{The states of a physical system can be represented by rays in
a complex Hilbert space H. Rays meaning that two vectors which differ by the
product by a complex number of module 1 shall be considered as representing
the same state.}

\bigskip

In Theorem 2 we have proven that in a model meeting precise conditions the
states of the system can be represented as vectors in an infinite dimensional,
separable, real Hilbert space. We have seen that it is always possible to
endow the Hilbert space with a complex structure, but this is not a necessity.
Moreover the Hilbert space is defined up to an isometry, so notably up to the
product by a fixed complex scalar of module 1.

The state $\psi$ (motion, kinematic and EM charge characteristics) of a
particle can be represented in a fiber bundle, with fiber a complex vector
space E. The gauge group for the EM field is $U(1)$ and the vectors $\psi$ are
defined up to a complex scalar of module 1.\ And this this the origin of rays
(see more in J.C.Dutailly \textquotedblleft Mathematics in Physics").

\bigskip

In Quantum Physics a great attention is given to the Principle of
Superposition. This Principle is equivalent to the condition that the
variables of the system (and then its state) belong to a vector space. There
is a distinction between pure states, which correspond to actual measures, and
mixed states which are linear combination of pure states, usually not actually
observed. There has been a great effort to give a physical meaning to these
mixed states. Here the concept of pure states appears only in the tensors
representing interacting systems, with the usual, but clear, explanation. In
Quantum Mechanics some states of a system cannot be achieved (through a
preparation for instance) as a combination of other states, and then
super-selection rules are required to sort out these specific states. Here
there is a simple explanation : because the set $H_{0}$ is not the whole of
$H$ it can happen that a linear combination of states is not inside $H_{0}.$
The remedy is to enlarge the model to account for other physical phenomena, if
it appears that these states have a physical meaning.

\bigskip

Actually the main difference comes from the precise conditions of the Theorem
2. The variables must be maps, but also belong to a vector space. Thus for
instance it does not apply to the model of a solid body represented by its
trajectory $x(t)$ and its speed $v(t)$ : the variable $x(t)$ is a map : $x:%
\mathbb{R}
\rightarrow M$ valued in a manifold (an affine space in Galilean geometry). So
it is necessary to adapt the model, using the fiber bundle formalism, and this
leads to a deep redefinition of the concept of motion (including rotation) and
to the spinors. And as it has been abundantly said, the state is defined by
maps over the evolution of the system, and not pointwise.

\subsection{Observables}

QM : 2. \textit{To any physical measure }$\Phi$\textit{, called an observable,
which can be done on the system, is associated a continuous, linear,
self-adjoint operator }$\widehat{\Phi}$\textit{\ on H.}

\bigskip

We have proven that this operator is also compact and trace-class. The main
result is that we have here a clear understanding of the concept of
observable, rooted in the practical way the data are analyzed and assigned to
the value of the variables, with the emphasize given to the procedure of
specification, an essential step in any statistical analysis and which is
usually overlooked. From primary observables it is possible to define von
Neumann algebras of operators, which are necessarily commutative when a fixed
basis has been chosen.\ As the choice of a privileged basis can always be
done, one can say that there is always a commutative von Neumann algebra
associated to a system. But, as it can be seen, these von Neumann algebras do
not play any role in the proofs of the theorems. Their introduction can be
useful, but they are not a keystone in our framework.

This is the opposite in the axiomatic interpretations of QM which define the
system itself from the existence of a von Neumann algebra. However such
interpretation is, eventually, based on the same assumption as any other
interpretation of QM : the postulate that for any system there is a quantity,
the state, which has a physical meaning and can be represented in a Hilbert
space. Nothing preclude the choice of a privileged Hilbertian basis for this
Hilbert space (as it is always done in any practical computation in QM), with
respect to which the operators can be defined, and then the algebra is
commutative. Which nullifies the emphasize given to the commutation of
operators.\ Or at least it should be given another interpretation than the
simultaneity of measures.

In QM a great emphasize if given to the commutation of observables, linked to
the physical possibility to measure simultaneously two variables. This concept
does not play any role here, for the strong reason that we consider maps with
a domain over the whole extension, spatial and temporal, of the system, there
is no assumption about how the measures are done, so the simultaneity of
measures is not considered. In our picture the variables and their properties
are the model, they are listed explicitly and it is assumed that there is some
way to estimate their value, without any consideration of the time at which
the measures are done. So the question of simultaneous measures does not
arise, and the product of observables itself has no clear meaning and no use.
If a variable is added, we have another model, the variable gets the same
status as the others, and it is assumed that it can be measured.

\subsection{Measure}

QM : 3. \textit{The result of any physical measure is one of the eigen-values
}$\lambda$\textit{\ of the associated operator }$\widehat{\Phi}. $
\textit{After the measure the system is in the state represented by the
corresponding eigen vector }$\psi_{\lambda}$

\bigskip

This is one of the most puzzling axiom.\ We have here a clear interpretation
of this result, with primary observables, and there is always a primary
observable which is at least as efficient than a secondary observable.

\bigskip

In our picture there is no assumption about how the measures are done, and
particularly if they have or not an impact on the state of the system. If it
is assumed that this is the case, a specific variable should be added to the
model.\ Its value can be measured directly or estimated from the value of the
other variables, but this does not make a difference : it is a variable as the
others. We will see an example in the following chapters.

There is no assumption about the times at which the measures are taken, when
the model represents a process the measures can be taken at the beginning,
during the process, or at the end. The variables which are estimated are maps,
and the estimation of maps requires more than one value of the arguments. The
estimation is done by a statistical method which uses all the available data.
From this point of view our picture is closer to what is done in the
laboratories, than to the idealized vision of simultaneous measures, which
should be taken all together at each time, and would be impossible because of
the perturbation caused by the measure.

Actually the importance granted to the simultaneity of measures, magnified by
Dirac, is somewhat strange. It is also problematic in the Relativist picture.
It is clear that some measures cannot be done, at the atomic scale, without
disturbing the state of the system that is studied, but this does not preclude
to use the corresponding variables in a model, or give them a special status.
Before the invention of radar the artillerymen used efficient models even if
they were not able to measure the speed of their shells. And in a collider it
is assumed that the speed and the location of particles are known when they collide.

\subsection{Probability}

QM : 4. \textit{The probability that the measure is }$\lambda$\textit{\ is
equal to }$\left\vert \left\langle \psi_{\lambda},\psi\right\rangle
\right\vert ^{2}$\textit{\ (with normalized eigen vectors). If a system is in
a state represented by a normalized vector }$\psi$\textit{\ , and an
experiment is done to test whether it is in one of the states }$\left(
\psi_{n}\right)  _{n=1}^{N}$\textit{\ which constitutes an orthonormal set of
vectors, then the probability of finding the system in the state }$\psi_{n}%
$\textit{\ is }$\left\vert \left\langle \psi_{n},\psi\right\rangle \right\vert
^{2}$\textit{\ .}

\bigskip

The first part is addressed by the theorem \ref{QMPRobPrimary}. The second
part has no direct equivalent in our picture but can be interpreted as follows
: a measure of the primary observable has shown that $\psi\in H_{J}$\ , then
the probability that it belongs to $H_{J^{\prime}}$\ for any subset
$J^{\prime}\subset J$\ is $\left\Vert \widehat{Y}_{J^{\prime}}\left(
\psi\right)  \right\Vert ^{2}.\ $It is a computation of conditional
probabilities :

\begin{proof}
The probability that $\psi\in H_{K}$ for any susbset $K\subset I$ is
$\left\Vert \widehat{Y}_{K}\left(  \psi\right)  \right\Vert ^{2}.$ The
probability that $\psi\in H_{J^{\prime}}$ knowing that $\psi\in H_{J}$ is :

$\Pr\left(  \psi\in H_{J^{\prime}}|\psi\in H_{J}\right)  =\frac{\Pr\left(
\psi\in H_{J^{\prime}}\wedge\psi\in H_{J}\right)  }{\Pr\left(  \psi\in
H_{J^{\prime}}|\psi\in H_{J}\right)  }=\frac{\Pr\left(  \psi\in H_{J^{\prime}%
}\right)  }{\Pr\left(  \psi\in H_{J^{\prime}}|\psi\in H_{J}\right)  }%
=\frac{\left\Vert \widehat{Y}_{J^{\prime}}\left(  \psi\right)  \right\Vert
^{2}}{\left\Vert \widehat{Y}_{J}\left(  \psi\right)  \right\Vert ^{2}%
}=\left\Vert \widehat{Y}_{J^{\prime}}\left(  \psi\right)  \right\Vert ^{2}$
because $\widehat{Y}_{J^{\prime}}\left(  \psi\right)  =\psi$ and $\left\Vert
\psi\right\Vert =1$
\end{proof}

Moreover we have seen how the concept of wave functions can be introduced, and
its meaning, for models where the variables are maps defined on the same set.
Of course the possibility to define such a function does not imply that it is
related to a physical phenomenon.

\subsection{Interacting systems}

QM : 5. \textit{When two systems interacts, the vectors representing the
states belong to the tensorial product of the Hilbert states.}

\bigskip

This is the topic of the theorem \ref{QMTensor}. We have seen how it can be
extended to N systems, and the consequences that entails for homogeneous
systems. If the number of microsystems is not fixed, the formalism of Fock
spaces can be used but would require a mathematical apparatus that is beyond
the scope of this book.

There is a fierce debate about the issue of locality in physics, mainly
related to the entanglement of states for interacting particles. It should be
clear that the formal system that we have built is global : more so, it is its
main asset. While most of the physical theories are local, with the tools
which have been presented we can deal with variables which are global, and get
some strong results without many assumptions regarding the local laws.

\subsection{Wigner's theorem}

QM : 6. \textit{If the same state is represented by two rays R, R', then there
is an operator }$\widehat{U}$\textit{, unitary or antiunitary, on the Hilbert
space }$\mathit{H}$\textit{\ such that if the state }$\psi$\textit{\ is in the
ray R then }$\widehat{U}\psi$\textit{\ is in the ray R'.}

\bigskip

This the topic of the theorem \ref{QMChangeVar}. The issue unitary /
antiunitary exists in the usual presentation of QM because of the rays. In our
picture the operator is necessarily unitary, which is actually usually the case.

\subsection{Schr\"{o}dinger equation}

QM : 7. \textit{The vector representing the state of a system which evolves
with time follows the equation} : $i\hbar\frac{\partial\psi}{\partial
t}=\widehat{H}\psi$ where $\widehat{H}$ \textit{is the Hamiltonian of the
system.}

\bigskip

This is actually the topic of the theorem \ref{QMEvolutSchrod} and the result
holds for the variables $X$ in specific conditions, including in the General
Relativity context. The imaginary $i$ does not appear because the Hilbert
space is real. As for Planck's constant of course it cannot appear in a formal
model. However as said before all quantities must be dimensionless, as it is
obvious in the equivalent expression $\psi\left(  t\right)  =\exp\frac
{t}{i\hbar}\widehat{H}\psi\left(  0\right)  .\ $Thus it is necessary either to
involve some constant, or that all quantities (including the time $t$) are
expressed in a universal system of units.\ This is commonly done by using the
Planck's system of units. Which is more important is that the theorems (and
notably the second) precise fairly strong conditions for their validity.\ In
many cases the Schr\"{o}dinger's equation, because of its linearity, seems
\textquotedblleft to good to be true\textquotedblright. We can see why.

\subsection{The scale issue}

The results presented here hold whenever the model meets the conditions 1.\ So
it is valid whatever the scale. But it is clear that the conditions are not
met in many models used in classic physics, notably in Analytic Mechanics (the
variables q are not vectorial quantities). Moreover actually in the other
cases it can often be assumed that the variables belong themselves to Hilbert
spaces. The results about observables and eigen values are then obvious, and
those about the evolution of the system, for interacting systems or for gauge
theories keep all their interest.

The \textquotedblleft Quantic World\textquotedblright, with its strange
properties does not come from specific physical laws, which would appear below
some scale, but from the physical properties of the atomic world
themselves.\ And of course these cannot be addressed in the simple study of
formal models.

\bigskip

So the results presented here, which are purely mathematical, give a
consistent and satisfying explanation of the basic axioms of Quantum
Mechanics, without the need for any exotic assumptions.\ They validate, and in
many ways make simpler and safer, the use of techniques used for many years.
Moreover, as it is easy to check, most of these results do not involve any
physics at all : they hold for any scientific theory which is expressed in a
mathematical formalism. From my point of view they bring a definitive answer
to the issue of the interpretation of QM : the interpretations were sought in
the physical world, but actually there is no such interpretation to be found.
There is no physical interpretation because QM is not a physical theory.

The results presented go beyond the usual axioms of QM : on the conditions to
detect an anomaly, on the quantization of a variable $Y=f(X)$, on the phases
transitions. And other results can probably be found.\ So the method should
give a fresh view of the foundations of QM in Physics.

\bigskip

jc.dutailly@free.fr

\section{BIBLIOGRAPHY}

\bigskip

R.D.Anderson \textit{Some open questions in infinite dimensional topology}
Proceeding of the 3d Prague symposium Praha (1972)

H.Araki \textit{Mathematical theory of quantum fields} Oxford Science
Publications (2000)

J.C. Baez, M.Stay \textit{Physics, Topology, Logic and Computation: A Rosetta
Stone} arXiv 0903.0340 (2009)

H.Baumgartel \textit{Operator Algebraic methods in Quantum Fields} Berlin
Akademie verl. (1995)

M.Le\ Bellac \textit{Physique quantique} CNRS (2003)

A.Berlinet, C.Thomas-Agnan \textit{Reproducing kernel, Hilbert spaces in
probability and statistics} Springer (2004)

A.Bird \textit{Philosophy of science} Rootledge (1998)

J.D.Bjorken,S.D.Drell \textit{Relativistic quantum fields} Mc Graw Hill (1965)

N.Bogolubov,A.A.Logunov, A.I.Ossak, I.T.Todorov \textit{General principles of
quantum fields theory} Kluwer (1990)

O.Bratelli, D.W.Robinson \textit{Operators algebras and quantum statistical
mechanics} Springer (2002)

B.Coecke, E.O.Paquette \textit{Categories for the practising physicist}
arXiv:0905-3010v1 [quant-ph] (16 may 2009)

B.Coecke \textit{New Structures for Physics} Lecture Notes in Physics vol.
813, Springer, Berlin, 2011

B.d'Espagnat \textit{Reality and the physicist} Cambridge University Press (1989)

P.A.M.Dirac \textit{The principles of Quantum Mechanics} Oxford Science
Publications (1958)

J.C.Dutailly \textit{Mathematics for theoretical physics} arXiv:1209-5665v2
[math-ph] (4 feb 2014)

J.C.Dutailly \textit{Estimation of the probability of transition between
phases} CNRS (http://hal.archives-ouvertes.fr/hal-01075940, 20 october 2014)

J.C.Dutailly \textit{Mathematics in Physics} CNRS
(http://hal.archives-ouvertes.fr/hal-01169985, 29 juin 2015)

R.P.Feynman, A.R.Hibbs \textit{Quantum Mechanics and Path Integrals} Dover (2005)

J.Finne \textit{Asymptotic study of canonical correlation analysis: from
matrix and analytic approach to operator and tensor approach }SORT 27 (2)
July-December 2003, 165-174

Francis C.E.H \textit{A construction of full QED\ using finite dimensional
Hilbert space} EJTP 10 N$%
{{}^\circ}%
$28 (2013)

Francis C.E.H \textit{The Hilbert space of conditional clauses}
arXiv:1205-4607 (2013)

Tepper L.Gill, G.R.Pantsulaia, W.W.Zachary \textit{Constructive analysis in
infinitely many variables} arXiv 1206-1764v2 [math-FA] (26 june 2012)

H.Halvorson \textit{Algebraic quantum fields theory} arXiv:math-ph/0602036v1
14 feb 2006

N.R. Hansen \textit{Non-parametric likelihood based estimation of linear
filters for point processes }arXiv:1304-0503v3 [stat:CO] (12 feb 2014)

D.W.Henderson \textit{Infinite dimensional manifolds are open subsets of
Hilbert spaces} (1969) Internet paper

S.S.Horuzhy \textit{Introduction to algebraic quantum field theory} Riedel (1989)

J.M.Jauch \textit{Foundations of Quantum Mechanics} AddisonWesley (1968)

Sir M.Kendall, A.Stuart \textit{The advanced theory of statistics} Charles
Griffin \& Co (1977)

A.W.Knapp \textit{Lie groups beyond an introduction} Birkh\"{a}user (2005)

F.Lalo\"{e} \textit{Comprenons-nous vraiment la m\'{e}canique quantique ?}
CNRS\ Editions (2011)

E.H.Lieb, M.Loss \textit{Analysis} American Mathematical Society (2000)

R.Haag \textit{Local quantum physics} 2nd Ed.Springer (1991)

G.Mackey \textit{The mathematical fundations of Quantum Mechanics}
W.A.Benjamin (1963)

J.von Neumann \textit{Mathematical Foundations of Quantum Mechanics}, Beyer,
R. T., trans., Princeton Univ. Press. (1996 edition)

R.Omn\`{e}s \textit{The interpretation of quantum mechanics} Princeton (1994)

J. E. Palomar Tarancon \textit{Conceptual systems, conceptual convergence
theory and and algebras of analogies}

K.Popper \textit{Quantum theory and the schism in physics} Routledge (1982)

K.Popper \textit{The logic of scientific discovery} Rootledge (1959)

I.Schnaid \textit{Wave function perturbations propagation in multi particles
system} arXiv 1307.2510v1 [physics-gen.ph] (9 july 2013)

A.Smola, A.Gretton, L.Song, B.Sch\"{o}lkop \textit{A Hilbert Space Embedding
for Distributions}

H.Torunczyk \textit{Characterizing Hilbert spaces topology} Fundamental
mathematica (1981)

A. Vourdas \textit{The complete Heyting algebra of subsystems and
contextuality }arXiv:1310-3604v1 [quant-ph] (14 oct 2013)

S.Weinberg \textit{The quantum theory of fields} Cambridge University Press (1995)

S.Weinberg \textit{Dreams of a Final Theory}\ Pantheon Books (1992)

H.Weyl \textit{The theory of groups and quantum mechanics} Dover (1931 / 1950)

\end{document}